\documentclass[traditabstract]{aa}

\usepackage{natbib}
\usepackage{graphicx}
\usepackage{txfonts}
\usepackage[bookmarks=false, colorlinks=true, citecolor=blue, linkcolor=blue]{hyperref}
\usepackage{subfig}
\usepackage{blindtext}
\usepackage{lipsum}
\usepackage{pdflscape}
\usepackage{afterpage}

\def\micron{\hbox{\,$\mu$m}}
\newcommand{\Lsun}{\hbox{$L_{\rm \odot}$}}
\newcommand{\Msun}{\hbox{$M_{\rm \odot}$}}

\newcommand{\degree}{\ensuremath{^\circ}}
\newcommand\nodata{ ~$\cdots$~ }

\bibpunct{(}{)}{;}{a}{}{,}

\DeclareCaptionListOfFormat{subreformat}{#1#2}
\titlerunning{PUMA II. The 220\,GHz continuum of local ULIRGs}
\authorrunning{Pereira-Santaella et al.}

\begin{document}

\title{Are local ULIRGs powered by AGN? The sub-kpc view of the 220\,GHz continuum. PUMA II}

\author{M.~Pereira-Santaella\inst{\ref{inst1}}
\and L.~Colina\inst{\ref{inst1}} \and S.~Garc\'ia-Burillo\inst{\ref{inst2}} \and I.~Lamperti\inst{\ref{inst1}} \and E.~Gonz\'alez-Alfonso\inst{\ref{inst3}} \and M.~Perna\inst{\ref{inst1},\ref{inst4}} \and S.~Arribas\inst{\ref{inst1}}  \and A.~Alonso-Herrero\inst{\ref{inst5}} \and S.~Aalto\inst{\ref{inst6}} \and F.~Combes\inst{\ref{inst7}} \and A.~Labiano\inst{\ref{inst5}} \and J.~Piqueras-L\'opez\inst{\ref{inst1}} \and D.~Rigopoulou\inst{\ref{inst8}} \and P.~van der Werf\inst{\ref{inst9}}}
\institute{Centro de Astrobiolog\'ia (CSIC-INTA), Ctra. de Ajalvir, Km 4, 28850, Torrej\'on de Ardoz, Madrid, Spain
\\ \email{miguel.pereira@cab.inta-csic.es}\label{inst1}
\and 
Observatorio Astron\'omico Nacional (OAN-IGN)-Observatorio de Madrid, Alfonso XII, 3, 28014, Madrid, Spain\label{inst2}
\and
Universidad de Alcal\'a, Departamento de F\'isica y Matem\'aticas, Campus Universitario, 28871 Alcal\'a de Henares, Madrid, Spain\label{inst3}
\and
INAF - Osservatorio Astrofisico di Arcetri, Largo Enrico Fermi 5, I-50125 Firenze, Italy\label{inst4}
\and
Centro de Astrobiolog\'{\i}a (CSIC-INTA), ESAC  Campus, E-28692 Villanueva de la Ca\~nada, Madrid, Spain\label{inst5}
\and
Department of Space, Earth and Environment, Onsala Space Observatory, Chalmers University of Technology, 439 92 Onsala, Sweden\label{inst6}
\and
LERMA, Obs. de Paris, PSL Univ., Coll\'ege de France, CNRS, Sorbonne Univ., Paris, France\label{inst7}
\and
Department of Physics, University of Oxford, Keble Road, Oxford OX1 3RH, UK\label{inst8}
\and
Leiden Observatory, Leiden University, PO Box 9513, 2300, RA Leiden, The Netherlands\label{inst9}
}

\abstract{We analyze new high-resolution (400\,pc) $\sim$220\,GHz continuum and CO(2--1) ALMA observations of a representative sample of {23} local (z$<$0.165) ULIRG systems ({34} individual nuclei) as part of the ``Physics of ULIRGs with MUSE and ALMA'' (PUMA) project. The deconvolved half-light radii of the $\sim$220\,GHz continuum sources, $r_{\rm cont}$, are between $<$60\,pc and 350\,pc (median 80--100\,pc). We associate these regions with the regions emitting the bulk of the infrared luminosity ($L_{\rm IR}$). The good agreement, within a factor of 2, between the observed $\sim$220\,GHz fluxes and the extrapolation of the infrared gray-body, and the small contributions from synchrotron and free-free emission support this assumption.
The cold molecular gas emission sizes, $r_{\rm CO}$, are between 60 and 700\,pc and are {similar in advanced mergers and early interacting systems. On average, $r_{\rm CO}$ are {$\sim$2.5} times larger than $r_{\rm cont}$.} Using these measurements, we derive the nuclear $L_{\rm IR}$ and cold molecular gas surface densities ($\Sigma_{L_{\rm IR}}=10^{11.5}-10^{14.3}$\,$L_\odot$\,kpc$^{-2}$ and $\Sigma_{\rm H_2}=10^{2.9}-10^{4.2}$\,$M_\odot$\,pc$^{-2}$, respectively). Assuming that the $L_{\rm IR}$ is produced by star-formation, the median $\Sigma_{L_{\rm IR}}$ corresponds to $\Sigma_{\rm SFR}=2500$\,$M_\odot$\,yr$^{-1}$\,kpc$^{-2}$. This $\Sigma_{\rm SFR}$ implies extremely short depletion times, $\Sigma_{\rm H_2}$\slash $\Sigma_{\rm SFR}<$1--15\,Myr, and unphysical star-formation efficiencies $>$1 for 70\% of the sample. Therefore, this favors the presence of an obscured AGN in these objects that could dominate the $L_{\rm IR}$.
We also classify the ULIRG nuclei in two groups: (a) compact nuclei ($r_{\rm cont}<$120\,pc) with high mid-IR excess emission ($\Delta L_{\rm 6-20\mu m}\slash L_{\rm IR}$) found in optically classified AGN; and (b) nuclei following a relation with decreasing $\Delta L_{\rm 6-20\mu m}\slash L_{\rm IR}$ for decreasing $r_{\rm cont}$.
The majority, {60\%}, of the nuclei in interacting systems lie in the low-$r_{\rm cont}$ end ($<$120\,pc) of this relation, while only {30\%} of the mergers do so. This suggests that in the early stages of the interaction, the activity occurs in a very compact and dust-obscured region while, in more advanced merger stages, the activity is more extended, unless an optically detected AGN is present.
Approximately two thirds of the nuclei have nuclear radiation pressures above the Eddington limit. This is consistent with the ubiquitous detection of massive outflows in local ULIRGs and supports the importance of the radiation pressure in the outflow launching process.}


\maketitle

\section{Introduction}\label{s:intro}

Ultraluminous infrared galaxies (ULIRGs; $L_{\rm IR}>10^{12}$\,\Lsun) are among the most luminous objects in the local Universe. The majority of local ULIRGs are major gas-rich mergers at different evolutionary stages: from interacting systems with two nuclei separated by few kpc to more advanced mergers with a single nucleus (see \citealt{Lonsdale2006} and references therein). A classic evolutionary scenario suggests that merging ULIRGs evolve into a quasar that quenches the star-formation (SF) and, after that, the merger remnant becomes an intermediate-mass elliptical galaxy (e.g., \citealt{Sanders1988, Springel2005}). However, recent observations and simulations indicate that mergers do not always quench the SF and also that disks can regrow in mergers remnants (e.g., \citealt{Ueda2014, Weigel2017, Weinberger2018}). This suggests that ULIRGs can have more varied evolutionary paths than that suggested by the classic scenario.
{Local ULIRGs might also be scaled down versions of the dusty star-forming mergers detected at z$>$2 (e.g., \citealt{Casey2014}).} Therefore, local ULIRGs are excellent targets for detailed studies of the physical processes that shape the, possibly diverse, evolutionary paths of merging gas-rich galaxies, which were important in the high-z universe.

One key property of local ULIRGs are their extremely compact ($<$1\,kpc; e.g., \citealt{Condon1991, Soifer2000}) and dust obscured nuclei ($A_{\rm v}$>1000\,mag in some cases based on their nuclear molecular gas column densities; e.g., \citealt{GonzalezAlfonso2015}). Because of this extreme obscuration most of the radiation produced in their nuclei, either by an active galactic nucleus (AGN) or by SF, is absorbed by dust and re-emitted in the infrared (IR) spectral range. For this reason, it is not straightforward to determine the dominant power source (AGN vs. SF) of local ULIRGs. Mid-IR studies, which are less affected by extinction than optical and near-IR works, suggest that ULIRGs are mostly powered by SF (e.g., \citealt{Genzel1998}), although the AGN contribution increases with increasing luminosities (e.g., \citealt{Nardini2008, Veilleux2009}). However, for $A_{\rm v}$>1000\,mag, the mid-IR extinction is still very high, $A_{\rm 15\mu m}$>50\,mag (e.g., \citealt{Jiang2006}), and a large part of the emission could be completely obscured even in the mid-IR which. {This} could prevent an accurate determination of the AGN and SF contributions to the $L_{\rm IR}$ of local ULIRGs {using mid-IR observations}.

Alternatively, it is possible to investigate what powers local ULIRGs by measuring the size of the region that emits the bulk of the $L_{\rm IR}$ as well as the molecular gas content (i.e., the fuel for SF) of this region. These quantities are needed to determine the nuclear IR luminosity and gas surface densities. Finding IR luminosity densities well above the limit of a maximal starburst (e.g., \citealt{Thompson2005}) can be used to infer the presence of an obscured AGN and to estimate its luminosity {(see e.g., \citealt{Downes2007, Imanishi2011, Sakamoto2017})}.

In this paper, we analyze high-resolution ($\sim$400\,pc) ALMA CO(2--1) and $\sim$220\,GHz continuum observations of sample of {23} local ULIRGs. We measure the size of the $\sim$220\,GHz ($\sim$1400\,\micron) continuum, which we link with the bulk of the $L_{\rm IR}$ in these sources, and we estimate the nuclear cold molecular gas content from the CO(2--1) emission. We use these results to calculate their nuclear luminosity and molecular gas densities.

These ALMA observations of a representative sample of local ULIRGs are part of the ``Physics of ULIRGs with MUSE and ALMA'' (PUMA) project. The main goals of this project are: a) to establish the impact of massive outflows in the evolution of ULIRGs (negative and positive feedback); and b) to determine what drives this feedback (AGN vs. SF) during the entire merging process (from early stages to advanced mergers).
To do so, we combine sub-kpc resolution {adaptive optics-assisted} VLT\slash MUSE {optical integral field spectroscopy} and {CO(2--1)} ALMA data to trace the multi-phase structure of the massive outflows as well as to investigate basic properties of the ULIRGs like their main power source. The first MUSE results on the spatially resolved stellar kinematics and the ionized outflow phase were presented by \citet{Perna2021} while the detailed analysis of the Arp~220 MUSE data was presented in \citet{Perna2020}. Likewise, \citet{Pereira2018} presented the first ALMA results on the spatially resolved cold molecular outflows detected in three of these local ULIRGs.

The paper is organized as follows. We briefly describe the PUMA sample in Sect.~\ref{s:sample}. The ALMA observations and data reduction are presented in Sect.~\ref{s:data}.
In Sect.~\ref{s:data_analysis}, we derive the spatial properties of the $\sim$220\,GHz continuum and the CO(2--1) emission, and fit the IR and radio spectral energy distributions (SEDs) of the ULIRGs. 
Sect.~\ref{s:discussion} investigates the origin of the high luminosity and molecular gas surface densities in the ULIRG nuclei,  the relation between the $\sim$220\,GHz continuum size with the mid-IR excess emission, the 9.7\micron\ silicate absorption, and the broad-band {IRAS} colors. We also estimate the radiation pressure in these nuclei. The main conclusions are summarized in Sect.~\ref{s:conclusions}.

Throughout this article we assume the following cosmology: $H_{\rm 0} = 70$\,km\,s$^{-1}$\,Mpc$^{-1}$, $\Omega_{\rm m}=0.3$, and $\Omega_{\rm \Lambda}=0.7$.

\begin{table*}[tp]
\caption{Sample of local ULIRGs}
\label{tbl:sample}
\centering
\begin{small}
\begin{tabular}{lcccccccccc}
\hline \hline
\\
IRAS name & Nucleus & R.A.\tablefootmark{a} & Dec.\tablefootmark{a} & v$_{\rm CO}$\tablefootmark{b} & z\,\tablefootmark{c} & $d_{\rm L}$\tablefootmark{d} & Scale\,\tablefootmark{d} & $\log L_{\rm IR}$\tablefootmark{e} & Class.\tablefootmark{f} & Morph.\tablefootmark{g} \\
& &  (ICRS) & (ICRS) & (km\,s$^{-1}$) &  & (Mpc) & (kpc\,arcsec$^{-1}$) & ($L_\odot$) \\
\hline
00091$-$0738 & & & & & 0.1181 & 550 & 2.13 & 12.34 & HII & I \\
 & S & 00 11 43.272  & $-$07 22 07.35 & 31637 &  &  &  &  & \nodata &  \\
 & N & 00 11 43.302  & $-$07 22 06.18 & 31686 &  &  &  &  & \nodata &  \\
\hline
00188$-$0856 & - & 00 21 26.513  & $-$08 39 25.99 & 34136 & 0.1285 & 602 & 2.29 & 12.42 & Sy2 & M \\
\hline
00509+1225 & - & 00 53 34.934  & $+$12 41 35.94 & 17265 & 0.0611 & 273 & 1.18 & 11.87 & Sy1 & M \\
\hline
01572+0009 & - & 01 59 50.251  & $+$00 23 40.88 & 42077 & 0.1633 & 782 & 2.80 & 12.65 & Sy1 & M \\
\hline
F05189$-$2524 & - & 05 21 01.400  & $-$25 21 45.30 & 12285 & 0.0427 & 188 & 0.84 & 12.10 & Sy2 & M \\
\hline
07251$-$0248 & & & & & 0.0878 & 400 & 1.64 & 12.45 &  & I \\
 & W & 07 27 37.532  & $-$02 54 54.38 & 24201 &  &  &  &  & HII &  \\
 & E & 07 27 37.613  & $-$02 54 54.25 & 24193 &  &  &  &  & HII &  \\
\hline
09022$-$3615 & - & 09 04 12.706  & $-$36 27 01.93 & 16856 & 0.0596 & 266 & 1.15 & 12.33 & HII & M \\
\hline
F10190$+$1322 & & & & & 0.0763 & 345 & 1.45 & 12.04 &  & I \\
 & W & 10 21 42.493  & $+$13 06 53.83 & 21336 &  &  &  &  & HII &  \\
 & E & 10 21 42.754  & $+$13 06 55.61 & 21167 &  &  &  &  & HII &  \\
\hline
11095$-$0238 & & & & & 0.1064 & 491 & 1.95 & 12.33 &  & I \\
 & SW & 11 12 03.359  & $-$02 54 23.29 & 28806 &  &  &  &  & LINER &  \\
 & NE & 11 12 03.383  & $-$02 54 22.94 & 28863 &  &  &  &  & LINER &  \\
\hline
F12072$-$0444 &  &  & & & 0.1284 & 601 & 2.29 & 12.48 & & I \\
 & S & 12 09 45.13 & $-$05 01 14.6$^\dagger$ & & & & & & Sy2 & \\
 & N & 12 09 45.13 & $-$05 01 13.5$^\dagger$ & & & & & & Sy2 & \\
\hline
F12112$+$0305 & & & & & 0.0730 & 329 & 1.39 & 12.32 & LINER & I \\
 & SW & 12 13 45.940  & $+$02 48 39.12 & 20448 &  &  &  &  & \nodata &  \\
 & NE & 12 13 46.057  & $+$02 48 41.55 & 20322 &  &  &  &  & \nodata &  \\
\hline
13120$-$5453 & - & 13 15 06.323  & $-$55 09 22.82 & 9046 & 0.0311 & 136 & 0.62 & 12.27 & Sy2 & M \\
\hline
F13451$+$1232 &  &  &  & & 0.1217 & 568 & 2.19 & 12.31 & & I \\
 & W & 13 47 33.36 & $+$12 17 24.2$^\dagger$ & & & & & & Sy2 & \\
 & E & 13 47 33.50 & $+$12 17 23.8$^\dagger$ & & & & & & LINER & \\
\hline
F14348$-$1447 & & & & & 0.0826 & 375 & 1.55 & 12.41 &  & I \\
 & SW & 14 37 38.281  & $-$15 00 24.23 & 22911 &  &  &  &  & LINER &  \\
 & NE & 14 37 38.397  & $-$15 00 21.27 & 22809 &  &  &  &  & LINER &  \\
\hline
F14378$-$3651 & - & 14 40 59.013  & $-$37 04 31.93 & 19113 & 0.0681 & 306 & 1.30 & 12.15 & Sy2 & M \\
\hline
F15327+2340 & & & & & 0.0181 & 78 & 0.37 & 12.19 &  & M \\
 & W & 15 34 57.224  & $+$23 30 11.44 & 5290 &  &  &  &  & LINER &  \\
 & E & 15 34 57.293  & $+$23 30 11.29 & 5380 &  &  &  &  & LINER &  \\
\hline
16090$-$0139 & - & 16 11 40.419  & $-$01 47 06.35 & 35352 & 0.1337 & 629 & 2.37 & 12.62 & HII & M \\
\hline
16155$+$0146 & & & & & 0.1330 & 625 & 2.36 & 12.24 &  & I \\
 & NW & 16 18 09.364  & $+$01 39 21.75 & 35188 &  &  &  &  & Sy2 &  \\
 & SE & 16 18 09.54  & $+$01 39 19.7$^\dagger$ & \nodata &  &  &  &  & \nodata &  \\
\hline
17208$-$0014 & - & 17 23 21.957  & $-$00 17 00.88 & 12304 & 0.0428 & 189 & 0.84 & 12.43 & LINER & M \\
\hline
F19297$-$0406 & & & & & 0.0856 & 390 & 1.61 & 12.45 &  & I \\
 & S & 19 32 22.30  & $-$04 00 01.8$^\dagger$ & 23685 &  &  &  &  & HII &  \\
 & N & 19 32 22.309  & $-$04 00 01.03 & 23589 &  &  &  &  & HII &  \\
\hline
19542$+$1110 & - & 19 56 35.785  & $+$11 19 05.03 & 17629 & 0.0625 & 280 & 1.20 & 12.09 & LINER & M \\
\hline
20087$-$0308 & - & 20 11 23.866  & $-$02 59 50.72 & 28600 & 0.1055 & 487 & 1.93 & 12.47 & LINER & M \\
\hline
20100$-$4156 & & & & & 0.1298 & 609 & 2.31 & 12.66 &  & I \\
 & NW & 20 13 29.48  & $-$41 47 32.6$^\dagger$ & 34428 &  &  &  &  & HII &  \\
 & SE & 20 13 29.556  & $-$41 47 35.21 & 34453 &  &  &  &  & LINER &  \\
\hline
20414$-$1651 & - & 20 44 18.159  & $-$16 40 16.82 & 23962 & 0.0869 & 396 & 1.63 & 12.24 & HII & M \\
\hline
F22491$-$1808 & & & & & 0.0776 & 352 & 1.47 & 12.23 &  & I \\
 & W & 22 51 49.24  & $-$17 52 23.7$^\dagger$ & \nodata &  &  &  &  & HII &  \\
 & E & 22 51 49.349  & $-$17 52 24.13 & 21600 &  &  &  &  & HII &  \\
\hline
\end{tabular}
\end{small}
\tablefoot{
\tablefoottext{a}{Coordinates of the $\sim$220--250\,GHz continuum detected by ALMA for each nucleus (see Sect.\,\ref{ss:alma_model}). The typical astrometric uncertainty is 25\,mas.}
\tablefoottext{$^\dagger$}{For the 4 nuclei undetected in the ALMA images and the 2 systems (F12072-0444 and F13451+1232) not observed by ALMA, we used near-IR and optical \textit{HST} images, whose astrometry was tied to Gaia DR2, to measure the nuclear position (see Sect.~3.1 of \citealt{Perna2021}).}
\tablefoottext{b}{CO(2--1) velocity of the nucleus using the radio definition in the kinematic local standard of rest (Lamperti et al. in prep.).}
\tablefoottext{c}{Redshift using the average velocity of the system.}
\tablefoottext{d}{Luminosity distance and scale for the assumed cosmology (see Sect.~\ref{s:intro}).}
\tablefoottext{e}{6--1500\micron\ IR luminosity derived from the SED fit. The typical uncertainty is 0.03\,dex (see Sect.~\ref{ss:sed_model}).}
\tablefoottext{f}{Nuclear activity classification based on optical spectroscopy (see \citealt{Perna2021}).}
\tablefoottext{g}{System morphology: I. Interacting system with nuclear separation $>$1\,kpc; M. Advanced merger with nuclear separation $<$1\,kpc (see \citealt{Perna2021}).\\[5mm]}
}
\end{table*}

\begin{table*}[t]
\caption{Summary of the continuum ALMA observations}
\label{tbl:alma_obs}
\centering
\begin{small}
\begin{tabular}{lcccccccc}
\hline \hline
\\
IRAS name & Synthesized beam & Beam FWHM\tablefootmark{b} & Sensitivity\tablefootmark{c} & Obs freq. & ALMA Band & ALMA Project ID\\
& (arcsec$\times$arcsec,\degree)\,\tablefootmark{a} &  (pc) & ($\mu$Jy\,beam$^{-1}$) & (GHz) & &\\
\hline
00091$-$0738 & 0.31x0.23, $-$83 & 570 &  45 & 194.2 & 5 & 2018.1.00699.S\\
00188$-$0856 & 0.13x0.12, $-$46 & 290 &  18 & 192.4 & 5 & 2018.1.00699.S\\
00509$+$1225 & 0.31x0.28, 24 & 340 &  24 & 232.6 & 6 & 2018.1.00699.S\\
01572$+$0009 & 0.16x0.13, 69 & 410 &  33 & 188.2 & 5 & 2018.1.00699.S\\
F05189$-$2524 & 0.52x0.42, $-$70 & 390 &  34 & 236.7 & 6 & 2018.1.00699.S\\
07251$-$0248 & 0.27x0.24, $-$50 & 420 &  23 & 228.3 & 6 & 2018.1.00699.S\\
09022$-$3615 & 0.30x0.27, $-$87 & 330 &  24 & 232.9 & 6 & 2018.1.00699.S\\
F10190$+$1322 & 0.30x0.27, 5 & 410 &  23 & 229.4 & 6 & 2018.1.00699.S\\
11095$-$0238 & 0.31x0.24, $-$86 & 540 &  29 & 196.2 & 5 & 2018.1.00699.S\\
F12112$+$0305 & 0.30x0.26, $-$76 & 390 &  21 & 231.1 & 6 & 2016.1.00170.S\\
13120$-$5453 & 0.65x0.65, 0 & 400 & 150 & 239.6 & 6 & 2016.1.00777.S\\
F14348$-$1447 & 0.29x0.25, 89 & 420 &  22 & 229.1 & 6 & 2016.1.00170.S\\
F14378$-$3651 & 0.36x0.24, 84 & 390 &  41 & 231.1 & 6 & 2018.1.00699.S\\
F15327$+$2340 & 1.27x0.81, 40 & 370 & 290 & 226.9 & 6 & 2015.1.00113.S \\
16090$-$0139 & 0.20x0.16, $-$85 & 420 &  37 & 193.3 & 5 & 2018.1.00699.S\\
16155$+$0146 & 0.26x0.14, $-$72 & 440 &  50 & 193.4 & 5 & 2018.1.00699.S\\
17208$-$0014 & 0.47x0.47, 0 & 400 & 240 & 237.8 & 6 & 2018.1.00486.S\\
F19297$-$0406 & 0.27x0.26, 79 & 420 &  18 & 228.5 & 6 & 2018.1.00699.S\\
19542$+$1110 & 0.35x0.30, $-$59 & 390 &  26 & 232.3 & 6 & 2018.1.00699.S\\
20087$-$0308 & 0.31x0.25, $-$70 & 540 &  17 & 196.3 & 5 & 2018.1.00699.S\\
20100$-$4156 & 0.18x0.12, 58 & 350 &  25 & 193.9 & 5 & 2018.1.00699.S\\
20414$-$1651 & 0.18x0.14, $-$54 & 260 &  27 & 228.4 & 6 & 2018.1.00699.S\\
F22491$-$1808 & 0.39x0.29, $-$89 & 500 &  26 & 229.6 & 6 & 2015.1.00263.S\\
\hline
\end{tabular}
\end{small}
\tablefoot{
\tablefoottext{a}{FWHM in arcsec and east of north Position Angle in degrees of the synthesized beam.}
\tablefoottext{b}{Average beam FWHM at the distance of the system (see Table~\ref{tbl:sample}).}
\tablefoottext{c}{1$\sigma$ continuum sensitivity.}}
\end{table*}

\section{Sample of local ULIRGs}\label{s:sample}

The PUMA sample is a volume-limited (z$<$0.165; $d<$800\,Mpc) representative sample of {25} local ULIRGs ({38} individual nuclei). These objects were selected to examine the most relevant parameters for the feedback processes: (1) the main power source (AGN vs. SF); (2) the interaction stage (from interacting pairs to advanced mergers); and (3) the IR luminosity. The parent sample is the 1\,Jy ULIRG sample \citep{Kim1998} extended to southern objects by \citet{Duc1997}. Our sample is limited to objects with Dec. between $-65\degree$ and $+25\degree$ which is appropriate for ALMA. We selected 12 interacting systems (nuclear separation $>$ 1\,kpc) and {13} mergers with nuclear separations $<$1\,kpc. Half of the objects in each interaction stage category were selected to be dominated by AGN based on mid-IR spectroscopy \citep{Veilleux2009, Spoon2013}. The selected objects uniformly cover the ULIRG luminosity range between 10$^{12.0}$ and 10$^{12.7}$\,$L_\odot$. See Table~\ref{tbl:sample} and \citet{Perna2021} for more details.

So far, we have obtained ALMA CO(2--1) and $\sim$220\,GHz continuum observations for 92\% of the systems in the sample ({23} systems with {34} individual nuclei). The CO(2--1) emission is detected in {33} nuclei and the continuum in {29} (see Sect.~\ref{s:data_analysis}). In addition to our VLT\slash MUSE-AO optical integral field spectroscopy \citep{Perna2020, Perna2021}, the majority of the targets have extensive ancillary multi-wavelength data which include mid- and far-IR (e.g., \citealt{Veilleux2009, Spoon2013, Pearson2016, Chu2017}), radio (e.g., \citealt{Condon1998, Helfand2015}), and X-ray (e.g., \citealt{Iwasawa2011, Teng2015}) observations.

\section{Observations and data reduction}\label{s:data}

\subsection{ALMA observations}\label{ss:data_alma}

We obtained ALMA 12-m array CO(2--1) 230.538\,GHz and continuum observations for {23 out of the 25} PUMA ULIRGs. ALMA observations for the remaining two ULIRGs have been scheduled but are not available at the time of writing.
These observations were mainly conducted as part of our programs 2015.1.00263.S, 2016.1.00170.S, and 2018.1.00699.S (PI: M. Pereira-Santaella). {For 13120$-$5453 and F15327+2340 (Arp~220), we used archive data from programs 2016.1.00777.S (PI: K. Sliwa) and 2015.1.00113.S (PI: N. Scoville), respectively.} In addition, we complemented this dataset with higher angular resolution data for 17208$-$0014 from program 2018.1.00486.S (PI: M. Pereira-Santaella). Observations of three of the ULIRGs in our sample (F12112$+$0305, F14348$-$1447, and F22491$-$1808) have been already presented in \citet{Pereira2018}, but we include them here for completeness.

We aimed to have a similar spatial resolution of $\sim$400\,pc in all the systems, so the synthesized beam full-width half-maximum (FWHM) varies between 0\farcs12 and {1\arcsec} depending on the distance of each target. 
We used a single 12-m array configuration with baselines set to achieve the required angular resolution. The maximum recoverable scale is about 10 times the beam FWHM (i.e., 4\,kpc). Depending on the redshift, the CO(2--1) transition lies in the ALMA Band 5 or Band 6. Details on the observations are listed in Table~\ref{tbl:alma_obs}.

We defined four 1.875\,GHz bandwidth spectral windows with 2 to 8\,MHz (3--10\,km\,s$^{-1}$) channels, depending on the targeted spectral feature. One spectral window was centered at the sky frequency of $^{12}$CO(2--1) 230.538\,GHz. The remaining spectral windows were centered at the frequency of nearby transitions (e.g., CS(5--4), H30$\alpha$, SiO(5--4)) when possible or at a ``line-free'' spectral range.

We used the ALMA reduction software CASA (v5.6.1; \citealt{McMullin2007}) to calibrate the data using the standard pipeline. {The absolute flux accuracy of Band 5 and 6 data is $\sim$10\% (ALMA Technical Handbook).}
For the CO(2--1) spectral window, we subtracted a constant continuum level estimated from the line emission free channels in the $uv$ plane.
The data were cleaned using the \textsc{tclean} CASA task and the Brigss weighting with robustness parameters between $-$0.5 and 2.0 to match the required $\sim$400\,pc spatial resolution. For two systems (13120$-$5453 and 17208$-$0014), the largest synthesized beam provides a spatial resolution better than 400\,pc, so we used the \textsc{imsmooth} task to convolve the cubes with a Gaussian and obtained the desired spatial resolution. {For F15327+2340 (Arp 220), we only used the compact configuration data which provide a $\sim$370\,pc spatial resolution comparable to that of the other ULIRGs in our sample.}
The channel width of the final cubes is $\sim$10\,km\,s$^{-1}$ and the pixel sizes are about a sixth of the beam FWHM (i.e., between 20 and 120\,mas). In addition to the line data cubes, we produced continuum images using spectral windows where no emission or absorption lines were present. The continuum sensitivities range from 18 to {290}\,$\mu$Jy\,beam$^{-1}$ with more sensitive data for the more distant objects (see Table~\ref{tbl:alma_obs}).

In this paper, we primarily focus on the analysis of the continuum {and CO(2--1) maps}. In a future paper (Lamperti et al. in prep.), we will present the {detailed analysis of the line data}.

\begin{figure*}[!ht]
\centering
\vspace{5mm}
\includegraphics[width=0.95\textwidth]{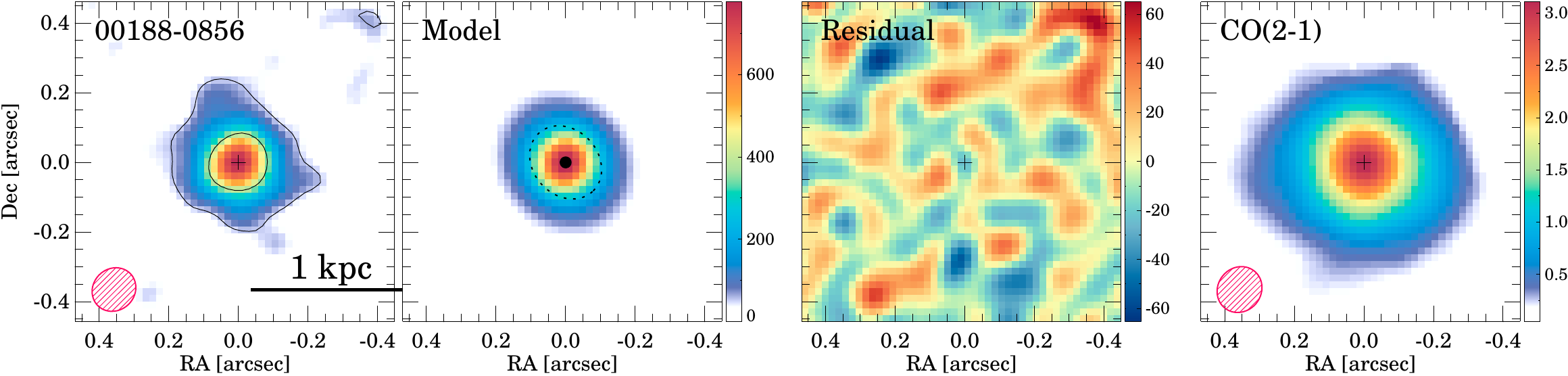}
\caption{ALMA continuum observation (first panel), best-fit model (second panel), residual emission after subtracting the continuum model (third panel), and the integrated CO(2--1) emission (moment 0) from Lamperti et al. in prep. (fourth panel) for 00188$-$0856 as an example. The two contour levels in the first panel indicate the 3$\sigma$ and the 0.5$\times$peak emission levels.
In the second panel, the individual components of the best-fit model are presented as a black circle (point source model) and as a dashed ellipse (deconvolved Gaussian model).
The black crosses in the first, third, and fourth panels mark the fitted location of the continuum peak.
The red hatched ellipses represent the beam FWHM. 
The units are $\mu$Jy\,beam$^{-1}$ for the continuum panels and Jy\,km\,s$^{-1}$\,beam$^{-1}$ for the CO(2--1) panel. The continuum model fits for the whole sample are shown in Fig.~\ref{fig:apx_alma_models}.\label{fig:alma_models}}
\end{figure*}

\subsection{Ancillary \textit{Spitzer} data}\label{ss:data_spitzer}

To complete the spectral energy distribution (SED) of the ULIRGs (Sect.~\ref{ss:sed_model}), we used mid-IR \textit{Spitzer} data. In particular we used the 5.2--38\,\micron\ low-resolution ($R\sim60-130$) spectra from the Infrared Spectrograph (IRS; \citealt{HouckIRS}) and the 70 and 160\,\micron\ images from the Multiband Imaging Photometer (MIPS; \citealt{Rieke2004MIPS}).

We downloaded the calibrated IRS spectra for all the systems in our sample from the Cornell Atlas of \textit{Spitzer}\slash Infrared Spectrograph Sources (CASSIS; \citealt{Lebouteiller2011}) and measured the flux at 34\micron, which is approximately at the middle point between the 24 and 70\micron\ photometric points in log scale and avoids the noisier long-wavelength edge of the IRS spectrum. 

We also downloaded the calibrated MIPS images for 5 systems from the \textit{Spitzer} Heritage Archive\footnote{https://sha.ipac.caltech.edu}. The ULIRG systems appear as point-sources at the MIPS angular resolution (18\arcsec and 40\arcsec\ at 70 and 160\micron, respectively). For the 70\micron\ image, we used a 35\arcsec\ radius aperture and a 39--65\arcsec\ background annulus and then multiplied the flux by 1.24 to account for the aperture correction factor (see Table 4.14 of the MIPS Instrument Handbook). For the 160\micron\ images, we subtracted a global background emission level and used a 60\arcsec\ radius aperture. We applied a 1.40 aperture correction factor which is appropriate for sources with temperatures between 30 and 150\,K (see Table 4.15 of the MIPS Instrument Handbook). The measured \textit{Spitzer} IRS and MIPS fluxes are listed in Appendix\,\ref{apx:spitzer_fluxes}.

\begin{figure}[t]
\centering
\includegraphics[width=0.37\textwidth]{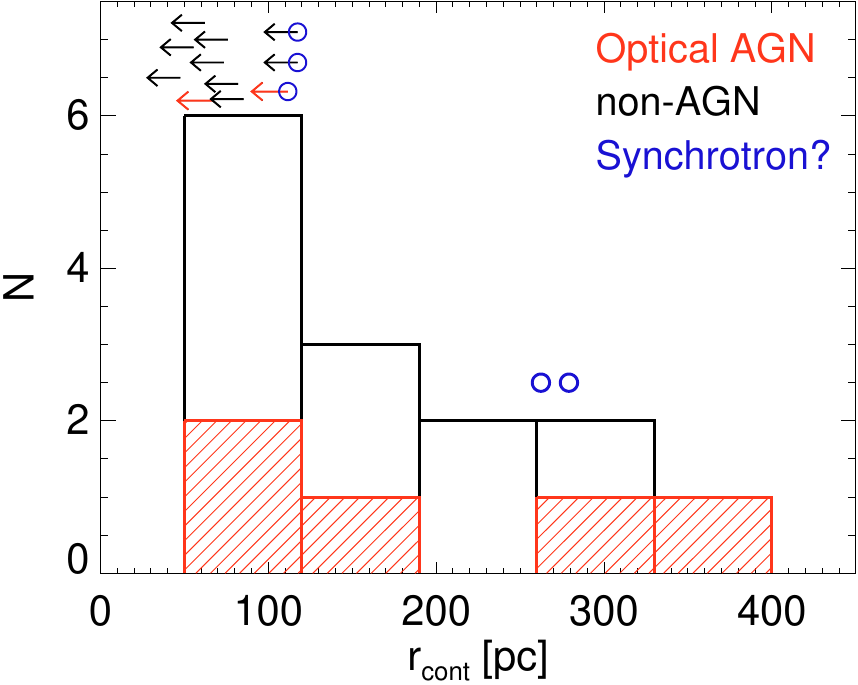}
\caption{Distribution of the $\sim$220\,GHz continuum half-light radius ($r_{\rm cont}$; see Table~\ref{tbl:alma_cont_model}). Upper limits are indicated with arrows. The red (black) histogram bars and arrows correspond to galaxies classified as AGN (non-AGN) from optical spectroscopy. The sizes of the systems whose ALMA flux might have high ($>$40\%) non-thermal synchrotron contributions are marked in blue (see Sect.~\ref{ss:disc_alma_comp}).\label{fig:histo_size}}
\end{figure}

\section{Data analysis}\label{s:data_analysis}

\subsection{ALMA continuum model. Size and flux}\label{ss:alma_model}

We modeled the ALMA 220--250\,GHz continuum images to determine the flux, size, and position of the detected emitting regions. In general, these regions are compact (FWHMs similar to the beam size) and their morphological structure is barely resolved. Therefore, we used simple models consisting of a point-source, a Gaussian, point-source $+$ Gaussian, or 2 Gaussians. 
These models were convolved with the beam and compared with the observations to determine a $\chi^2$ value. Then, we minimized the $\chi^2$ by varying the fluxes, sizes, and positions of the model components.

We tried these 4 models for each nucleus and selected that with the lowest reduced $\chi^2$. The best-fit models reproduce quite well the observed emission. The median (mean) reduced $\chi^2$ is 1.1 (1.4), the maximum is 3.1, and no significant structures are seen in the residual images (see Fig.~\ref{fig:alma_models}). This figure also shows the best-fit model whose parameters are listed in Table~\ref{tbl:alma_cont_model}. The best-fit positions are presented in Table~\ref{tbl:sample}.
Based on these parameters, we computed the half-light radius of the $\sim$220\,GHz continuum, $r_{\rm cont}$, which is defined as the radius of the region that contains 50\% of the observed flux.

For {11} nuclei whose model includes a point-source, the half-light radius is not well defined because the point source contributes $>$50\% to the total flux. Therefore, to estimate the size upper limit in these cases, we performed a series of simulations.
First, we subtracted, when present, the extended Gaussian component of the model.
Then, we used circular Gaussian models with fixed FWHM from 2 to 6\,pixels\footnote{6\,pixels approximately corresponds to the beam FWHM.}, that were convolved with the beam, and obtained the $\chi^2$ variation as function of the model FWHM. 
Finally, we estimated the 3$\sigma$ FWHM upper limit as the FWHM at which the $\chi^2$ increases by 9.0~\footnote{The 9.0 constant corresponds to the value at which the cumulative distribution function of a $\chi^2$ distribution with one degree of freedom is equivalent to a 3$\sigma$ confidence interval of a normal distribution ($\sim$0.997).} with respect to the minimum $\chi^2$. 
The FWHM upper limits are also included in Table~\ref{tbl:alma_cont_model}.

Fig.~\ref{fig:histo_size} shows the distribution of the half-light radius and the upper limits. The measured $r_{\rm cont}$ range from {$<$50\,pc} to 350\,pc with a median value of {80--100}\,pc.
The $r_{\rm cont}$ of AGN and non-AGN objects, based on optical spectroscopy, are similar. 

Mid-IR observations already indicate that ULIRGs are very compact ($<$1\,kpc; \citealt{Soifer2000, DiazSantos2010, AAH2014, AAH2016, Imanishi2020}). Our higher angular resolution ALMA continuum data suggest that they are even more compact.

\afterpage{
\begin{landscape}
\begin{table}
\caption{ALMA continuum models}
\label{tbl:alma_cont_model}
\centering
\begin{tabular}{lccccccccccc}
\hline \hline
\\
& & &  Total & Point & \multicolumn{2}{c}{Gaussian 1} & \multicolumn{2}{c}{Gaussian 2} \\
IRAS name & Nucleus & Rest freq. & flux\tablefootmark{a} & Flux & Flux & FWHM\tablefootmark{b} & Flux & FHWM\tablefootmark{b}  & \multicolumn{2}{c}{$r_{\rm cont}$\,\tablefootmark{c}} & $\chi_{\rm red}^2$\\
& & (GHz) &  (mJy) & (mJy) & (mJy) & (mas, mas) & (mJy) & (mas, mas) & (mas) & (pc) \\
\hline
00091$-$0738 & S & 217.10 & 4.83$\pm$0.05 & 4.83$\pm$0.05 &  \nodata & \nodata & \nodata & \nodata & <40 & <85 & 0.91\\
 & N &  & $<$0.15 & \nodata & \nodata & \nodata & \nodata & \nodata & \nodata & \nodata & \nodata  \\
00188$-$0856 & - & 217.12 & 1.74$\pm$0.10 & 0.47$\pm$0.03 & 1.26$\pm$0.09 & 227$\pm$24,187$\pm$13 & \nodata & \nodata & 75$\pm$2 & 173  & 0.98\\
00509$+$1225 & - & 246.83 & 1.03$\pm$0.07 &  \nodata  & 1.03$\pm$0.07 & 191$\pm$20,104$\pm$22 & \nodata & \nodata & 70$\pm$4  & 83  & 1.05\\
01572$+$0009 & - & 218.90 & 1.12$\pm$0.13 & 0.53$\pm$0.05 & 0.59$\pm$0.11 & 331$\pm$100,89$\pm$70 & \nodata & \nodata & 22$\pm$4 & 63  & 1.05\\
F05189$-$2524 & - & 246.86 & 6.29$\pm$0.21 & 3.90$\pm$0.12 & 2.39$\pm$0.18 & 488$\pm$42,449$\pm$24 & \nodata & \nodata & <80 & <67 & 1.13\\
07251$-$0248 & W & 248.39 & 0.97$\pm$0.05 &  \nodata  & 0.97$\pm$0.05 & 122$\pm$16,95$\pm$18 & \nodata & \nodata & 53$\pm$3 & 88 & 0.73\\
 & E &  & 9.29$\pm$0.15 & 7.74$\pm$0.14 & 1.55$\pm$0.07 & 328$\pm$51,199$\pm$21 & \nodata & \nodata & <50 & <82 & 1.20\\
09022$-$3615 & - & 246.80 & 6.98$\pm$0.59 &  & 2.50$\pm$0.32 & 319$\pm$107,18$\pm$93 & 4.48$\pm$1.05 & 1050$\pm$160,540$\pm$133 & 227$\pm$66 & 262  & 2.17\\
F10190$+$1322 & W & 246.86 & 0.39$\pm$0.10 &  \nodata  & 0.39$\pm$0.10 & 483$\pm$133,258$\pm$104 & \nodata & \nodata & 176$\pm$22 & 255 & 1.01 \\
 & E &  & 2.80$\pm$0.11 & 0.79$\pm$0.06 & 2.00$\pm$0.09 & 618$\pm$37,294$\pm$26 & \nodata & \nodata & 153$\pm$4 & 222 & 1.25 \\
11095$-$0238 & SW & 217.13 & 0.41$\pm$0.13 &  \nodata  & 0.41$\pm$0.13 & 275$\pm$112,48$\pm$76 & \nodata & \nodata & <60 & <117 & 1.11 \\
 & NE &  & 0.89$\pm$0.03 & 0.89$\pm$0.03 &  \nodata & \nodata & \nodata & \nodata & <60 & <117 & 1.11 \\
F12112$+$0305 & SW & 247.99 & 0.70$\pm$0.09 & 0.13$\pm$0.04 & 0.57$\pm$0.08 & 273$\pm$93,178$\pm$64 & \nodata & \nodata & 93$\pm$11 & 129 & 0.73 \\
 & NE &  & 6.84$\pm$0.11 & 4.78$\pm$0.07 & 2.06$\pm$0.08 & 430$\pm$22,369$\pm$24 & \nodata & \nodata & <40 & <55 & 1.13 \\
13120$-$5453 & - & 247.05 & 32.14$\pm$2.23 &  \nodata  &  15.04$\pm$1.16 & 668$\pm$143,518$\pm$115 & 17.09$\pm$1.37 & 1504$\pm$223,1328$\pm$196 & 448$\pm$89 & 279 & 1.03 \\
F14348$-$1447 & SW & 248.00 & 2.82$\pm$0.11 & 1.51$\pm$0.04 & 1.31$\pm$0.10 & 601$\pm$75,482$\pm$44 & \nodata & \nodata & <40 & <62 & 0.96 \\
 & NE &  & 1.69$\pm$0.13 & 0.71$\pm$0.11 & 0.98$\pm$0.06 & 326$\pm$49,239$\pm$28 & \nodata & \nodata & 64$\pm$3 & 100 & 1.07 \\
F14378$-$3651 & - & 246.84 & 2.66$\pm$0.18 & 0.48$\pm$0.07 & 2.18$\pm$0.16 & 714$\pm$54,540$\pm$34 & \nodata & \nodata & 262$\pm$6 & 342 & 1.12 \\

F15327+2340 & W & 230.98 & 135.50$\pm$1.48 & 117.36$\pm$0.97 & 18.13$\pm$1.11 & 1739$\pm$205,1119$\pm$181 & \nodata & \nodata & <130 & <47 & 1.22\\
 & E &  & 56.16$\pm$1.03 &  \nodata  & 56.16$\pm$1.03 & 509$\pm$34,357$\pm$20 & \nodata & \nodata & 213$\pm$5 & 78 \\

16090$-$0139 & - & 219.09 & 3.11$\pm$0.19 & 0.38$\pm$0.13 & 2.73$\pm$0.13 & 355$\pm$35,186$\pm$28 & \nodata & \nodata & 116$\pm$4  & 275 & 1.37 \\
16155$+$0146 & NW &  219.14 & 0.60$\pm$0.08 & 0.60$\pm$0.08 &  \nodata & \nodata & \nodata & \nodata & <47 & <111 & 0.87 \\
 & SE &  & $<$0.15 & \nodata & \nodata & \nodata & \nodata & \nodata & \nodata & \nodata & \nodata \\
17208$-$0014 & - & 247.99 & 42.60$\pm$1.12 & 20.45$\pm$0.56 & 22.15$\pm$0.97 & 842$\pm$58,636$\pm$44 & \nodata & \nodata & <90 & <75 & 1.33 \\
F19297$-$0406 & S & 248.01 & $<$0.06 & \nodata & \nodata & \nodata & \nodata & \nodata & \nodata & \nodata & \nodata \\
 & N &  & 5.99$\pm$0.45 &  \nodata  &  3.40$\pm$0.28 & 247$\pm$15,83$\pm$13 & 2.58$\pm$0.34 & 718$\pm$82,640$\pm$76 & 113$\pm$4 & 182 & 2.74 \\
19542$+$1110 & - & 246.81 & 3.50$\pm$0.21 & 1.28$\pm$0.12 & 2.22$\pm$0.19 & 354$\pm$32,276$\pm$33 & \nodata & \nodata & 91$\pm$3 & 110 & 1.05 \\
20087$-$0308 & - & 217.05 & 5.56$\pm$0.14 & 2.47$\pm$0.07 & 3.09$\pm$0.12 & 457$\pm$19,332$\pm$12 & \nodata & \nodata & 78$\pm$2 & 151 & 2.11\\
20100$-$4156 & NW & 219.11 & $<$0.08 & \nodata & \nodata & \nodata & \nodata & \nodata & \nodata & \nodata & \nodata \\
 & SE &  & 3.74$\pm$0.21 & 1.66$\pm$0.15 & 2.08$\pm$0.14 & 211$\pm$26,152$\pm$15 & \nodata & \nodata &  34$\pm$2 & 80 & 2.39 \\
20414$-$1651 & - & 248.28 & 4.57$\pm$0.44 & 1.82$\pm$0.30 & 2.75$\pm$0.32 & 191$\pm$45,131$\pm$38 & \nodata & \nodata & 41$\pm$4  & 66 & 3.10 \\
F22491$-$1808 & W & 247.48 & $<$0.08 & \nodata & \nodata & \nodata & \nodata & \nodata & \nodata & \nodata & \nodata \\
 & E &  & 5.08$\pm$0.26 & 3.90$\pm$0.14 & 1.17$\pm$0.22 & 628$\pm$196,383$\pm$44 & \nodata & \nodata & <50 & <73 & 2.89 \\
\hline
\end{tabular}
\tablefoot{{In addition to the statistical uncertainties listed in this table, the absolute flux accuracy is $\sim$10\%.}
\tablefoottext{a}{Total flux including all the model components (point source, Gaussian 1, and Gaussian 2). For undetected sources, we indicate the 3$\sigma$ upper limit.}
\tablefoottext{b}{Deconvolved FWHM of the Gaussian models.}
\tablefoottext{c}{Deconvolved half-light radius.}
}
\end{table}
\end{landscape}
}

\subsubsection{Higher resolution observations}\label{ss:alma_model_hires}

{The CO(2--1) and 230\,GHz continuum emission of F15327+2340 (Arp~220) have been observed by ALMA at much higher resolution (8\,pc) than the data used in this paper. However, these data are still unpublished. Instead, we can compare with the 20--40\,pc 2.7\,mm ($\sim$110\,GHz) continuum observations presented by \citet{Scoville2017} and \citet{Sakamoto2017}. These authors measure deconvolved Gaussian FWHM of 160--180\,mas for the East nucleus and 74--114\,mas for the West nucleus. From the low resolution data, we obtained 2$\times$\,$r_{\rm cont}$=400\,mas and $<$260\,mas for the East and West nuclei, respectively. Therefore, we recover a reliable upper limit size for the bright and compact source in the West nucleus. For the East nucleus, we derive a size 2 times larger. However, it is possible that part of extended emission detected in the low-resolution data is filtered out, or too faint, in the ten times higher resolution published observations.}

{In addition, 17208$-$0014} was observed as part of another program which aimed to obtain $\sim$100\,pc spatial resolution CO(2--1) and continuum data. These observations will be analyzed in detail in a future paper. However, here we use the high resolution continuum data (120\,pc vs. 400\,pc), to test whether the source size derived from the low-resolution data ($r_{\rm cont}<75$\,pc for this object) is consistent with the size measured in the high-resolution data.

\begin{figure}[t]
\centering
\vspace{5mm}
\includegraphics[width=0.48\textwidth]{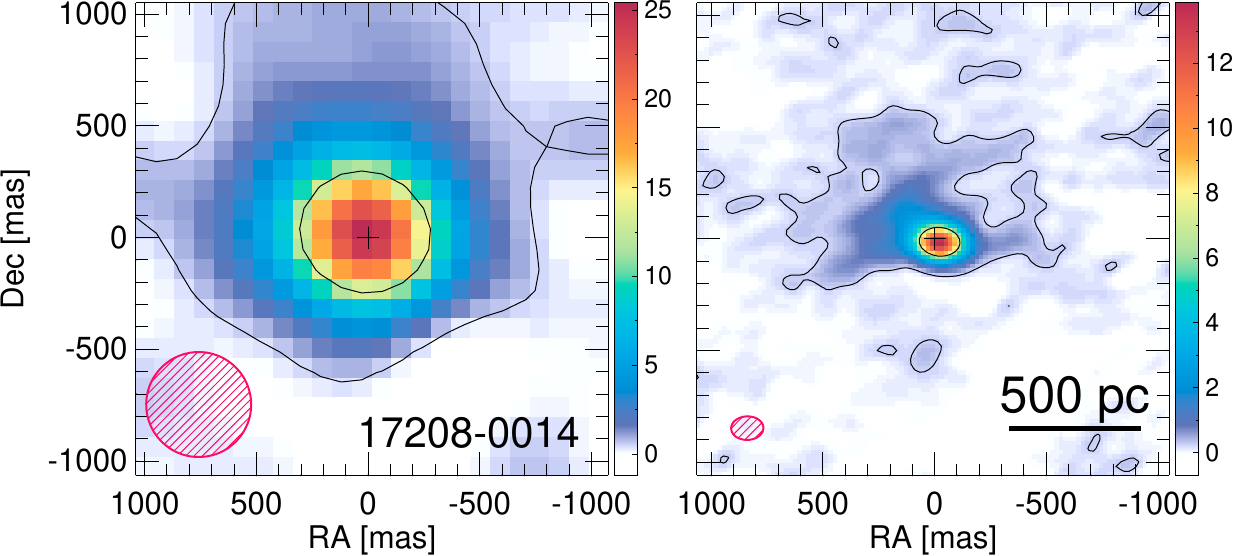}
\caption{Comparison between the 400\,pc resolution 247\,GHz continuum image analyzed in Sect.~\ref{ss:alma_model} (left panel) and the higher resolution (120\,pc) data (rigth panel) available for 17208$-$0014. The black cross is the position of the center measured on the 400\,pc image .The contours are as in Fig.~\ref{fig:alma_models}. The hatched red ellipses correspond to the beam FWHM of each image. The color scales are in mJy\,beam$^{-1}$.\label{fig:comp_hi_lo}}
\end{figure}

Fig.~\ref{fig:comp_hi_lo} compares the low- and the high-resolution maps. The difference between the continuum fluxes in both images measured using a 1\arcsec\ radius aperture is $<$9\% (42.7$\pm$0.2 vs. 46.3$\pm$0.3\,mJy). This indicates that the higher resolution data do not miss significant low surface brightness emission.
This figure also shows that the continuum emission peak measured in the original low-resolution data (Table~\ref{tbl:sample}) appears slightly shifted (30\,mas or 26\,pc) in the high-resolution image. This is possibly because now we start to spatially resolve the inner structure of the nucleus and multiple smaller regions appear.

We applied the same model fitting procedure described in Sect.~\ref{ss:alma_model} to the higher resolution image. The original model consisted of a point source plus a Gaussian (see Table~\ref{tbl:alma_cont_model} and Fig.~\ref{fig:apx_alma_models}). For the high-resolution data, we used 2 Gaussians since the emission core is resolved. 
This ``core'' Gaussian has a flux of 17.6$\pm$1.1\,mJy and a circularized FWHM of 74$\pm$3\,mas (62\,pc; $r_{\rm cont}=31$\,pc). It contains about 40\% of the 247\,GHz continuum emission from 17208$-$0014, so the $r_{\rm cont}<$75\,pc upper limit we estimated from the low-resolution data (Table~\ref{tbl:alma_cont_model}) seems to be consistent with what is observed at higher angular resolution. 

{The result for these two objects} supports that our method to estimate the $r_{\rm cont}$ can produce realistic values, even below the beam size. %

\begin{figure}[t]
\centering
\vspace{5mm}
\includegraphics[trim=23mm 130mm 93mm 98mm, clip=true, width=0.43\textwidth]{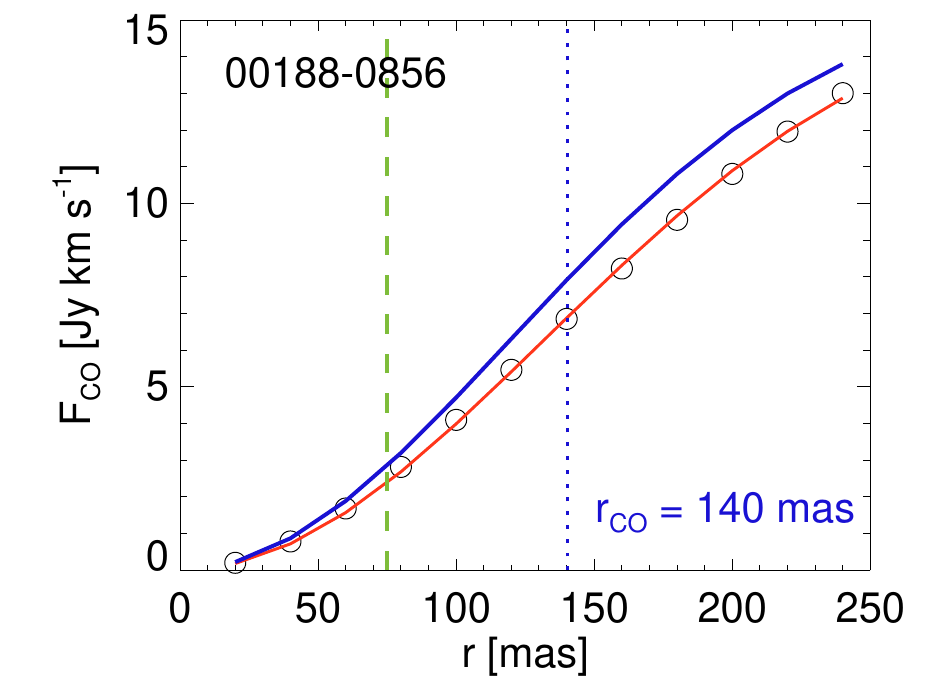}
\caption{Growth curve of the CO(2--1) moment 0 map for 00188$-$0856 as an example. The circles correspond to the observed flux within a circular aperture of radius $r$. The red line is the best fit model (Sect.~\ref{ss:nuclear_mol}). The deconvolved best fit profile is shown in blue and its effective radius, $r_{\rm CO}=$\,FWHM/2, is indicated by the vertical blue dotted line. The dashsed green line marks the $\sim$220\,GHz continuum radius, $r_{\rm cont}$ for comparison. The CO(2--1) model fits for the whole sample are shown in Fig.~\ref{fig:apx_co_models}.\label{fig:co_models}}
\end{figure}

\begin{figure}[t]
\centering
\vspace{5mm}
\includegraphics[width=0.4\textwidth]{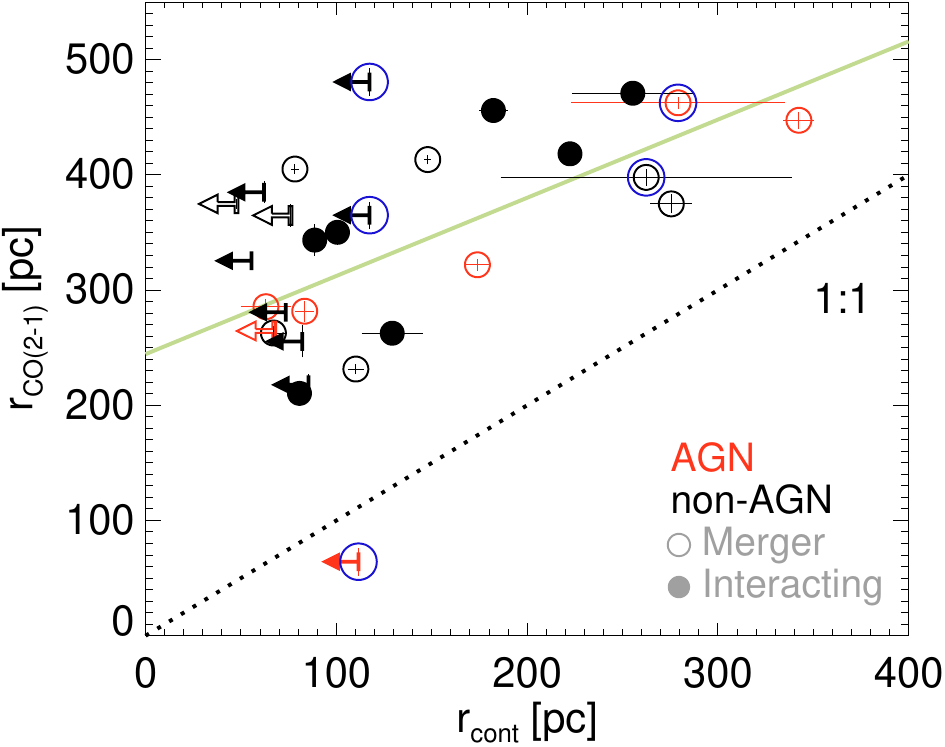}
\caption{Half light radius of the 220\,GHz continuum $r_{\rm cont}$ vs. 0.5$\times$FWHM of the CO(2--1) emission $r_{\rm CO}$. Red (black) symbols mark systems classified as AGN (non-AGN) based on optical spectroscopy. {Filled symbols correspond to nuclei in interacting systems and empty symbols to nuclei in mergers (see Table~\ref{tbl:sample}).}
{Blue} encircled symbols are galaxies with excess non-thermal emission whose continuum size estimates might be inaccurate (Sect.~\ref{ss:disc_alma_comp}). The green line is best linear fit excluding the nuclei with $r_{\rm cont}$ upper limits. The dotted line indicates the 1:1 relation.\label{fig:size_cont_co}}
\end{figure}

\begin{table*}[t]
\caption{{Nuclear} CO(2--1) emission and cold molecular gas mass}
\label{tbl:alma_co21}
\centering
\begin{small}
\begin{tabular}{lcccccccc}
\hline \hline
\\
IRAS name & Nucleus & $r_{\rm CO}$\,\tablefootmark{a} & $S_{\rm CO}$\,\tablefootmark{b} & $\log M_{\rm H_2}$\,\tablefootmark{c} & $\log \Sigma_{\rm H_2}$\,\tablefootmark{d} \\
 & & (pc) & (Jy\,km\,s$^{-1}$) & ($M_\odot$) & ($M_\odot$\,pc$^{-2}$) \\
\hline
00091$-$0738 & S & 217$\pm$9 & 11.6$\pm$0.1 & 9.22$\pm$0.04 & 3.74$\pm$0.05 \\
 & N & 421$\pm$9 & 8.00$\pm$0.41 & 9.05$\pm$0.05 & 3.01$\pm$0.06 \\
00188$-$0856 & - & 321$\pm$5 & 15.9$\pm$0.2 & 9.43$\pm$0.04 & 3.61$\pm$0.05 \\
00509$+$1225 & - & 281$\pm$9 & 22.7$\pm$0.4 & 8.92$\pm$0.04 & 3.23$\pm$0.06 \\
01572$+$0009 & - & 285$\pm$6 & 5.12$\pm$0.07 & 9.15$\pm$0.04 & 3.44$\pm$0.04 \\
F05189$-$2524 & - & 264$\pm$8 & 71.3$\pm$0.9 & 9.11$\pm$0.04 & 3.46$\pm$0.06 \\
07251$-$0248 & W & 343$\pm$13 & 12.5$\pm$1.7 & 8.99$\pm$0.07 & 3.12$\pm$0.09 \\
 & E & 255$\pm$13 & 31.7$\pm$0.4 & 9.39$\pm$0.04 & 3.78$\pm$0.06 \\
09022$-$3615 & - & 397$\pm$7 & 153$\pm$2 & 9.73$\pm$0.04 & 3.74$\pm$0.04 \\
F10190$+$1322 & W & 470$\pm$9 & 17.0$\pm$0.8 & 9.00$\pm$0.05 & 2.85$\pm$0.05 \\
 & E & 418$\pm$9 & 52.6$\pm$0.8 & 9.49$\pm$0.04 & 3.44$\pm$0.05 \\
11095$-$0238 & SW & 480$\pm$11 & 20.1$\pm$0.8 & 9.36$\pm$0.05 & 3.20$\pm$0.05 \\
 & NE & 365$\pm$11 & 22.5$\pm$0.5 & 9.41$\pm$0.04 & 3.49$\pm$0.05 \\
F12112$+$0305 & SW & 262$\pm$6 & 17.2$\pm$0.5 & 8.96$\pm$0.05 & 3.33$\pm$0.05 \\
 & NE & 325$\pm$6 & 72.2$\pm$0.9 & 9.58$\pm$0.04 & 3.76$\pm$0.05 \\
13120$-$5453 & - & 462$\pm$5 & 472$\pm$5 & 9.65$\pm$0.04 & 3.52$\pm$0.05 \\
F14348$-$1447 & SW & 384$\pm$9 & 53.9$\pm$0.8 & 9.57$\pm$0.04 & 3.60$\pm$0.04 \\
 & NE & 350$\pm$9 & 33.3$\pm$1.3 & 9.36$\pm$0.05 & 3.47$\pm$0.05 \\
F14378$-$3651 & - & 447$\pm$7 & 40.9$\pm$0.6 & 9.28$\pm$0.04 & 3.18$\pm$0.05 \\
F15327$+$2340\,$^\dagger$ & - & 370$\pm$4 & 1360$\pm$7 & 9.63$\pm$0.04 & 3.69$\pm$0.05 \\
16090$-$0139 & - & 374$\pm$8 & 43.9$\pm$0.6 & 9.91$\pm$0.04 & 3.96$\pm$0.06 \\
16155$+$0146 & NW & 64$\pm$12 & 2.24$\pm$0.03 & 8.61$\pm$0.04 & 4.19$\pm$0.16 \\
 & SE & \nodata & $<$0.04 & $<$6.81 & \nodata \\
17208$-$0014 & - & 364$\pm$9 & 338$\pm$6 & 9.79$\pm$0.04 & 3.86$\pm$0.05 \\
F19297$-$0406 & S & \nodata & 1.2$\pm$0.3 & 7.93$\pm$0.11 & \nodata \\
 & N & 455$\pm$8 & 81.7$\pm$1.3 & 9.78$\pm$0.04 & 3.66$\pm$0.05 \\
19542$+$1110 & - & 231$\pm$4 & 35.4$\pm$0.2 & 9.14$\pm$0.04 & 3.61$\pm$0.04 \\
20087$-$0308 & - & 413$\pm$4 & 59.7$\pm$0.6 & 9.83$\pm$0.04 & 3.80$\pm$0.05 \\
20100$-$4156 & NW & 364$\pm$8 & 0.939$\pm$0.042 & 8.21$\pm$0.05 & 2.29$\pm$0.05 \\
 & SE & 210$\pm$4 & 21.1$\pm$0.1 & 9.56$\pm$0.04 & 4.12$\pm$0.05 \\
20414$-$1651 & - & 262$\pm$5 & 32.2$\pm$0.5 & 9.39$\pm$0.04 & 3.75$\pm$0.06 \\
F22491$-$1808 & W & \nodata & 0.41$\pm$0.02 & 7.38$\pm$0.02 & \nodata \\
 & E & 280$\pm$5 & 45.4$\pm$0.3 & 9.44$\pm$0.04 & 3.74$\pm$0.05 \\
\hline
\end{tabular}
\end{small}
\tablefoot{
\tablefoottext{a}{Deconvolved radius (0.5$\times$FWHM) of the nuclear CO(2--1) Gaussian emission model.}
\tablefoottext{b}{Nuclear CO(2--1) flux derived from the Gaussian model. This value does not include extended CO(2--1) emission beyond $r>$0.7\,kpc. {In addition to the statistical uncertainties listed in this column, the absolute flux accuracy is $\sim$10\%.}}
\tablefoottext{c}{Molecular gas mass calculated using a ULIRG-like $\alpha_{\rm CO}$ conversion factor (0.78\,$M_\odot$\,${\rm (K\,km\,s^{-1}\,pc^{-2})^{-1}}$) and a CO 2--1 to 1--0 ratio $r_{\rm 21}=0.91$ (see Sect.~\ref{ss:nuclear_mol}).}
\tablefoottext{d}{CO(2--1) surface density within $r_{\rm CO}$ calculated as 0.5$\times M_{\rm H_2}$\slash ($\pi r_{\rm CO}^2$).}
\tablefoottext{$^\dagger$}{At the resolution of these observations (370\,pc) it is not possible to disentangle the CO(2--1) emission from the two nuclei of F15327+2340 (see Fig.~\ref{fig:apx_alma_models}).}
}
\end{table*}

\subsection{Nuclear molecular gas}\label{ss:nuclear_mol}

Figs.~\ref{fig:alma_models} and \ref{fig:apx_alma_models} show that the CO(2--1) emission is more extended than the continuum and also that it has a more complex morphology. As a consequence, the simple set of models used to fit the continuum (Sect.~\ref{ss:alma_model}) does not reproduce the CO(2--1) emission properly. Therefore, we considered a different approach to determine the size and flux of the nuclear CO(2--1) emission.

We used the CO(2--1) moment 0 maps (Lamperti et al. in prep.) to extract the flux in concentric apertures centered at the continuum peak. This produces an azimuthally averaged growth curve for the CO(2--1) emission (see Fig.~\ref{fig:co_models}). To fit this curve, we simulated a 2D circular Gaussian model, which was convolved with the beam, and we compared the model growth curve with the observed one. From this fit, we obtained the deconvolved circularized FWHM of the CO(2--1) emission and the nuclear flux. Then, we used a ULIRG-like $\alpha_{\rm CO}$ conversion factor (0.78\,$M_\odot$\,${\rm (K\,km\,s^{-1}\,pc^{-2})^{-1}}$) and a CO 2--1 to 1--0 ratio $r_{\rm 21}=0.91$ \citep{Bolatto2013} to estimate the molecular gas mass. We limited this growth curve to the central $\sim$1.5\,kpc, so the most extended emission of the systems is not included in the measured flux. Nevertheless, we are mostly interested in the CO(2--1) sizes and fluxes of the nuclear regions detected in the continuum and these are well covered by the apertures used ($\sim$700\,pc aperture radius vs. $r_{\rm cont}<$350\,pc).
The nuclear CO(2--1) deconvolved sizes, $r_{\rm CO}$, fluxes, molecular gas masses, and molecular gas surface densities, $\Sigma_{\rm H_2}$, are presented in Table\,\ref{tbl:alma_co21}. 

The CO(2--1) radius $r_{\rm CO}$ ranges from 60 to 500\,pc (median 320\,pc). Fig.~\ref{fig:size_cont_co} shows that $r_{\rm CO}$ is larger then the continuum size $r_{\rm cont}$. The median $r_{\rm CO}$\slash $r_{\rm cont}$ ratio is {2.5$\pm$1.1}. As for the continuum size, we do not find significant differences between the $r_{\rm CO}$ of AGN and non-AGN nuclei. If we exclude the nuclei with upper limits for $r_{\rm cont}$, there is a good correlation between the CO and continuum sizes {(Spearman's rank correlation coefficient $r_{\rm s}=0.72$, probability of no correlation $p = 7\times10^{-4}$).  The best linear fit is $r_{\rm CO}=(0.70\pm 15)\times r_{\rm cont}$ + (240$\pm$30)\,pc.}

\subsection{Spectral energy distribution fit}\label{ss:sed_model}

\begin{figure}
\centering
\vspace{5mm}
\includegraphics[width=0.4\textwidth]{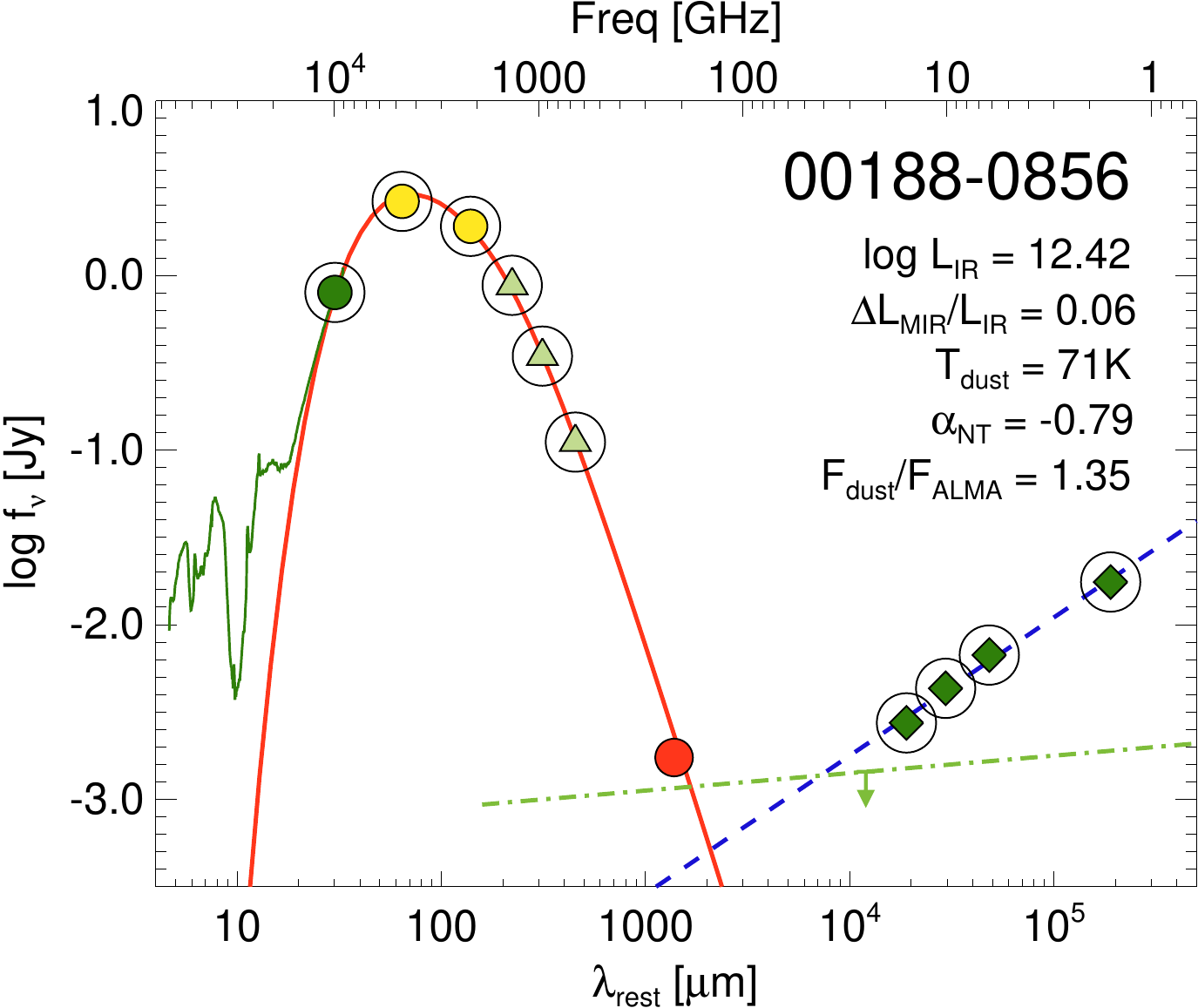}
\caption{SED fit for 00188$-$0856 as an example. The data points correspond to the radio (green diamonds), ALMA $\sim$220\,GHz from this paper (red circle), and IR (remaining points) observations. The solid green line is the 5--38\micron\ {\it Spitzer}\slash IRS spectrum.
The IR observations are color coded as follows: {Spitzer}\slash IRS synthetic photometry at 34\micron\ (green circle); \textit{Spitzer}\slash MIPS (yellow circles); and  \textit{Herschel}\slash SPIRE (green triangles). The solid red line is the best gray-body fit and the dashed blue line represents the best power-law fit to the non-thermal radio emission.  Only the encircled symbols have been used for the fits (i.e., the ALMA point is excluded from the SED fit).
The dot-dashed green line represents the expected maximum free-free emission {assuming that all the $L_{\rm IR}$ is produced by SF} (see Sect.~\ref{ss:sed_model}).
The SED fits for the whole sample are shown in Fig.~\ref{fig:apx_sed_models}.\label{fig:sed_models}}
\end{figure}

\begin{table}
\caption{Far-IR and radio fluxes references}
\label{tbl:bib_fluxes}
\centering
\begin{small}
\begin{tabular}{lcccccccc}
\hline \hline
\\
IRAS name & IR  & Radio \\
\hline
00091$-$0738  & M90      &   H15, H21  \\
00188$-$0856  & PS21, P16  &  H15, H21   \\
00509$+$1225  & P15      &   H15, B89 \\
01572$+$0009  & P15, P16  &  H15, B89   \\
F05189$-$2524 & P16, C17  &  C90, C91   \\
07251$-$0248  & C17      &   C98  \\
09022$-$3615  & P16, C17  &  C98   \\
F10190$+$1322 & M90, PS21  &  H15  \\
11095$-$0238  & P16, PS21  &  H15  \\
F12072$-$0444 & M90, P16  &  H15   \\
F12112$+$0305 & C17      &   C90, C91  \\
13120$-$5453  & P16, C17  &  W94, M07   \\
F13451$+$1232 & PS21      &   S98  \\
F14348$-$1447 & P16, C17  &  C90, C91   \\
F14378$-$3651 & P16, C17  &  C98, M03   \\
F15327$+$2340 & M90, C17  &  C91, C98, BM15  \\
16090$-$0139  & M90, K01, P16  &  C98   \\
16155$+$0146  & M90      &   H15  \\
17208$-$0014  & P16, C17  &  B06   \\
F19297$-$0406 & P16, C17  &  C98   \\
19542$+$1110  & C17      &   C98, L11  \\
20087$-$0308  & M90, P16  &  W98, C98, M17   \\
20100$-$4156  & PS21, P16  &  C96   \\
20414$-$1651  & M90, P16  &  C98, N03   \\
F22491$-$1808 & P16, C17  &  C90, C91, H21   \\
\hline
\end{tabular}
\end{small}
\tablebib{IR: (C17) \citealt{Chu2017}; (K01) \citealt{Klaas2001}; (M90) \citealt{Moshir1990}; (P15) \citealt{Petric2015}; (P16) \citealt{Pearson2016}; (S03) \citealt{SandersRBGS}; and {(PS21)} this work Appendix~\ref{apx:spitzer_fluxes}.
Radio: (B89) \citealt{Barvainis1989};  (B06) \citealt{Baan2006}; {(BM15) \citealt{BarcosMunoz2015}}; (C90) \citealt{Condon1990}; (C91) \citealt{Condon1991}; (C96) \citealt{Condon1996}; (C98) \citealt{Condon1998}; (H15) \citealt{Helfand2015}; (H21) \citealt{Hayashi2021}; (L11) \citealt{Leroy2011radio}; (M03) \citealt{Mauch2003}; (M07) \citealt{Murphy2007}; (M17) \citealt{Meyers2017}; (N03) \citealt{Nagar2003}; (S98) \citealt{Stanghellini1998}; and (W94) \citealt{Wright1994}; 
}
\end{table}

\subsubsection{Infrared SED}\label{ss:ir_sed}
We fitted the IR spectral energy distribution of these ULIRGs to determine the expected dust emission at the ALMA frequency (220--250\,GHz$\equiv$1400--1200\,\micron). We used published IR photometry from \textit{Herschel}, \textit{ISO}, and \textit{IRAS} (see Table~\ref{tbl:bib_fluxes}), as well as \textit{Spitzer} IRS and MIPS data (see Sect.~\ref{ss:data_spitzer} and Appendix\,\ref{apx:spitzer_fluxes}). For observation at similar wavelengths, we gave preference to the data with the highest angular resolution.

We fitted the far-IR SED using a single-temperature gray-body (e.g., equations 1 and 2 of \citealt{Kovacs2010}). We assumed a fixed $\beta=1.8$ \citep{Planck2011}, but allowed the optical depth to vary. This model reproduces well the observed far-IR SED between 30 and 500\micron\ for most objects.
At shorter wavelengths, some of these ULIRGs have excess mid-IR emission which has been associated with warmer dust due to an AGN (e.g., \citealt{Nardini2009, Veilleux2009}). Since we are interested in the longer wavelength emission to compare with the ALMA observation, we only used in the fit the photometric points between 34 and 500\micron\ to avoid any bias due to this mid-IR excess. We note that we excluded the ALMA continuum flux from the SED fit.
For four systems (00509+1225, 01572+0009, F05189-2524, and F13451+1232), we started the fit at 70\micron\ because the excess mid-IR emission was clear even at the 34\micron\ photometric point. Also, for F13451+1232 (which hosts the radio source 4C$+$12.50), we excluded the 500\micron\ flux because it has a noticeable contribution from non-thermal emission.
We show the best-fit models in Figs.~\ref{fig:sed_models} and \ref{fig:apx_sed_models} and the model parameters are presented in Table~\ref{tbl:sed_result}.
To compute the total IR luminosity between 6 and 1500\micron\ (rest-frame), we first integrated the gray body emission. Then we subtracted the gray body model to the IRS spectrum and obtained the 6--20\micron\ mid-IR excess, $\Delta L_{\rm 6-20\mu m}$. The total $L_{\rm IR}$ is the addition of the gray body emission and the mid-IR excess. The total $L_{\rm IR}$ and the $\Delta L_{\rm 6-20\mu m}$\slash $L_{\rm IR}$ ratio are listed in Tables~\ref{tbl:sample} and \ref{tbl:sed_result}, respectively.

\subsubsection{Non-thermal synchrotron emission}
At the frequency of the ALMA observations, it is possible to have a significant contribution from non-thermal synchrotron emission. To estimate this contribution, we used published radio observations of our sample of ULIRGs with frequencies between 1 and 40\,GHz (Table~\ref{tbl:bib_fluxes}). All the systems have been observed at least at 1.4\,GHz, except 13120$-$5453 which only has 4.85 and 0.843\,GHz data. For objects with more than one radio observation ({16} systems), we fitted a power law to the radio data. We obtained spectral indexes between $-$0.3 and $-$1.0 with a mean index of $-$0.62 (see Table~\ref{tbl:sed_result}), which are similar to the spectral indexes found in a sample of 31 local ULIRGs by \citet{Clemens2008}. For the remaining objects with just one radio observation at 1.4\,GHz (9 systems), we assumed the mean spectral index between 1.4 and 22.5\,GHz ($\alpha_{22.5}^{1.4}=-$0.671) measured in local ULIRGs \citep{Clemens2008}. We find that the non-thermal emission contributes between 4 and 55\% (median 20\%) of the ALMA continuum flux (Table~\ref{tbl:sed_result}). In this fit, we ignored that the free-free emission (see below) can also affect (flatten) the spectral index (e.g., \citealt{Hayashi2021}). Therefore, if the free-free emission is strong compared to the synchrotron, our ``non-thermal'' contribution estimate would be closer to the combined free-free$+$synchrotron total emission. 

For 20087$-$0308, the predicted non-thermal flux is 2.8 times higher than the observed ALMA flux. However, this source shows radio fluxes between 1.4 and 4.85\,GHz that are not fully compatible with a power-law (Fig.~\ref{fig:apx_sed_models}). This suggests that either the power-law model is not adequate for this source, that it presents variable radio emission, or that some of the radio fluxes are not reliable. 

\subsubsection{Free-free thermal emission}

The ALMA continuum measurements can also include a contribution from thermal free-free emission. The free-free emission is related to the ionizing photon rate from young stars and can trace the SFR (e.g., \citealt{Condon2016}). We used the relation between the SFR and the free-free radio emission to estimate its contribution at the ALMA frequency (equation~11 from \citealt{Murphy2011}). The SFR of the ULIRGs was derived from the total IR luminosity using the \citet{Kennicutt2012} calibration. The free-free contributions are presented in Table~\ref{tbl:sed_result}. However, we note that this free-free emission estimate is an upper limit because we ignored the potential AGN contribution to the $L_{\rm IR}$, and also because, in the dusty nuclear regions of ULIRGs, a fraction of the ionizing photons can be absorbed by dust grains instead of ionizing H atoms and, therefore, reduce the actual free-free emission (e.g., \citealt{Abel2009}). Actually, with these assumptions, six systems (five classified as AGN and one as LINER) have predicted free-free upper limits above 90--100\% of the observed ALMA flux. This indicates that a large part of their $L_{\rm IR}$ likely comes from an AGN and {it} cannot be directly translated into SFR and subsequently into free-free emission.

\begin{table*}[t]
\caption{IR and radio SED fit results}
\label{tbl:sed_result}
\centering
\begin{small}
\begin{tabular}{lccccccccc}
\hline \hline
\\
IRAS name & $T_{\rm dust}\slash$K\tablefootmark{a} & $\tau_{\rm 350\mu m}$\,\tablefootmark{b} & $\log \Delta L_{\rm 6-20\mu m}\slash L_{\rm IR}$ & $\alpha_{\rm non-thermal}$\tablefootmark{c} & $F_{\rm dust}\slash F_{\rm ALMA}$\tablefootmark{d} & $F_{\rm free-free}\slash F_{\rm ALMA}$\tablefootmark{d} & $F_{\rm non-thermal}\slash F_{\rm ALMA}$\tablefootmark{d} \\
\hline
00091$-$0738 & 78  &  0.40 & $-$1.45$\pm$0.19   & $-$0.33$\pm$0.05 & 0.22$\pm$0.04 & $<$0.24 & 0.20$\pm$0.03 \\
00188$-$0856 & 71  &  0.71$\pm$0.09 & $-$1.23$\pm$0.10   & $-$0.79$\pm$0.05 & 1.35$\pm$0.14 & $<$0.67 & 0.21$\pm$0.05 \\
00509$+$1225 & 57  &  0.66$\pm$0.17 & $-$0.34$\pm$0.01   & $-$0.68$\pm$0.05 & 4.39$\pm$0.56 & $<$1.77 & 0.16$\pm$0.03 \\
01572$+$0009 & 77  &  0.42$\pm$0.17 & $-$0.73$\pm$0.11   & $-$1.02$\pm$0.05 & 0.90$\pm$0.08 & $<$1.16 & 0.15$\pm$0.04 \\
F05189$-$2524 & 66 &  0.28$\pm$0.17 & $-$0.72$\pm$0.10   & $-$0.54$\pm$0.10 & 1.31$\pm$0.10 & $<$0.99 & 0.28$\pm$0.12 \\
07251$-$0248 & 75  &  0.50$\pm$0.08 & $-$1.54$\pm$0.25   & $-$0.671 & 0.61$\pm$0.06 & $<$0.27 & 0.04 \\
09022$-$3615 & 65  &  0.30$\pm$0.04 & $-$1.10$\pm$0.06   & $-$0.671 & 1.39$\pm$0.10 & $<$0.70 & 0.40 \\
F10190$+$1322 & 61 &  0.40 & $-$1.33$\pm$0.06   & $-$0.671 & 1.49$\pm$0.18 & $<$0.46 & 0.17 \\
11095$-$0238 & 85  &  0.27$\pm$0.05 & $-$1.26$\pm$0.20   & $-$0.671 & 0.53$\pm$0.04 & $<$1.06 & 0.55 \\
F12072$-$0444 & 82 & 0.36$\pm$0.06 & $-$0.93$\pm$0.08 & $-$0.671 &  \nodata & \nodata & \nodata \\
F12112$+$0305 & 63 &  0.36$\pm$0.05 & $-$1.63$\pm$0.16   & $-$0.51$\pm$0.07 & 1.14$\pm$0.09 & $<$0.41 & 0.23$\pm$0.07 \\
13120$-$5453 & 58  &  0.37$\pm$0.05 & $-$1.45$\pm$0.07   & $-$0.63$\pm$0.05 & 1.82$\pm$0.15 & $<$0.50 & 0.38$\pm$0.10 \\
F13451$+$1232 & 70 & 0.64$\pm$0.24 & $-$0.66$\pm$0.05 & $-$0.54$\pm$0.03 &  \nodata & \nodata & \nodata \\
F14348$-$1447 & 64 &  0.35$\pm$0.04 & $-$1.62$\pm$0.14   & $-$0.75$\pm$0.06 & 1.62$\pm$0.12 & $<$0.65 & 0.17$\pm$0.06 \\
F14378$-$3651 & 66 &  0.32$\pm$0.05 & $-$1.48$\pm$0.18   & $-$1.00$\pm$0.25 & 1.95$\pm$0.14 & $<$0.91 & 0.08$\pm$0.10 \\
F15327$+$2340 & 63 & 0.57$\pm$0.10 & $-$1.86$\pm$0.24 & $-$0.53$\pm$0.05 & 0.73$\pm$0.06 & $<$0.26 & 0.13$\pm$0.02 \\
16090$-$0139 & 69  &  0.39$\pm$0.05 & $-$1.31$\pm$0.08   & $-$0.671 & 0.70$\pm$0.05 & $<$0.54 & 0.25 \\
16155$+$0146 & 95  &  0.40 & $-$0.86$\pm$0.13   & $-$0.671 & 0.56$\pm$0.05 & $<$1.20 & 0.49 \\
17208$-$0014 & 62  &  0.38$\pm$0.06 & $-$1.73$\pm$0.17   & $-$0.47$\pm$0.09 & 0.89$\pm$0.08 & $<$0.28 & 0.20$\pm$0.06 \\
F19297$-$0406 & 67 &  0.48$\pm$0.07 & $-$1.50$\pm$0.12   & $-$0.671 & 1.46$\pm$0.11 & $<$0.49 & 0.16 \\
19542$+$1110 & 69  &  0.35$\pm$0.05 & $-$1.40$\pm$0.16   & $-$0.57$\pm$0.04 & 1.47$\pm$0.09 & $<$0.72 & 0.32$\pm$0.04 \\
20087$-$0308 & 61  &  0.46$\pm$0.08 & $-$1.45$\pm$0.09   & $-$0.42$\pm$0.08 & 0.82$\pm$0.06 & $<$0.37 & 2.79$\pm$1.24 \\
20100$-$4156 & 80  &  0.48$\pm$0.09 & $-$1.49$\pm$0.22   & $-$0.671 & 0.56$\pm$0.05 & $<$0.52 & 0.15 \\
20414$-$1651 & 66  &  0.54$\pm$0.08 & $-$1.68$\pm$0.21   & $-$0.70$\pm$0.06 & 1.32$\pm$0.12 & $<$0.38 & 0.15$\pm$0.04 \\
F22491$-$1808 & 75 &  0.33$\pm$0.04 & $-$1.66$\pm$0.26   & $-$0.39$\pm$0.08 & 0.65$\pm$0.04 & $<$0.43 & 0.18$\pm$0.05 \\
\hline
Mean & 70$\pm$9 & 0.43$\pm$0.13 & $-$1.30$\pm$0.08 & \nodata & 1.21$\pm$0.83 & $<$0.77 & 0.34$\pm$0.55 \\
Median & 67$\pm$7 & 0.38$\pm$0.12 & $-$1.45$\pm$0.07 & \nodata & 1.14$\pm$0.62 & $<$0.65 & 0.20$\pm$0.08 \\
\hline
\end{tabular}
\end{small}
\tablefoot{
\tablefoottext{a}{The typical dust temperature uncertainty is $\pm$3\,K.}
\tablefoottext{b}{Dust optical depth at 350\micron. For objects with less than four IR photometric points, we assumed a fixed $\tau_{\rm 350\mu m}$ of 0.4 based on the average $\tau_{\rm 350\mu m}$ of the sample.}
\tablefoottext{c}{We assumed a spectral index of $-$0.671 for the objects with only one radio observation (see Sect.~\ref{ss:sed_model}).}
\tablefoottext{d}{Ratio between the flux densities predicted by the gray-body, upper limit free-free, and non-thermal synchrotron models and the observed $\sim$220\,GHz ALMA continuum flux densities, respectively.}
}
\end{table*}

\section{Discussion}\label{s:discussion}

\subsection{ALMA continuum as tracer of the IR luminosity}\label{ss:alma_bulk_ir}

We aim to determine the physical size and luminosity surface density of the regions that emit the bulk of the IR luminosity in local ULIRGs. Far-infrared telescopes, that detect the peak of the IR emission, lack the angular resolution to spatially resolve it, although it is possible to infer the size of the far-IR emission through indirect methods like the modeling of far-IR OH absorptions (e.g., \citealt{GonzalezAlfonso2015}).
In this section, we investigate if the $\sim$220\,GHz ALMA continuum, which provides much higher angular resolutions, can be used as a proxy of the IR emission to obtain direct estimates of the IR emitting region sizes.

However, using the $\sim$220\,GHz continuum to trace the size of ULIRGs is not straightforward. At this frequency, the continuum includes emission from dust, which is connected to the IR luminosity, but it may also include contributions from free-free and synchrotron emissions \citep{Condon2016}, which might not be directly related to the IR luminosity. This is important because, even if the IR luminosity of local ULIRGs is thought to be dominated by SF, the AGN contribution increases with increasing $L_{\rm IR}$ \citep{Veilleux2009, Nardini2009}, and the synchrotron AGN emission could affect the $\sim$220\,GHz source sizes. In addition, ALMA is an interferometer, so part of the emission might be filtered out. In this section, we study the impact of these effects on the measured source sizes.

\subsubsection{Filtered out flux}

In general, due to the limited coverage of the $uv$ plane, interferometric observations filter out extended large scale emission. Therefore, it might be possible that extended continuum emission from our ULIRGs is missing in the measured fluxes (Sect.~\ref{ss:alma_model}). 
To evaluate this possibility we compare the maximum recoverable scale of our observations (about 4\,kpc; Sect.~\ref{ss:data_alma}) with the sizes of the detected sources. The radii range from $<$60 to 300\,pc (Sect.~\ref{ss:alma_model} and Table~\ref{tbl:alma_cont_model}), which are $\ll$4\,kpc. If $>$4\,kpc structures were actually present in these ULIRGs, we would expect to detect, in addition to these very compact sources, intermediate size structures ($\sim$2\,kpc FHWM) which are not seen in the continuum images.
This suggests that the 220\,GHz continuum of ULIRGs is intrinsically compact and that we can recover most of the continuum emission with these data.

It also possible that we do not detect extended low-surface brightness continuum emission due to the observations sensitivity. To quantify its possible impact, we assume an emitting area with a 4$\times$4\,kpc$^2$ size (the typical H$\alpha$ effective radius of ULIRGs is $<$2\,kpc; \citealt{Arribas2012}). From the continuum sensitivity, Table~\ref{tbl:alma_obs}, we estimated extended emission 3$\sigma$ upper limits which are on average 8\% of the measured fluxes and up to 20--75\% for the 4 faint nuclei with $f_{\rm 220\,GHz}<0.7$\,mJy (F10190+1322 W; 11095$-$0238 SW, F12112+0305 SW, and 16155+0146 NW; see Table~\ref{tbl:alma_cont_model}). Therefore, if low-surface brightness emission is present, its contribution would be small for the great majority of the nuclei (at least for {25} out of {29}).

\subsubsection{Dust, free-free, and synchrotron contributions}\label{ss:disc_alma_comp}

At the frequencies of the ALMA observations (190--250\,GHz), in addition to the Rayleigh-Jeans tail of the IR dust emission, a contribution from thermal free-free, and non-thermal synchrotron emission is possible. In particular, we explore whether the free-free or synchrotron emissions could bias the measured sizes toward more compact sizes in the case of a starburst nucleus.

\paragraph{Dust}
We first estimate the dust contribution. Based on the IR SED modeling (Sect.~\ref{ss:sed_model}), we found that the ALMA flux densities are just slightly lower than the extrapolation of the IR gray-body fit (median ratio of {1.14}; Table~\ref{tbl:sed_result}). We note that we did not use the ALMA flux in the gray-body fit. Thus, a possible interpretation for the good agreement between the data and the model prediction is that the ALMA flux comes from the long-wavelength tail of the dust gray-body  emission. If this is the case, we could use the high-resolution ALMA data to determine the size of the IR emitting regions. This good agreement between the ALMA continuum flux and the IR SED extrapolation was also found  by \citet{Imanishi2019} using 260\,GHz observations of local ULIRGs at comparable spatial resolutions.

\paragraph{Free-Free}
The ALMA emission can include free-free emission as well. The free-free emission is produced by ionized hydrogen usually associated with star-forming regions (see Sect.~\ref{ss:ir_sed}). We estimated an upper limit for the free-free emission assuming that all the $L_{\rm IR}$ is produced by SF (i.e., ignoring the possible AGN contribution). This assumption also implies that dust does not absorb any ionizing photon (Sect.~\ref{ss:ir_sed}). The upper limit for the free-free contribution has a median value of {$<$65\%} (Table~\ref{tbl:sed_result}). But even if the free-free emission dominates the ALMA flux, it should not affect the region size estimates since, for this free-free upper limit estimation, both IR and free-free emissions should be co-spatial as they have a common star-formation origin. Moreover, taking into account the AGN contribution to the $L_{\rm IR}$ and the effect of absorption of UV photons by dust would reduce this upper limit and, therefore, the possible impact of the free-free emission on the source size measurements.

\paragraph{Synchrotron}
Some contribution from synchrotron emission is possible too. If this synchrotron emission is produced by supernovae (i.e., related to star-forming regions), the ALMA regions sizes should not be affected since the IR emission and the supernovae (SNe) should have similar spatial distributions.
Alternatively, AGN can produce strong synchrotron emission. In our sample, only F13451+1232 (4C$+$12.50) has excess radio emission with respect to the radio-IR relation (see \citealt{Perna2021}). Actually, the F13451+1232 $\sim$220\,GHz emission is dominated by synchrotron radiation (see Fig.~\ref{fig:apx_sed_models}) and, therefore, cannot be directly used as a proxy of the IR emitting region.
For the remaining objects, we estimated a median non-thermal contribution, which includes both SNe and AGN emission, of 20\% and up to 40--60\% in four objects: one of the starbursts (09022$-$3615), one LINER (11095$-$0238), and 2 out of the 7 systems classified as AGN in the optical (13120$-$5453, and 16155$+$0146). The latter suggests that the optical detection of an AGN does not imply that the $\sim$220\,GHz emission is always dominated by synchrotron AGN emission in local ULIRGs.
However, how the AGN synchrotron emission affects the ALMA source sizes is unclear. AGN jets producing synchrotron emission have sizes ranging from pc to few kpc in radio-quiet AGN \citep{Hardcastle2020}. For instance, the synchrotron radio jet emission from the AGN ULIRG 01572+0009 (PG~0157+001) has a $\sim$7\,kpc diameter \citep{Leipski2006}, although according to our SED modeling, the synchrotron contribution at 220\,GHz is small, about 0.15, and should not affect the estimated size in this object.

For the 5 systems with a high non-thermal contribution, we find that their sizes and luminosity and molecular gas surface densities do not differ from those of the rest of the sample (see Fig.~\ref{fig:histo_size}).

\subsubsection{Summary}
It seems likely that these $\sim$220\,GHz continuum ALMA observations trace the IR emitting region for the majority of local ULIRGs and that the filtered out flux due to the interferometric observations is small. The good agreement between the IR gray-body extrapolation and the ALMA fluxes supports this. Free-free and SNe synchrotron emissions could contribute to the ALMA flux, but since they have a star-formation origin they should not affect the size estimates for a starburst ULIRG. Synchrotron emission from AGN could bias the size measurements, although we do not find significant differences in size between the 5 systems with high synchrotron emission and the rest of the sample.
Therefore, in the following sections we assume that the size of the $\sim$220\,GHz continuum is equivalent to the size of the region which emits the bulk of the IR luminosity in these ULIRGs.

\begin{figure*}[t]
\centering
\vspace{5mm}
\includegraphics[width=0.6\textwidth]{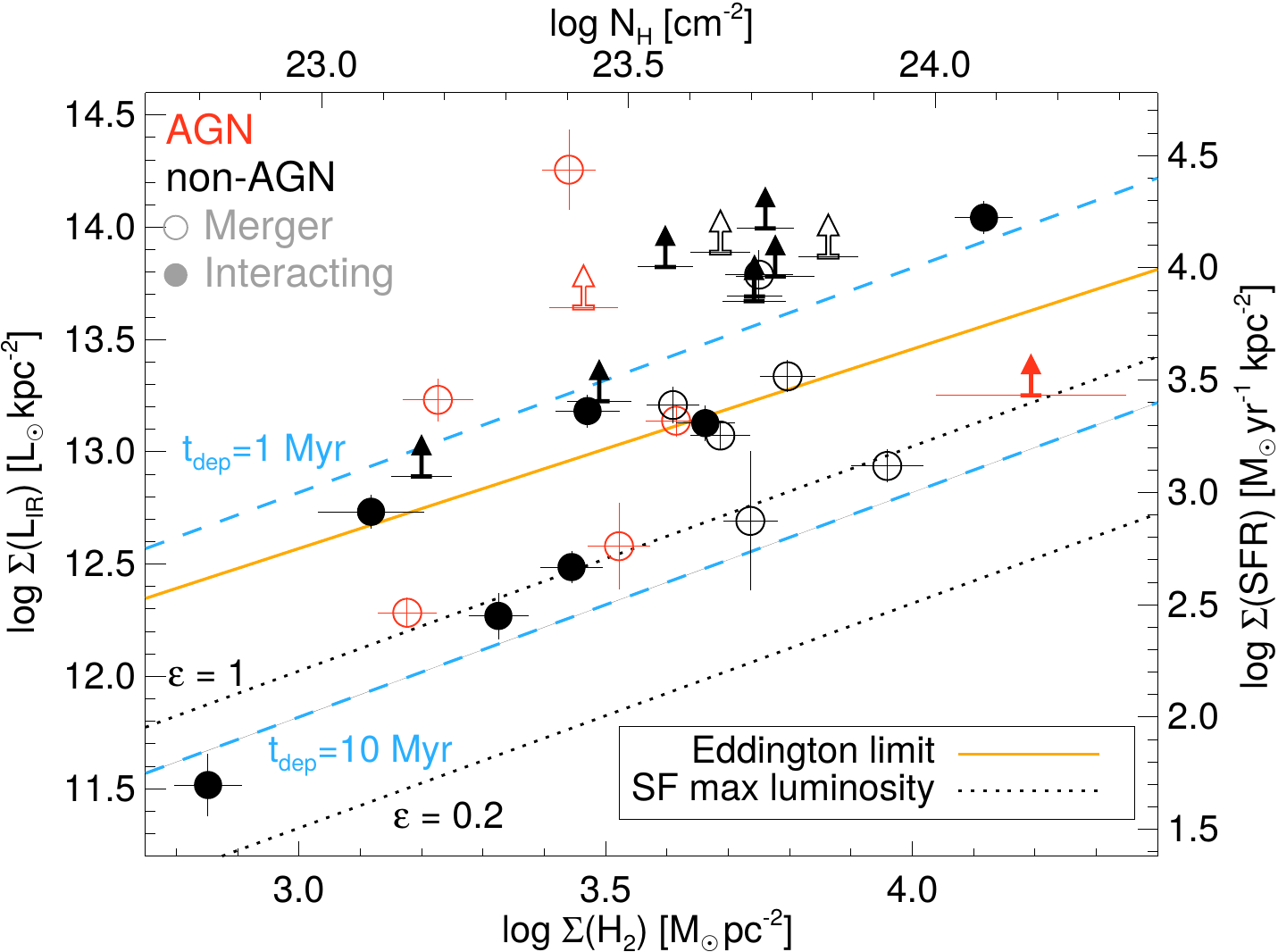}
\caption{Cold molecular gas surface density ($\Sigma_{\rm H_2}$) vs. IR luminosity surface density ($\Sigma_{L_{\rm IR}}$). {Galaxy symbols are as in Fig.~\ref{fig:size_cont_co}.} The dotted black lines indicate the maximum luminosity from an instantaneous starburst using 100\% ($\epsilon = 1$) or 20\% ($\epsilon = 0.2$) of the available cold molecular gas (see Sect.~\ref{ss:thick_SB}). 
The orange solid line is the Eddington luminosity limit. For points above this line, the radiation pressure is stronger than gravity (Sect.~\ref{ss:Edd_limit}). The dashed blue lines indicate the 1 and 10\,Myr depletion times for reference assuming that the IR luminosity is produced by SF. The column density ($N_{\rm H}$) axis is calculated from $\Sigma_{\rm H_2}$ assuming a uniform mass distribution (i.e., $N_{\rm H} = 2 \times \Sigma_{\rm H_2}\slash m({\rm H_2})$ where $m({\rm H_2})$ is the H$_2$ molecular weight).\label{fig:lir_density}}
\end{figure*}

\subsection{Extreme nuclear IR luminosity densities}

Using the half-light radius, $r_{\rm cont}$, of the ALMA continuum and half of the IR luminosity (Sects.~\ref{ss:alma_model} and \ref{ss:sed_model}), we calculated the luminosity surface density, $\Sigma_{L_{\rm IR}}$, in the nuclear regions of the ULIRGs. 
For systems with 2 nuclei, we estimated their IR luminosity fraction using their relative ALMA continuum fluxes. In 70\% of the interacting systems, the luminosity is completely dominated ($>$90\% of the total luminosity) by one of the nuclei. These fractions and the resulting surface densities are listed in Table~\ref{tbl:lir_density}.
We find $\log \Sigma_{L_{\rm IR}}\slash (L_\odot$\,kpc$^{-2})$ between 11.5 and 14.3 with a median value of  13.2. If this IR luminosity is produced by SF, it corresponds to $\Sigma_{\rm SFR}=2500$\,$M_\odot$\,yr$^{-1}$\,kpc$^{-2}$ using the \citet{Kennicutt2012} SFR calibration.
These values are much higher (1--2 orders of magnitude) than the densities found in local starburst LIRGs, even when they are observed at higher angular resolutions of $\sim$100\,pc (e.g., \citealt{Xu2015, Pereira2016, Michiyama2020}). Similarly, z$\sim$3--6 sub-mm galaxies have lower surface densities, 150--1300\,$M_\odot$\,yr$^{-1}$\,kpc$^{-2}$  when observed at $\sim$kpc resolutions (e.g., \citealt{Riechers2017, GomezGuijarro2018}). At higher resolutions, $\sim$200\,pc, these sub-mm galaxies have $\Sigma_{\rm SFR}$ between 100 and {3000}\,$M_\odot$\,yr$^{-1}$\,kpc$^{-2}$ which are still lower than the majority of the local ULIRGs (e.g., \citealt{Oteo2017, Gullberg2018}).

The observed $\Sigma_{L_{\rm IR}}$ range is comparable to that measured in local ULIRGs using radio and IR observations at similar angular resolution. Using 33\,GHz radio data, \citet{BarcosMunoz2017} found a median surface density of 10$^{12.8}$\,$L_\odot$\,kpc$^{-2}$ in a sample of 22 local interacting\slash merging systems with $\log L_{\rm IR}\slash L_\odot>11.6$. Similarly, \citet{GonzalezAlfonso2015} estimated nuclear $\Sigma_{L_{\rm IR}}>$10$^{12.8}$\,$L_\odot$\,kpc$^{-2}$ for 10 local ULIRGs based on the modeling of the far-IR OH absorptions. In addition, {mid-IR ground-based studies of ULIRGs derive maximum luminosity densities between 10$^{12.1}$ and 10$^{14.6}$\,$L_\odot$\,kpc$^{-2}$ \citep{Soifer2000, Imanishi2011}.}
{For the western nucleus of F15327+2340 (Arp~220 W), using higher resolution data ($<50$\,pc) a luminosity density about  10$^{14.3-15.5}$\,$L_\odot$\,kpc$^{-2}$ has been estimated \citep{Downes2007, Wilson2014, Sakamoto2017}.}
Therefore, our results are compatible with previous findings and confirm that the luminosity density in the nucleus of local ULIRGs is much higher than in other local and high-z starbursts when measured at 100--1000\,pc scales.

\begin{table*}[t]
\caption{Nuclear properties}
\label{tbl:lir_density}
\centering
\begin{tabular}{lcccccccc}
\hline \hline
\\
IRAS name & Nucleus & Frac.\tablefootmark{a} & $\log \Sigma_{L_{\rm IR}}$\,\tablefootmark{b} & $t_{\rm dep}$\,\tablefootmark{c} & Above Eddington & Optical AGN?\,\tablefootmark{e} & CON\,\tablefootmark{f} & HCN\,14\micron\,\tablefootmark{g} \\
& & & $(L_\odot$\,kpc$^{-2})$  &  (Myr) & limit?\,\tablefootmark{d} \\
\hline
00091$-$0738 & S & 0.97 & $>$13.67 &  $<$0.78 & Y & N & \nodata & Y\\
 & N & $<$0.03 & \nodata & \nodata & \nodata & \nodata & \nodata\\
00188$-$0856 & - & 1.00 & 13.15$\pm$0.08 &  2.0$\pm$0.1  & Y & Y & \nodata & N\\
00509$+$1225 & - & 1.00 & 13.30$\pm$0.09 & 0.65$\pm$0.09  &  Y & Y & \nodata & N\\
01572$+$0009 & - & 1.00 & 14.30$\pm$0.18 & 0.10$\pm$0.01  &  Y & Y& \nodata & N\\
F05189$-$2524 & - & 1.00 & $>$13.70 & $<$0.4  &  Y & Y & \nodata & Y\\
07251$-$0248 & W & 0.09 & 12.74$\pm$0.07  & 1.6$\pm$0.3  & Y & N & \nodata & Y\\
& E & 0.91 & $>$13.78 & $<$0.65  &  Y & N & \nodata\\
09022$-$3615 & - & 1.00 & 12.70$\pm$0.24 & 7.3$\pm$0.7  & N & N & N  & N\\
F10190$+$1322 & W & 0.12 & 11.53$\pm$0.11 & 14.3$\pm$1.8  &  N & N & \nodata & N\\
& E & 0.88 & 12.50$\pm$0.08 & 6.0$\pm$0.7 & N & N & \nodata\\
11095$-$0238& SW & 0.32 & $>$12.88  & $<$1.3  &  Y & N & \nodata & Y\\
& NE & 0.68 & $>$13.22 & $<$1.2  &  Y & N & \nodata\\
F12112$+$0305 & SW & 0.09 & 12.27$\pm$0.12 & 7.5$\pm$0.8  & N & N & \nodata & N\\
& NE & 0.91 & $>$14.00 & $<$0.38  &  Y & N & ? \\
13120$-$5453 & - & 1.00 & 12.59$\pm$0.23 & 5.8$\pm$0.7  & N & Y & N  & N\\
F14348$-$1447 & SW & 0.63 & $>$13.83 & $<$0.39  &  Y & N & N  & N\\
 & NE & 0.37 & 13.19$\pm$0.08 & 1.3$\pm$0.2  &  Y & N & N \\
F14378$-$3651 & - & 1.00 & 12.29$\pm$0.07 & 5.2$\pm$0.6  & N & Y & ?  & N\\
F15327$+$2340 & W & 0.71 & $>$13.89 & 0.42$\pm$0.05 &  Y & N & Y & Y \\
 & E & 0.29 & 13.07$\pm$0.07 & 2.7$\pm$0.3 &  N & N & ? &\\
16090$-$0139 & - & 1.00 & 12.94$\pm$0.07 & 7.0$\pm$0.9  & N & N& \nodata  & Y\\
16155$+$0146 & NW & 0.80 & $>$13.26  & $<$5.8 & ? & Y& \nodata & N\\
 & SE & $<0.2$ & \nodata & \nodata  & \nodata & \nodata & \nodata\\
17208$-$0014 & - & 1.00 & $>$13.88 & $<$0.65 &  Y & N & Y & Y\\
F19297$-$0406 & S & $<0.01$ & \nodata & \nodata &  \nodata & \nodata & \nodata & Y\\
 & N & 0.99 & 13.13$\pm$0.07 & 2.3$\pm$0.2 & N & N & \nodata\\
19542$+$1110 & - & 1.00 & 13.22$\pm$0.07  & 1.7$\pm$0.2 & Y & N & \nodata & N\\
20087$-$0308 & - & 1.00 & 13.32$\pm$0.06 & 2.0$\pm$0.2 & Y & N & \nodata & N\\
20100$-$4156 & NW & $<$0.02 & \nodata & \nodata &  \nodata & \nodata & \nodata & Y\\
 & SE & 0.98 & 14.04$\pm$0.07 & 0.78$\pm$0.09 &  Y & N & \nodata\\
20414$-$1651 & - & 1.00 & 13.80$\pm$0.11  & 0.60$\pm$0.08 &  Y & N & \nodata & N\\
F22491$-$1808 & W & $<$0.02 & \nodata & \nodata  & \nodata & \nodata & \nodata & Y\\
 & E & 0.98 & $>$13.69 & $<$0.74  & Y & N & Y\\
\hline
\end{tabular}
\tablefoot{
\tablefoottext{a}{Fraction of the total IR luminosity (Table~\ref{tbl:sample}) which is assigned to each nucleus based on their relative ALMA continuum fluxes.}
\tablefoottext{b}{Logarithm of the IR luminosity surface density assuming the source size of the ALMA continuum (Table~\ref{tbl:alma_cont_model}).}
\tablefoottext{c}{Depletion time using the molecular gas surface density from Table~\ref{tbl:alma_co21} and the IR luminosity surface density in this Table assuming that the latter is completely produced by SF.}
\tablefoottext{d}{Indicates if the nucleus is above (Y), or not (N), the Eddington limit estimated in Sect.~\ref{ss:Edd_limit}.}
\tablefoottext{e}{Indicates whether an AGN is detected (Y), or not (N), from optical spectroscopy. See also Table~\ref{tbl:sample}.}
\tablefoottext{f}{Objects classified as CONs based on their HCN-vib luminosity (see \citealt{Falstad2021}). ``?'' indicates that HCN-vib emission is detected but below the CON threshold ($\Sigma_{\rm HCN-vib}$<1\,\Lsun\,pc$^{-2}$).}
\tablefoottext{g}{Indicates if the mid-IR HCN-vib\,14\micron\ absorption is detected in the system integrated  \textit{Spitzer}\slash IRS spectra from the archive (see also \citealt{Lahuis2007}).}
}
\end{table*}

To investigate the origin of this high $\Sigma_{L_{\rm IR}}$ values, we discuss two alternatives: an optically thick starburst and the presence of an obscured AGN.

\subsubsection{Optically thick starburst}\label{ss:thick_SB}

In Fig.~\ref{fig:lir_density}, we plot the $\Sigma_{L_{\rm IR}}$ vs. $\Sigma_{\rm H_2}$ (Sect.~\ref{ss:nuclear_mol}) relation for our sample of local ULIRGs. Based on theoretical models, a $\Sigma_{L_{\rm IR}}$ of $\sim$10$^{13}$\,$L_\odot$\,kpc$^{-2}$ has been suggested as the maximum $\Sigma_{L_{\rm IR}}$ of warm ($T<200$\,K) optically thick starbursts \citep{Thompson2005}. The maximum $\Sigma_{L_{\rm IR}}$ for these warm starbursts is similar to the median value found in our sample. Hot ($T > 200$\,K) optically thick starbursts could have  $\Sigma_{L_{\rm IR}}\sim10^{15}$\,$L_\odot$\,kpc$^{-2}$ when $\Sigma_{\rm H_2}>10^6$\,\Msun\,pc$^{-2}$ \citep{Andrews2011}.
However, we measure $\Sigma_{\rm H_2}\ll10^6$\,\Msun\,pc$^{-2}$, so these local ULIRGs might be more similar to the warm optically thick starburst models.

Although the $\Sigma_{L_{\rm IR}}$ are similar to the maximum for a warm optically thick starburst, when combined with the $\Sigma_{\rm H_2}$, the resulting depletion times, $\Sigma_{\rm H_2}$\slash $\Sigma_{\rm SFR}$, would be shorter ($<$1--15\,Myr; see Table~\ref{tbl:lir_density}) than those measured in other LIRG starbursts ($>$30--100\,Myr; \citealt{Xu2015, Pereira2016}).

We also estimated the maximum $\Sigma_{L_{\rm IR}}$ that a starburst can produce for a given $\Sigma_{\rm H_2}$.
Using \textsc{starburst}99 \hbox{\citep{Leitherer1999}}, we find that the maximum luminosity produced by a solar metallicity instantaneous burst, assuming a \citet{Kroupa2001} initial mass function (IMF), is $\sim$1060\,$L_\odot$\,$M_\odot^{-1}$ at an age of $\sim$2.2\,Myr {(see also \citealt{Sakamoto2013})}. In Fig.~\ref{fig:lir_density}, we show this limit ($\epsilon=1$ dotted black line). This limit assumes that all the molecular gas is instantaneously transformed into stars (i.e., 100\% efficient SF). In reality, stellar feedback  dissipates the molecular clouds before a 100\% efficiency is achieved, so the maximum luminosity from a starburst would be lower than this limit. Based on magneto-hydrodynamic simulations, the maximum efficiency per free-fall time is about 20\% (e.g., \citealt{Padoan2012}) which is shown in Fig.~\ref{fig:lir_density} too ($\epsilon=0.2$).
This figure shows that 70\% of the nuclei are above the 100\% efficiency limit ($\epsilon = 1$) and all of them are above this 20\% efficiency limit.

In addition, if the nuclear luminosity is produced by a compact and intense starburst, a continuous supply of molecular gas to the nucleus would be required to sustain the ongoing SFR level. Otherwise, it would not be possible to achieve the observed $\Sigma_{L_{\rm IR}}$ which implies depletion times $<$1\,Myr in 30\% of the sample and $<$15\,Myr in all of them. 
However, the high radiation pressure in the nucleus, which is compatible with being above the Eddington limit for {67\%} of the nuclei (see Sect.~\ref{ss:Edd_limit}), could prevent these massive gas inflows.

\subsubsection{Or obscured AGN?}

We find that 70\% of the ULIRGs have $\Sigma_{L_{\rm IR}}$\slash $\Sigma_{\rm H_2}$ ratios above the limit for a $\epsilon=1$ efficient starburst. Also, 65\% have $\Sigma_{L_{\rm IR}}$ above the theoretical value of an optically thick warm starburst ($\sim$10$^{13}$\,$L_\odot$\,kpc$^{-2}$). 
These fractions are similar for systems optically classified as AGN and starbursts (see Fig.~\ref{fig:lir_density}). These results suggest that what produces the bulk of the IR luminosity in these local ULIRGs is not a standard starburst.

Alternatively, an AGN could dominate the $L_{\rm IR}$ of these objects. {This possibility has also been suggested because of the high $\Sigma_{L_{\rm IR}}$ in the ULIRG nuclei derived from mid-IR data (e.g., \citealt{Imanishi2011}) or from the mm continuum in Arp~220 (e.g., \citealt{Downes2007, Wilson2014, Scoville2017, Sakamoto2017}). From $\Sigma_{\rm H_2}$, we estimate that} the nuclear H column densities are moderate, between 10$^{23}$ and 10$^{24}$\,cm$^{-2}$ (see Fig.~\ref{fig:lir_density}). These values are lower than the Compton thick limit ($\sim$2$\times$10$^{24}$\,cm$^{-2}$), so we would expect the AGN X-ray emission not to be completely absorbed.
\citet{Iwasawa2011} observed a sample of local U\slash LIRGs with \textit{Chandra} at 0.5--7\,keV. Twelve out of our {25} systems are part of their sample. They found AGN evidence in 6 out of the {12}, but they estimated low AGN contributions to the $L_{\rm IR}$ (3--20\%).
However, it is possible that the actual $N_{\rm H}$ that obscures these AGN is actually higher and could absorb the 0.5--7\,keV X-ray emission. Our $N_{\rm H}$ estimates are based on molecular gas observations at $\sim$400\,pc resolution, but the obscuring molecular torus could be smaller, as observed in local Seyfert galaxies (median diameter of 40\,pc; \citealt{GarciaBurillo2021}), so our $N_{\rm H}$ values might be underestimated.

To minimize the effects of the obscuring column density, we also considered the \textit{Swift}-BAT 105-Month 14--195\,keV survey \citep{Oh2018}. This higher energy X-ray band is less affected by $N_{\rm H}$ than the \textit{Chandra} 0.5--7\,keV range. The 5$\sigma$ sensitivity of the survey is 8.4$\times$10$^{-12}$\,erg\,s$^{-1}$\,cm$^{-2}$. Only one source in our sample, F05189$-$2524, is detected at 14--195\,keV. For the rest of the targets, the \textit{Swift}-BAT survey implies $L_{\rm 14-195\,keV} <$10$^{~43.3-44.8}$\,erg\,s$^{-1}$, depending on the distance. Assuming that the bolometric AGN luminosity is $L_{\rm AGN}\sim12\times L_{\rm 14-195\,keV}$ \citep{Marconi2004}, the 5$\sigma$ upper limits would correspond to $L_{\rm AGN}<$10$^{44.3-45.9}$\,erg\,s$^{-1}=10^{10.7-12.3}$\,$L_\odot$. This would result in an AGN contribution $L_{\rm AGN}$\slash $L_{\rm IR}<$0.45 for all but one of these ULIRGs and a median upper limit of $<$0.25.
{More sensitive NuSTAR $>$10\,keV observations were presented by \hbox{\citet{Teng2015}}. Their sample contains four of our ULIRGs, three of them already classified as Sy in the optical. One is undetected (F14378$-$3651), in F15327+2340 (Arp~220) no AGN evidence is found, although a very deeply buried AGN is still possible, and F05189$-$2524 and 13120$-$5453 are Compton-thin and -thick AGN, respectively.}

ULIRGs are known to be {hard} X-ray underluminous \citep{Imanishi2004c, Teng2015} and AGN in mergers are also heavily obscured \citep{Ricci2017}. The combination of these two factors could explain why the AGN in these sources, if present, remain {mostly} undetected in X-ray observations.

As discussed before, the high $\Sigma_{L_{\rm IR}}$\slash $\Sigma_{\rm H_2}$ nuclear ratios cannot be easily explained by a starburst, even if an optically thick one is considered. These ALMA data would be consistent with an AGN dominating the IR luminosity, but it is not possible to confirm that an AGN is present in the nuclei of the majority of these ULIRGs. The non-detection of these possible AGN in ultra-hard X-ray observations could indicate extremely high obscuring column densities. Higher angular resolution ALMA data could spatially resolve the obscuring material and establish its actual column density as in nearby Seyfert galaxies (e.g., \citealt{GarciaBurillo2021}).

\begin{figure*}[t]
\centering
\vspace{0.5cm}
\includegraphics[width=0.38\textwidth]{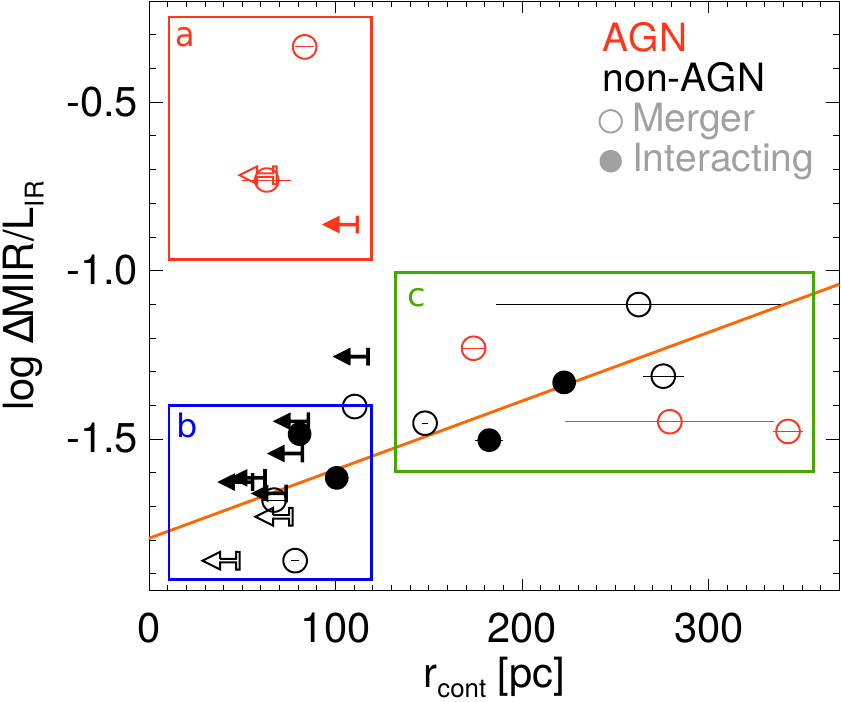}\hspace{1.5cm}
\includegraphics[width=0.38\textwidth]{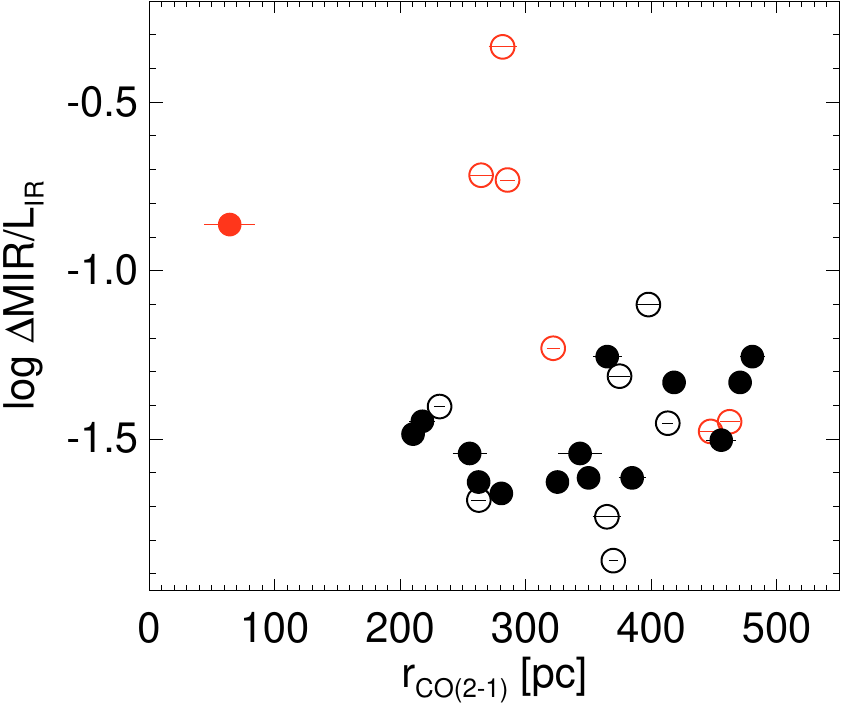}
\caption{Logarithm of the excess mid-IR emission vs. size of the 220\,GHz continuum ({left} panel) and size of the CO(2--1) emission ({right} panel). {Galaxy symbols are as in Fig.~\ref{fig:size_cont_co}}. The solid {orange} line is the best linear fit to the non-AGN (black) points. In the {left} panel, the red box (a) marks a region of this diagram solely occupied by optically detected AGN. The blue box (b) indicates the location of very compact ULIRG nuclei (mostly unresolved by our data) with negligible excess mid-IR emission. {The green box shows the location of more extended nuclei with higher mid-IR excess.}\label{fig:emir_r50}}
\end{figure*}

\subsubsection{Systematic uncertainties}

There are several assumptions which could bias the $\Sigma_{L_{\rm IR}}$ and $\Sigma_{\rm H_2}$ values presented in Fig.~\ref{fig:lir_density}. For instance, the $r_{\rm CO}$ size used to estimate the nuclear $\Sigma_{\rm H_2}$ is larger than the  $r_{\rm cont}$ used for $\Sigma_{L_{\rm IR}}$. This is because the CO(2--1) emission is more extended than the $\sim$220\,GHz continuum (\hbox{$r_{\rm CO}$\slash $r_{\rm cont}$}={2.5$\pm$1.1}; see Sect.~\ref{ss:nuclear_mol}) and it is not possible to exactly determine the amount of CO(2--1) within the $r_{\rm cont}$ region since, in most cases, both $r_{\rm cont}$ and $r_{\rm CO}$ are smaller than the beam size. Therefore, $\Sigma_{\rm H_2}$ are averaged over larger regions than $\Sigma_{\rm L_{\rm IR}}$ and the true nuclear $\Sigma_{\rm H_2}$ could be higher. Consequently, the real star-formation efficiency (depletion time) could be lower (longer).
Actually, \citet{GonzalezAlfonso2015}, by modeling the far-IR OH absorptions, estimated nuclear $\Sigma_{\rm H_2}$ between 10$^{3.8}$ and 10$^{4.7}$\,\Msun\,pc$^{-2}$ for 10 local ULIRGs. These values are $\sim$5 times higher than our average $\Sigma_{\rm H_2}$ derived from CO(2--1) {using $\sim$400\,pc resolution data}. In addition, {8} of our targets are part of a survey to detect vibrationally excited HCN emission to identify compact obscured nuclei (CONs). The HCN-vib (3--2) 267.199\,GHz line is detected in {5} of these systems ({63$\pm$18\%}; see Table~\ref{tbl:lir_density} and \citealt{Imanishi2016, Imanishi2019, Falstad2021}). Similarly, the mid-IR HCN-vib~14\micron\ absorption, which populates the levels originating the HCN-vib emission, is detected in {10 (6 interacting systems and 4 mergers) out of the 23 systems (43$\pm$10\% globally, 60$\pm$15\% of the interacting systems and 31$\pm$13\% of the advanced mergers; Table~\ref{tbl:lir_density})}. The presence of these HCN-vib spectral features suggests the presence of extreme nuclear column densities too.
Higher resolution CO(2--1) data will help us to establish the cold molecular gas density more accurately. The already available high resolution CO(2--1) data for 17208$-$0014 (120\,pc resolution; see Sect.~\ref{ss:alma_model_hires}), suggest a factor of 2 higher nuclear $\Sigma_{\rm H_2}$ (Pereira-Santaella in prep.). However, for this galaxy, the nuclear $\Sigma_{L_{\rm IR}}$ also increases by a factor of $\sim$5 using these data, so the resulting $\Sigma_{\rm SFR}$\slash $\Sigma_{\rm H_2}$ ratio would be even higher and reinforce the need for an obscured AGN in this object.

Another uncertainty related to the $\Sigma_{\rm H_2}$ is the $\alpha_{\rm CO}$ conversion factor. We assumed a ULIRG-like factor, which is relatively low compared to the conversion factor used for normal galaxies (see e.g., \citealt{Bolatto2013}). High nuclear column densities can produce self-absorbed CO(2--1) line profiles as seen in the compact nuclei of some local LIRGs (e.g., \citealt{Sakamoto2013}, \hbox{\citealt{Pereira2017Water}}, \citealt{ GonzalezAlfonso2021}). If the nuclear CO(2--1) emission of these ULIRGs is self-absorbed, the assumed ULIRG-like $\alpha_{\rm CO}$ conversion factor could result in underestimated $\Sigma_{\rm H_2}$ values. For this paper, we opt to use the standard ULIRG-like $\alpha_{\rm CO}$, as it is typically done in local ULIRG studies, but we will investigate the presence of self-absorbed CO(2--1) profiles in these targets in a future paper.
The depletion times depend on the assumed $\alpha_{\rm CO}$ conversion factor. If the actual $\alpha_{\rm CO}$ of these objects is similar to the Milky Way factor (i.e., 5 times higher; \citealt{Bolatto2013}), the molecular gas mass will be 5 times higher, and the depletion times 5 times longer. However, the median $t_{\rm dep}$ would still be very short ($<$6\,Myr).

Finally, the maximum luminosity for a starburst (dotted lines in Fig.~\ref{fig:lir_density}) is calculated using a \citet{Kroupa2001} IMF. If the IMF in ULIRGs is top-heavy as suggested by some works (see e.g., \citealt{Sliwa2017, Brown2019}), the maximum starburst luminosity per unit of molecular gas could be higher. For example, the maximum luminosity for an IMF truncated at 30\,\Msun would produce $\sim$10 times more luminosity per unit of molecular gas than the standard \citet{Kroupa2001} IMF. For an $\epsilon=$0.2 efficiency, this top-heavy IMF could explain the $\Sigma_{L_{\rm IR}}$\slash $\Sigma_{\rm H_2}$ ratios observed in about half of the nuclei, although these starbursts would have radiation pressures above the Eddington limit and they would not be stable on short timescales of few Myr (see Sect.~\ref{ss:Edd_limit}).

\begin{figure}
\centering
\vspace{5mm}
\includegraphics[width=0.37\textwidth]{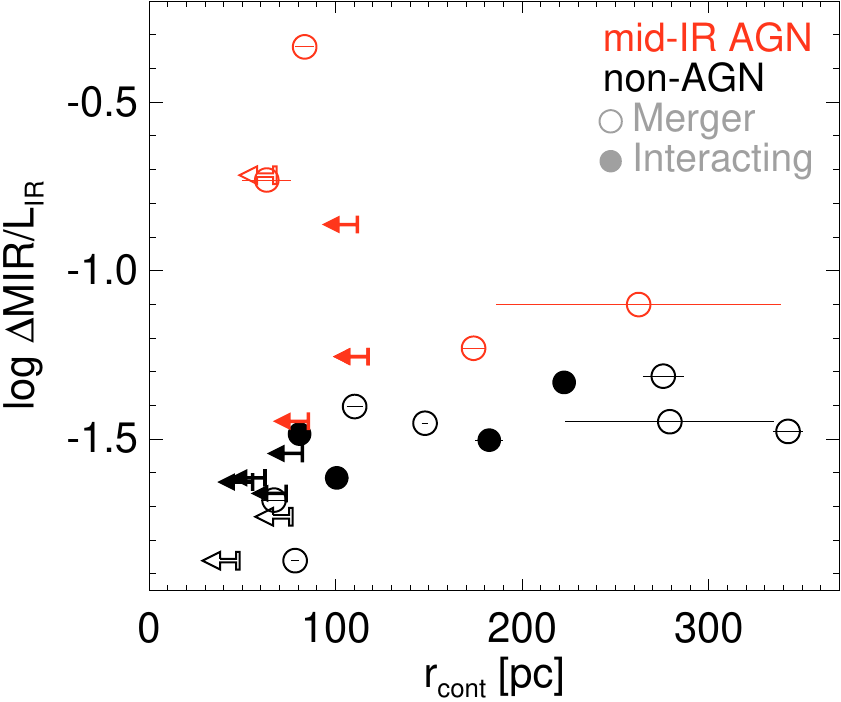}
\caption{Same as left panel of Fig.~\ref{fig:emir_r50}, but galaxies with more than 50\%\ AGN contribution based on mid-IR observations are indicated in red \citep{Veilleux2009,Nardini2010}.\label{fig:emir_r50_mir}}
\end{figure}

\begin{figure}
\centering
\vspace{5mm}
\includegraphics[width=0.25\textwidth]{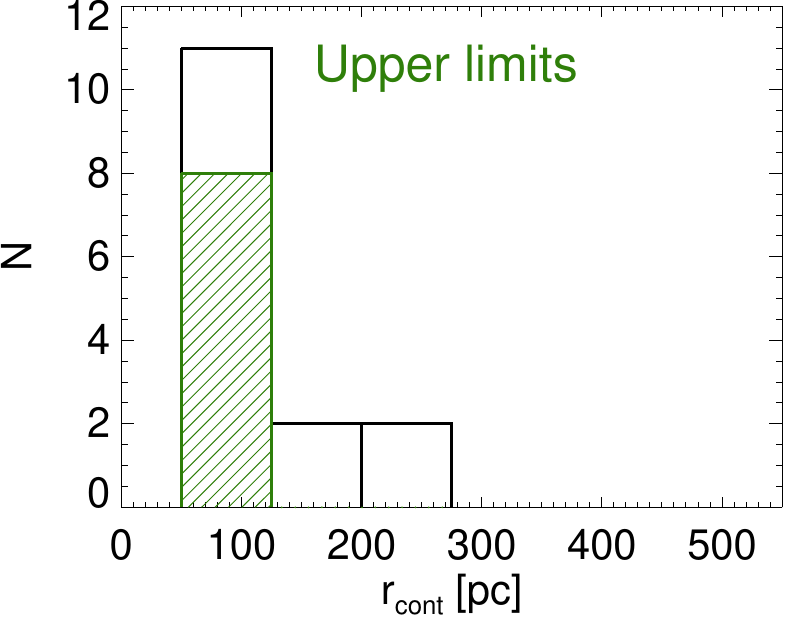}
\hspace{-7mm}
\includegraphics[width=0.25\textwidth]{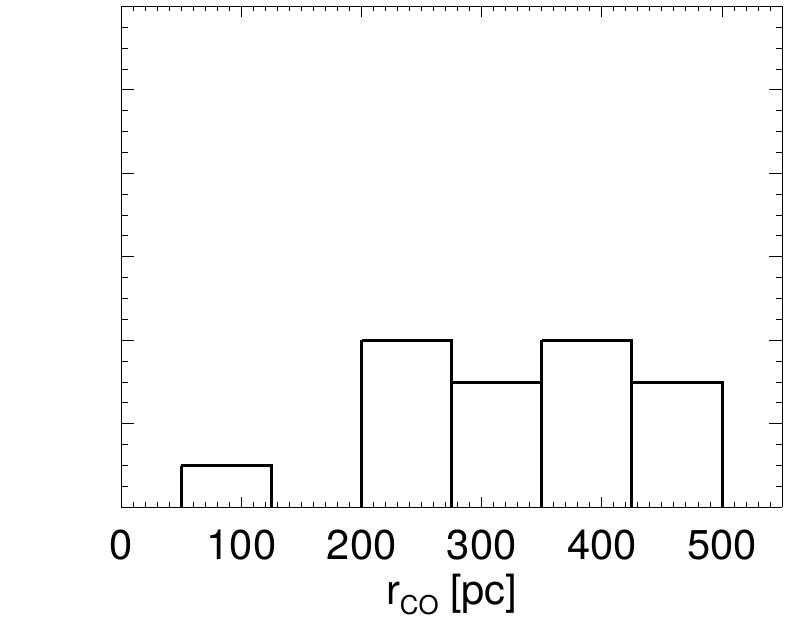}
\caption{Distribution of the 220\,GHz continuum radius $r_{\rm cont}$ (left) and the CO(2--1) emission radius $r_{\rm CO}$ (right) for the nuclei in interacting ULIRGs. The upper limits for the 220\,GHz continuum are included in the first bin and indicated by the green shaded area.\label{fig:histo_inter_sizes}}
\end{figure}

\begin{figure*}[t]
\centering
\vspace{0.5cm}
\includegraphics[width=0.31\textwidth]{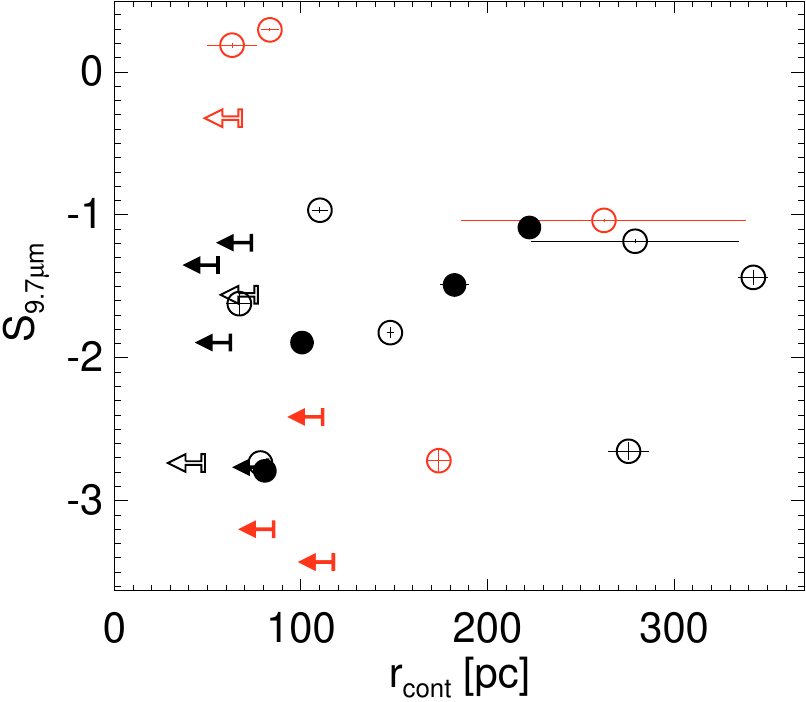}\hspace{2mm}
\includegraphics[width=0.31\textwidth]{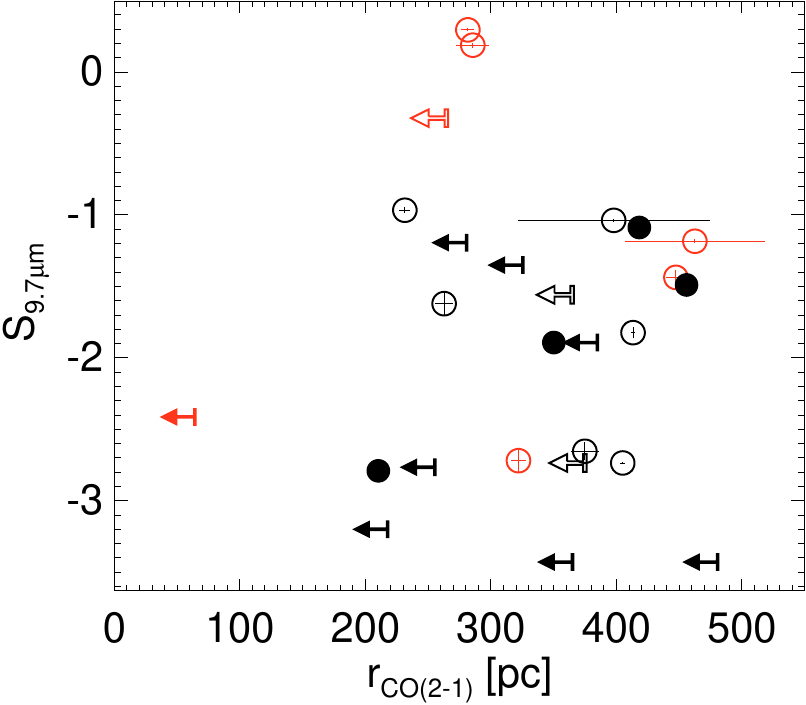}\hspace{2mm}
\includegraphics[width=0.324\textwidth]{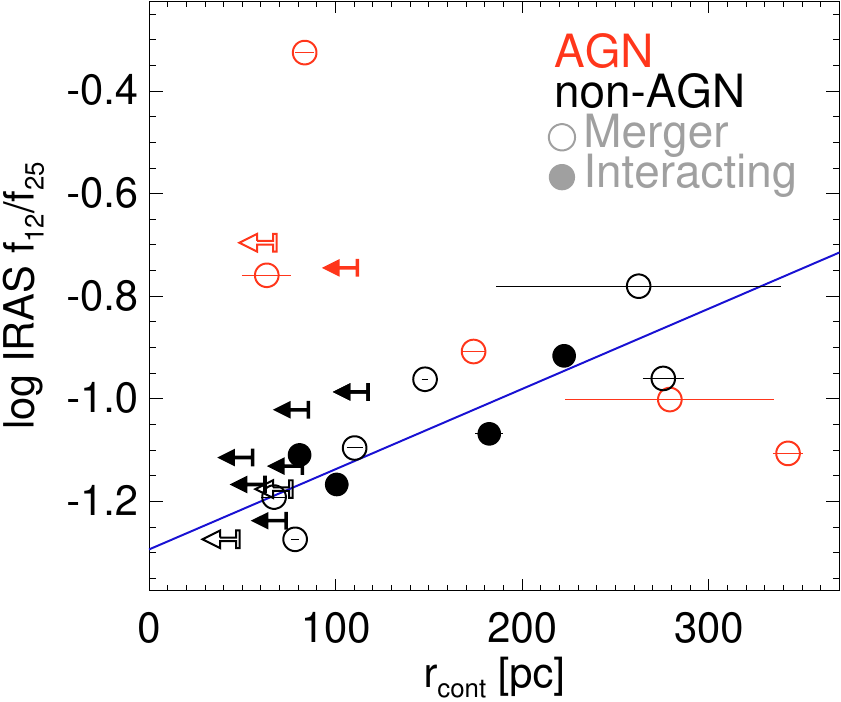}
\caption{{Relation between the 9.7\micron\ silicate absorption and the 220\,GHz continuum (left) and the CO(2-1) (middle) sizes. The right panel shows the 220\,GHz continuum size vs. IRAS 12\micron\slash 25\micron\ color relation}. Symbols as in Fig.~\ref{fig:size_cont_co}. {For the IRAS 12\micron\slash 25\micron\ color relation (right),} the best linear fit to the non-AGN objects (black symbols) {is $\log f_{\rm 12}$\slash $f_{\rm 25} = (-1.29\pm0.04) + (1.6\pm0.3)\times 10^{-3}\times r_{\rm cont}\slash {\rm pc}$.} 
\label{fig:iras_r50}}
\end{figure*}

\subsection{Continuum and CO(2--1) sizes vs. mid-infrared excess}\label{ss:mir_excess}

In the {left} panel of Fig.~\ref{fig:emir_r50}, we show the relation between the excess mid-IR emission (Sect.~\ref{ss:ir_sed}) and the 220\,GHz continuum size, $r_{\rm cont}$. {We note that the excess mid-IR emission is estimated from the system integrated {\it Spitzer}\slash IRS spectra. Therefore, in interacting systems where a nuclei dominates the total $L_{\rm IR}$, the mid-IR excess of the secondary nucleus cannot be directly estimated from this integrated spectrum. Thus, we excluded the 8 nuclei in interacting systems with IR luminosity fractions $<$20\% (see Table~\ref{tbl:lir_density}). In this diagram,}  we can classify the ULIRG nuclei in two main groups: compact objects ($r_{\rm cont}<$120\,pc) with high mid-IR excess (logarithm $>-0.9$; red box); and objects following a linear relationship between the two
properties with decreasing excess mid-IR emission for decreasing $r_{\rm cont}$ (blue line).

The first group exclusively contains objects classified as AGN in the optical. We note that the 2 interacting systems in our sample with no ALMA data, F12072$-$0444 and F13451$+$1232, are optical AGN with $\log \Delta L_{\rm 6-20\mu m}$\slash $L_{\rm IR}>-0.9$ and could lie in this region of the diagram as well. The location of these AGN in this diagram (red box) is what is expected if warm dust (which emits in the mid-IR) from a compact torus produces a significant part of their total $L_{\rm IR}$. Actually, the mid-IR $f_{15\mu m}\slash f_{30\mu m}$ color, which behaves similar to the mid-IR excess, have been used to estimate the AGN luminosity contribution in ULIRGs (e.g., \citealt{Veilleux2009}).

For the remaining nuclei there is a {correlation  ($r_{\rm s}=0.87$ and $p = 1\times10^{-3}$)} between the continuum size and the mid-IR excess with increasing sizes for increasing mid-IR excess emission. The size of these sources ranges from $<$70\,pc to $\sim$300\,pc and the {best linear fit  is $\log \Delta L_{\rm 6-20\mu m}$\slash $L_{\rm IR} = (-1.8\pm0.1) + (2.2\pm0.6)\times 10^{-3}\times r_{\rm cont}\slash {\rm pc}$.}

In Fig.~\ref{fig:emir_r50} ({left} panel), we highlighted (blue box) sources with $r_{\rm cont}<120$\,pc. This box contains {46\% of the sample (12 nuclei out of 26)}. They have sizes comparable to the AGN (red box), although most of them only have an upper limit $r_{\rm cont}$, so their real size could be lower than that of the AGN.
Opposite to most AGN, which have excess mid-IR emission between 15 and 70\% of the $L_{\rm IR}$, these objects have negligible mid-IR excess ($<$3\%). A majority of the nuclei in interacting systems {(60\%)} lie in this blue box while it only includes {30\%} of the mergers. Actually, mergers following this relation have larger radii {(median 160\,pc)} than the interacting systems (median 90\,pc). This suggests that during the early phases of the interaction (nuclear separation $>1$\,kpc), most of the activity {(AGN and\slash or SF)} occurs in compact regions where most of the mid-IR emission is absorbed by dust and then re-emitted in the far-IR. 
Then, in more advanced merger stages (nuclear separation $<1$\,kpc), the activity appears on more extended regions {(green box in Fig.~\ref{fig:emir_r50})}, unless an optically detected AGN is present (red box sources).

{In Fig.~9, we indicate the systems with AGN contributions $>$50\% based on mid-IR observations \citep{Veilleux2009, Nardini2010}. The majority of the optical AGN are also classified as AGN by these mid-IR diagnostics (5 out of 7) and only 3 non-Sy are identified as mid-IR AGN. In addition, this figure shows that there is a good correlation between the mid-IR AGN classification and the mid-IR excess. This is because these mid-IR diagnostics are based on the detection of warm dust which is closely related to the mid-IR excess definition used here. Therefore, if the mid-IR emission of the AGN in these ULIRGs is absorbed (see also Sect.~\ref{ss:cont_sil_iras}), these mid-IR diagnostics could fail to detect them.}

The {right} panel of Fig.~\ref{fig:emir_r50} shows the excess mid-IR emission as function of the molecular gas emission size $r_{\rm CO}$. We do not find a significant correlation between the molecular gas size and the mid-IR excess {($r_{\rm s}=0.34$ and $p = 0.12$).}
The molecular gas emission size is similar in the interacting systems and advanced mergers (median $\sim$350\,pc).
{On the contrary, hydrodynamic simulations predict that strong torques during the first passage can pile up the gas in $\sim$kpc-scale starbursts and later, during the final-coalescence, in more compact sub-kpc starbursts (e.g., \citealt{Hopkins2013}).}

We find that, for the interacting systems, the molecular gas and the continuum size distributions are very different (see Fig.~\ref{fig:histo_inter_sizes}). The radii of the molecular gas emission is uniformly distributed over the whole observed range (mostly between 200 and 500\,pc; right panel of Fig.~\ref{fig:histo_inter_sizes}) while the continuum size distribution peaks at compact radii ($<$100\,pc; left panel of Fig.~\ref{fig:histo_inter_sizes}). This suggests that, even if there is a global correlation between the continuum and the CO emission sizes for the whole sample of ULIRGs (see Sect.~\ref{ss:nuclear_mol} and Fig.~\ref{fig:histo_size}), for the interacting systems, the two sizes seem to be decoupled. An obscured AGN, which do not need large amounts of molecular gas to produce high luminosities, could explain why these interacting nuclei have very compact continuum sources but a more extended molecular gas distribution.

\subsection{Continuum size vs. silicate absorption and IR colors}\label{ss:cont_sil_iras}

In this section, we explore possible relations between the 220\,GHz continuum size and other IR tracers. {As in Sect.~\ref{ss:mir_excess}, we excluded from this analysis nuclei with contributions $<$20\% to the total IR luminosity of the systems}.
We first consider the 9.7\micron\ silicate absorption. {We use the 9.7\micron\ silicate strengths from the Infrared Database of Extragalactic Observables from Spitzer (IDEOS; Table~\ref{tbl:spitzer_fluxes}; \citealt{HernanCaballero2020}, Spoon et al. in prep.).}
Deep silicate absorptions have been associated with an evolutionary phase in which the obscuring molecular cocoon created during the early interaction phases have not been yet shed (e.g., \citealt{Spoon07}). Therefore, we could expect a relation between the silicate absorption and the size of the cold molecular or dust continuum emissions.
However, we find no significant correlation between them {(see left and middle panels of Fig.~\ref{fig:iras_r50})}. To explain this absence of correlation, we argue that to measure deep 9.7\micron\ silicate absorptions, some mid-IR radiation must escape the nuclear molecular cocoon. As shown in Sect.~\ref{ss:mir_excess}, the fraction of the total IR emission that is emitted in the mid-IR is greatly reduced in the most compact nuclei (i.e., the mid-IR emission is absorbed and then re-emitted in the far-IR). Therefore, the 9.7\micron\ silicate absorption is not necessarily extreme in these compact sources since their nuclear mid-IR emission is possibly absorbed and what is observed in the mid-IR is likely produced by less obscured external regions. The presence of several CONs in our sample (see \citealt{Falstad2021}), is also consistent with the mid-IR emission of the hot nucleus being obscured. Likewise, ground-based mid-IR spectroscopy of local AGN ULIRGs showed that their silicate absorptions are produced by dust not directly associated with the AGN \citep{AAH2016}. This scenario is also supported by radiative transfer models of compact nuclei where the mid-IR emission is completely absorbed for objects with very high column densities ($N_{\rm H_2}>$10$^{25}$\,cm$^{-2}$; see fig. 2 of \citealt{GonzalezAlfonso2019}).

We also studied possible correlations between the broad-band IRAS colors and the size of the continuum and cold molecular emissions. Most of the ULIRGs are undetected by IRAS at 12\micron, for this reason we computed synthetic fluxes for the four IRAS bands (12, 25, 60, and 100\micron) form the {\it Spitzer}\slash IRS spectrum and the IR SED model (Sect.~\ref{ss:sed_model}). We tried the 6 IRAS color combinations. Only the $f_{\rm 12}$\slash $f_{\rm 25}$ color shows a significant correlation with the continuum size (Fig.~\ref{fig:iras_r50} {right}). The $f_{\rm 12}$\slash $f_{\rm 25}$ ratio is actually related to the mid-IR excess emission. The 12\micron\ flux would trace the mid-IR emission while at 25\micron, the emission is dominated by the IR gray-body for most of the non-AGN ULIRGs. Therefore, this relation would be equivalent to that presented in Sect.~\ref{ss:mir_excess} with the mid-IR excess emission.

\subsection{Nuclear radiation pressure. Eddington limit}\label{ss:Edd_limit}

We find very high $\Sigma_{L_{\rm IR}}$ in the nuclei of the ULIRGs, so it is important to determine if the radiation pressure can overcome the gravity attraction in these objects. 
To do so, we calculated the Eddington limit in the $\Sigma_{\rm H_2}$-$\Sigma_{L_{\rm IR}}$ plane by fitting the model results reported by \cite{GonzalezAlfonso2019}. These models assume spherical symmetry and accurately determine the force due to radiation pressure once the equilibrium $T_{\rm dust}$ profile across the source is calculated. The models assume a density profile $\sim r^{-1}$, such that $\Sigma_{\rm H_2}$ does not depend on the source radius. Our fitting, shown in Fig.~\ref{fig:lir_density} with an orange line, is
approximately valid for $\Sigma_{\rm H_2}\lesssim3\times10^4$\,M$_{\odot}$\,pc$^{-2}$, and assumes an intermediate molecular gas fraction with respect to the total mass, $f_{\rm g} = 0.3$, and a gas-to-dust ratio of $f_{\rm gd}=100$ by mass. 
In the optically thick limit, the Eddington luminosity is proportional to $f_{\rm g}^{-1/2}\times f_{\rm gd}$ \citep{Andrews2011}.

Fig.~\ref{fig:lir_density} shows that half of the nuclei, {19 out of the 29}, are above the estimated Eddington limit (see also Table~\ref{tbl:lir_density}). This suggests that the radiation pressure can be stronger than gravity in these nuclei and, therefore, massive gas outflows are expected.  This result is consistent with the detection of massive molecular outflows in local ULIRGs \citep{Sturm2011, Cicone2014, GonzalezAlfonso2017, Pereira2018, Lutz2020} and also supports that radiation pressure plays a relevant role as a potential launching mechanism of the outflows in local ULIRGs.

\section{Conclusions}\label{s:conclusions}

We have analyzed new high-resolution ALMA $\sim$220\,GHz and CO(2--1) observations of a representative sample of {23 local ULIRGs (34 individual nuclei)} as part of the Physics of ULIRGs with MUSE and ALMA (PUMA) project (see also \citealt{Perna2021}).
The main results of this work are the following:

\begin{enumerate}
\item We modeled the $\sim$220\,GHz (190--250\,GHz) continuum emission of these ULIRGs. We find that the median deconvolved half light radius ($r_{\rm cont}$) is {80--100\,pc} and about 40\% ({11\slash 29} with continuum detection) of the nuclei are not resolved by these data. From the IR and radio SED modeling, we obtain that the ALMA $\sim$220\,GHz continuum fluxes are in good agreement, within a factor of 2 {(median ratio of 1.1$\pm$0.7)}, with the extrapolation of the dust far-IR gray body emission. This suggests that the $\sim$220\,GHz continuum traces the regions emitting the bulk of the IR luminosity in these objects. We estimate that the contributions from synchrotron ($\sim$20\%) and free-free emission {($<$65\%)} to the ALMA flux are not likely to bias the measured sizes.
Using the $\sim$220\,GHz continuum size, we calculate IR luminosity densities, $\Sigma_{L_{\rm IR}}$, in the range 10$^{11.5}-10^{14.3}$\,$L_\odot$\,kpc$^{-2}$ (median 10$^{13.2}$\,$L_\odot$\,kpc$^{-2}$), which is equivalent to $\Sigma_{\rm SFR}=2500$\,$M_\odot$\,yr$^{-1}$\,kpc$^{-2}$. This is similar to the range derived from previous radio and ground-based mid-IR observations and 1--2 orders of magnitude brighter than local and high-z starbursts.

\item Similarly, we measure deconvolved CO(2--1) emission sizes, $r_{\rm CO}$, between 60 and 700\,pc. These are on average {2.5$\pm$1.1} times larger than the $\sim$220\,GHz continuum size. We find no differences between systems optically classified as AGN or starburst {or between interacting systems and advanced mergers}. Using a ULIRG-like $\alpha_{\rm CO}$ conversion factor, we find nuclear molecular gas surface densities, $\Sigma_{\rm H_2}$, in the range 10$^{2.9}-10^{4.2}$\,$M_\odot$\,pc$^{-2}$. 

\item If the $L_{\rm IR}$ is produced by SF, the $\Sigma_{L_{\rm IR}}$\slash $\Sigma_{\rm H_2}$ ratios imply extremely short molecular gas depletion times ($<$1--15\,Myr). In addition, 70\% of the nuclei would have SF efficiencies above the maximum for a starburst ($\epsilon=1$) and all of them would have $\epsilon>0.2$, which is the maximum efficiency per free-fall time predicted by simulations. 
These findings suggests that the bulk of the IR luminosity of these ULIRGs does not originate in a nuclear starburst. An obscured AGN with $L_{\rm AGN}$\slash $L_{\rm IR}> 0.5$ would be an alternative energy source.

\item For the compact nuclei ($r_{\rm cont}<$120\,pc) in interacting system with low mid-IR excess emission, the $r_{\rm cont}$ is not correlated with the cold molecular gas emission, $r_{\rm CO}$, which varies between 200 and 500\,pc. This could support the presence of a deeply embedded AGN which, opposed to star-formation, would not require large amounts of cold molecular gas to produce the observed high IR luminosities.

\item The presence of compact and extremely embedded nuclei is supported by the detection of the HCN-vib 14\micron\ absorption in {43$\pm$10\%} of the sample. The detection rate is higher, 60$\pm$15\%, in interacting systems than in advanced mergers,  {31$\pm$13\%}.

\item The ULIRG nuclei can be classified in two groups in the 220\,GHz continuum size, $r_{\rm cont}$, vs. 
mid-IR excess emission, $\Delta L_{\rm 6-20\mu m}$, diagram. $\Delta L_{\rm 6-20\mu m}$ is the 6--20\,$\mu$m emission excess after subtracting the far-IR gray-body contribution in this wavelength range.
These two groups are: (a) compact ($r_{\rm cont}<$120\,pc) nuclei with high mid-IR excess emission (log\,$\Delta L_{\rm 6-20\mu m}$\slash $L_{\rm IR}$>$-0.9$) which are optically classified AGN; and (b) objects that follow a relation with decreasing $r_{\rm cont}$ for decreasing mid-IR excess emission. A majority of the nuclei in interacting systems {(60\%)} that follow this relation have $r_{\rm cont}<$120\,pc, which are at the lower end of the $r_{\rm cont}$-mid-IR excess relation, while only {30\%} of the mergers have these compact continuum emission. Mergers following this relation have larger sizes on average. This suggest that in the early stages of the interaction (nuclear separation $>$ 1\,kpc) most of the activity occurs in very compact regions while, in more advanced merger stages, the activity is more extended unless and optically detected AGN is present. 

\item We find no correlation between the 9.7\micron\ silicate absorption and the $\sim$220\,GHz continuum or CO(2--1) sizes. The relatively faint mid-IR emission of the most compact nuclei could prevent the presence of deep silicate absorptions in their mid-IR spectra and this could hinder the use of this absorption {feature} to find some obscured nuclei.

\item We find that 67\% {(19\slash 29)} of the nuclei have nuclear radiation pressures above the estimated Eddington limit. This is consistent with the presence of massive molecular outflows in ULIRGs and supports that radiation pressure can have a relevant role in the outflow launching process.

\end{enumerate}

\begin{acknowledgements}
{We thank the referee for the useful comments and suggestions. We are grateful to A.~Hern\'an-Caballero and H.~Spoon for providing  measurements from the IDEOS database.}
MPS and IL acknowledge support from the Comunidad de Madrid through the Atracci\'on de Talento Investigador Grant 2018-T1/TIC-11035 and PID2019-105423GA-I00 (MCIU/AEI/FEDER,UE).
AA-H and SG-B acknowledge support through grant PGC2018-094671-B-I00 (MCIU/AEI/FEDER,UE).
MP is supported by the Programa Atracción de Talento de la Comunidad de Madrid via grant 2018-T2/TIC-11715.
AL acknowledges the support from Comunidad de Madrid through the Atracción de Talento Investigador Grant 2017-T1/TIC-5213.
SA, LC, MP, and AL  acknowledge support from the Spanish Ministerio de Economía y Competitividad through grants ESP2017-83197-P and PID2019-106280GB-I00.
This work was done under project No. MDM-2017-0737 Unidad de Excelencia "Mar\'ia de Maeztu"- Centro de Astrobiolog\'ia (INTA-CSIC).
D. Rigopoulou acknowledges support from STFC through grant ST/S000488/1.
This paper makes use of the following ALMA data: {ADS/JAO.ALMA\#2015.1.00113.S,} ADS/JAO.ALMA\#2015.1.00263.S, ADS/JAO.ALMA\#2016.1.00170.S, ADS/JAO.ALMA\#2016.1.00777.S, ADS/JAO.ALMA\#2018.1.00486.S, and ADS/JAO.ALMA\#2018.1.00699.S.
ALMA is a partnership of ESO (representing its member states), NSF (USA) and NINS (Japan), together with NRC (Canada) and NSC and ASIAA (Taiwan) and KASI (Republic of Korea), in cooperation with the Republic of Chile. The Joint ALMA Observatory is operated by ESO, AUI/NRAO and NAOJ.
The National Radio Astronomy Observatory is a facility of the National Science Foundation operated under cooperative agreement by Associated Universities, Inc.

\end{acknowledgements}

\appendix
\section{Mid-IR \textit{Spitzer} observations of the ULIRGs}\label{apx:spitzer_fluxes}

\begin{table}[h]
\caption{{\textit{Spitzer} IRS and MIPS fluxes and 9.7\micron\ silicate strength}}
\label{tbl:spitzer_fluxes}
\centering
\begin{small}
\begin{tabular}{lcccccccc}
\hline \hline
\\
IRAS name & $S_{\rm 9.7\mu m}$\,\tablefootmark{a} & IRS 34\micron\tablefootmark{b}  & MIPS 70\micron & MIPS 160\micron\\
\hline
00091$-$0738  & $-$3.20 $\pm$  0.12  & 1.07 &  \nodata &  \nodata \\
00188$-$0856  & $-$2.72 $\pm$  0.08  & 0.80 & 2.65 & 1.91 \\
00509$+$1225  &    0.30 $\pm$  0.01  & 1.34 &  \nodata &  \nodata \\
01572$+$0009  &    0.19 $\pm$  0.02 & 1.17 &  \nodata &  \nodata \\
F05189$-$2524 & $-$0.32 $\pm$  0.01 & 7.30 &  \nodata &  \nodata \\
07251$-$0248  & $-$2.77 $\pm$  0.06 & 2.62 &  \nodata &  \nodata \\
09022$-$3615  & $-$1.04 $\pm$  0.01  & 3.53 &  \nodata &  \nodata \\
F10190$+$1322 & $-$1.09 $\pm$  0.02 & 0.85 & 3.55 &  \nodata \\
11095$-$0238  & $-$3.43 $\pm$  0.07 & 1.54 & 2.95 & 1.28 \\
F12072$-$0444 & $-$1.42 $\pm$  0.02  & 1.20 &  \nodata &  \nodata \\
F12112$+$0305 & $-$1.35 $\pm$  0.03  & 2.06 &  \nodata &  \nodata \\
13120$-$5453  & $-$1.18 $\pm$  0.01  & 10.1 &  \nodata &  \nodata \\
F13451$+$1232 & $-$0.32 $\pm$  0.02  & 1.03 & 2.04 & 1.41 \\
F14348$-$1447 & $-$1.89 $\pm$  0.03   & 1.98 &  \nodata &  \nodata \\
F14378$-$3651 & $-$1.44 $\pm$  0.05  & 1.85 &  \nodata &  \nodata \\
F15327$+$2340 & $-$2.74 $\pm$  0.01 &  34.4 &  \nodata &  \nodata \\
16090$-$0139  & $-$2.65 $\pm$  0.06  & 1.07 &  \nodata &  \nodata \\
16155$+$0146  & $-$2.41 $\pm$  0.07  & 0.75 &  \nodata &  \nodata \\
17208$-$0014  & $-$1.56 $\pm$  0.01  & 8.67 &  \nodata &  \nodata \\
F19297$-$0406 & $-$1.49 $\pm$  0.04  & 2.12 &  \nodata &  \nodata \\
19542$+$1110  & $-$0.97 $\pm$  0.02  & 2.19 &  \nodata &  \nodata \\
20087$-$0308  & $-$1.82 $\pm$  0.04  & 0.99 &  \nodata &  \nodata \\
20100$-$4156  & $-$2.79 $\pm$  0.06  & 1.81 & 4.56 & 2.33 \\
20414$-$1651  & $-$1.62 $\pm$  0.08   & 1.25 &  \nodata &  \nodata \\
F22491$-$1808 & $-$1.19 $\pm$  0.03  & 2.20 &  \nodata &  \nodata \\
\hline
\end{tabular}
\end{small}
\tablefoot{Fluxes are in Jy. The flux uncertainties are dominated by the $\sim$10\% calibration uncertainty.
\tablefoottext{a}{{9.7\micron\ silicate strength from the IDEOS database (\citealt{HernanCaballero2020}, Spoon et al. in prep.) based on the IRS spectroscopy of these ULIRGs.}}
\tablefoottext{b}{34\micron\ {observed wavelength} flux measured in the \textit{Spitzer}\slash IRS spectrum.}
}
\end{table}

\clearpage
\onecolumn
\section{ALMA continuum models}\label{apx:alma_cont_models}

\begin{figure}[h]
\centering
\vspace{5mm}
\includegraphics[width=0.95\textwidth]{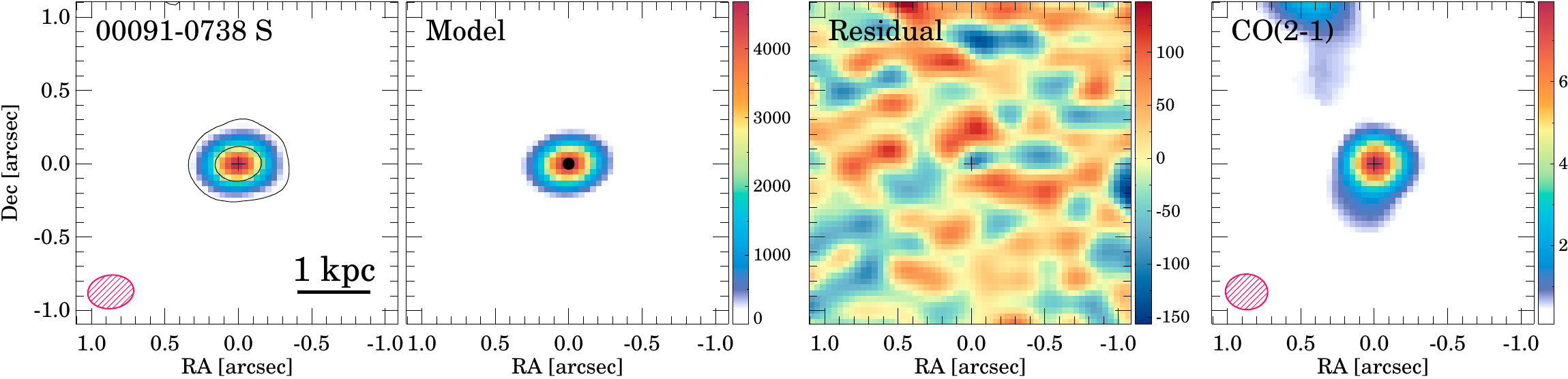}
\includegraphics[width=0.95\textwidth]{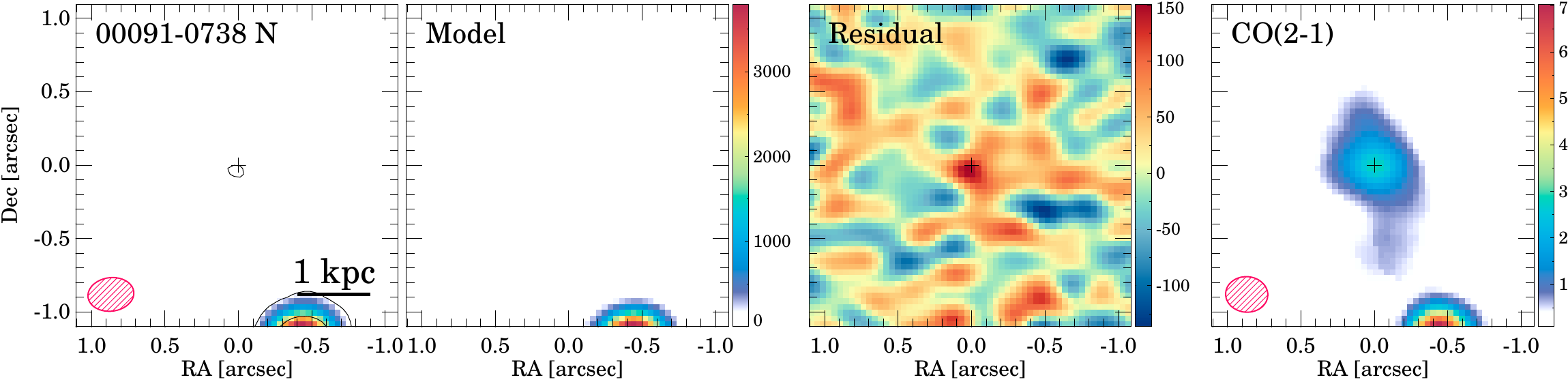}
\includegraphics[width=0.95\textwidth]{cont_model_00188-0856.pdf}
\includegraphics[width=0.95\textwidth]{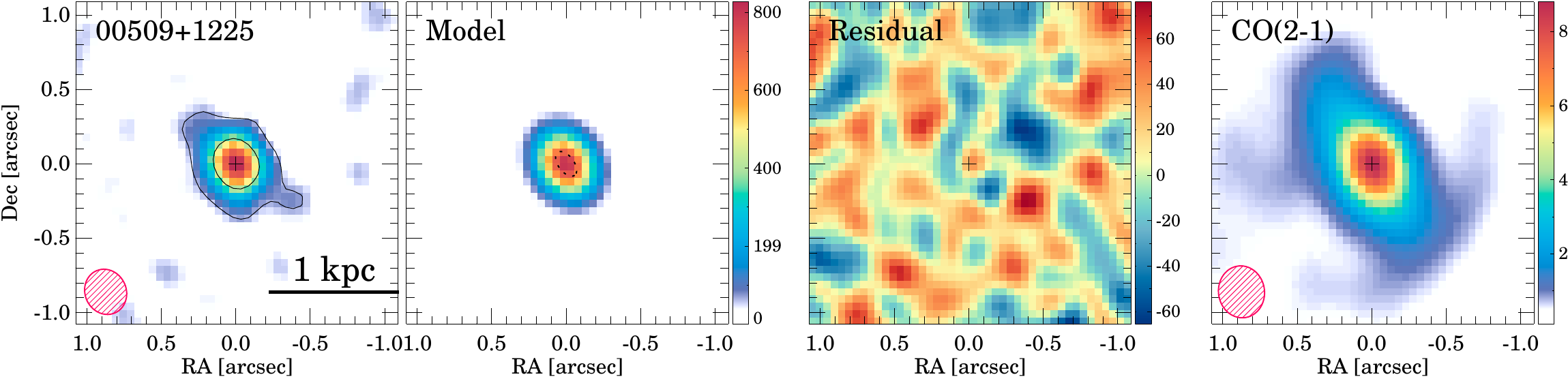}
\includegraphics[width=0.95\textwidth]{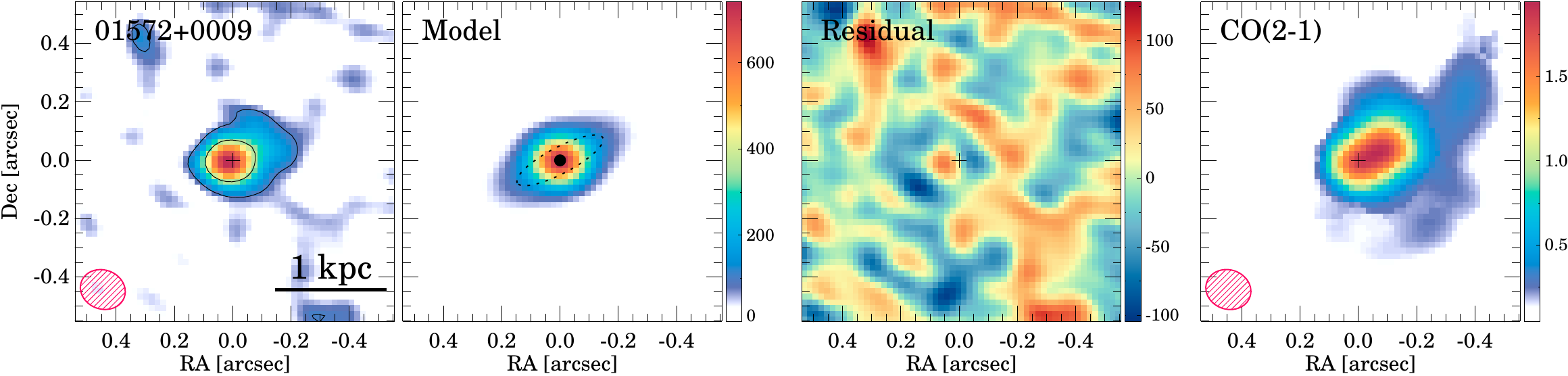}
\caption{Same as Fig.~\ref{fig:alma_models}.\label{fig:apx_alma_models}}
\end{figure}

\addtocounter{figure}{-1}
\begin{figure}[h]
\centering
\vspace{5mm}
\includegraphics[width=0.95\textwidth]{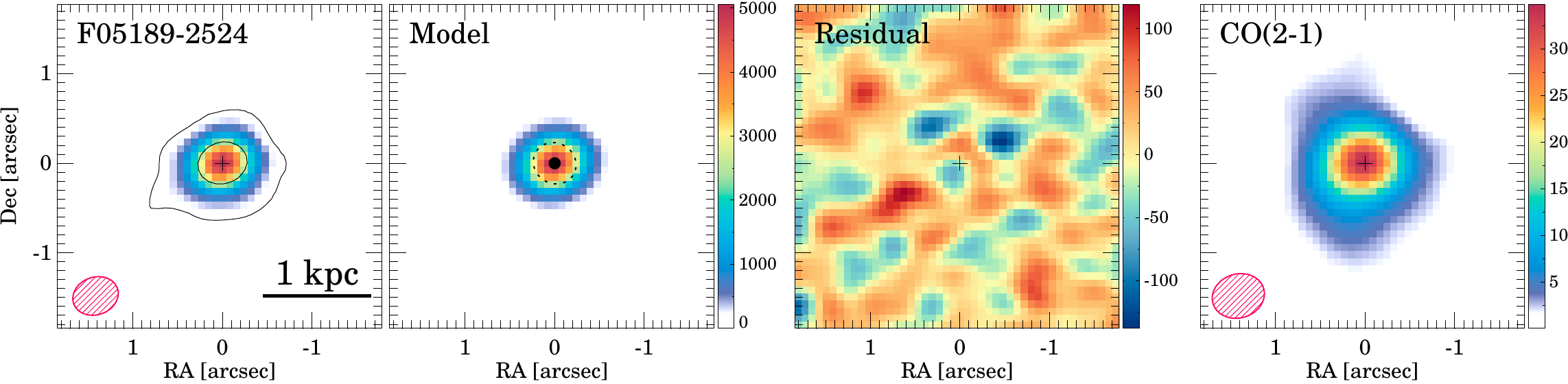}
\includegraphics[width=0.95\textwidth]{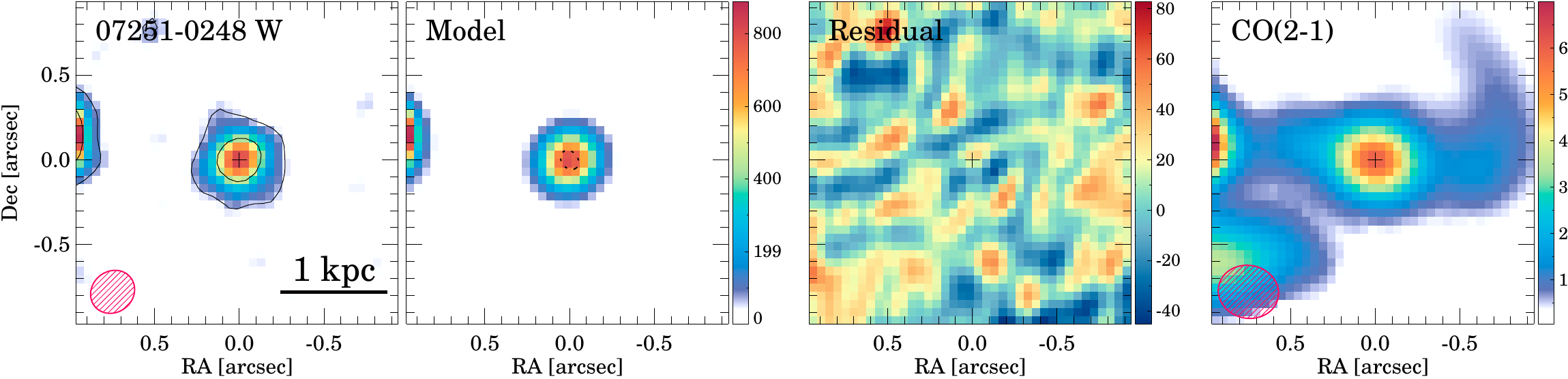}
\includegraphics[width=0.95\textwidth]{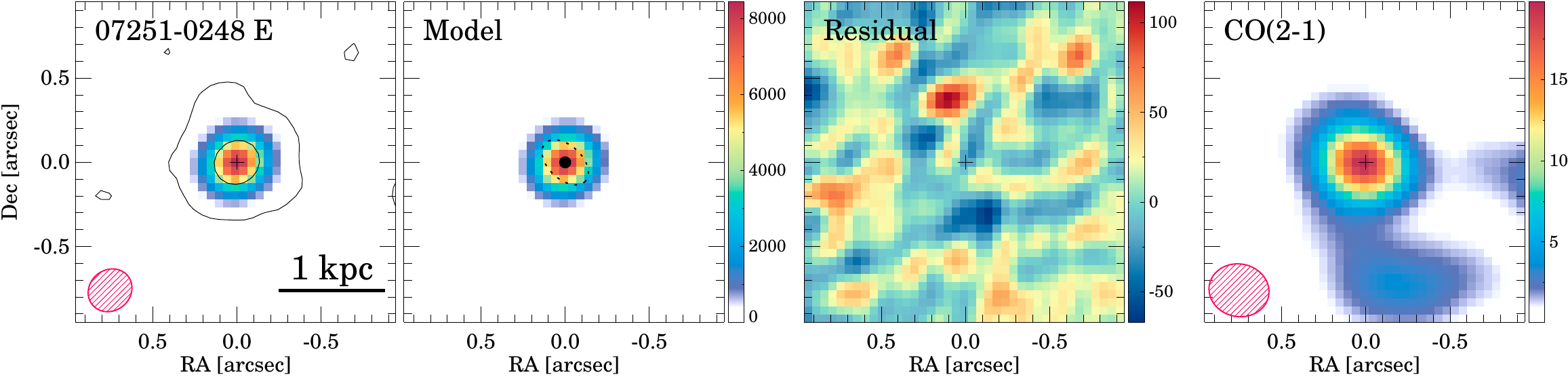}
\includegraphics[width=0.95\textwidth]{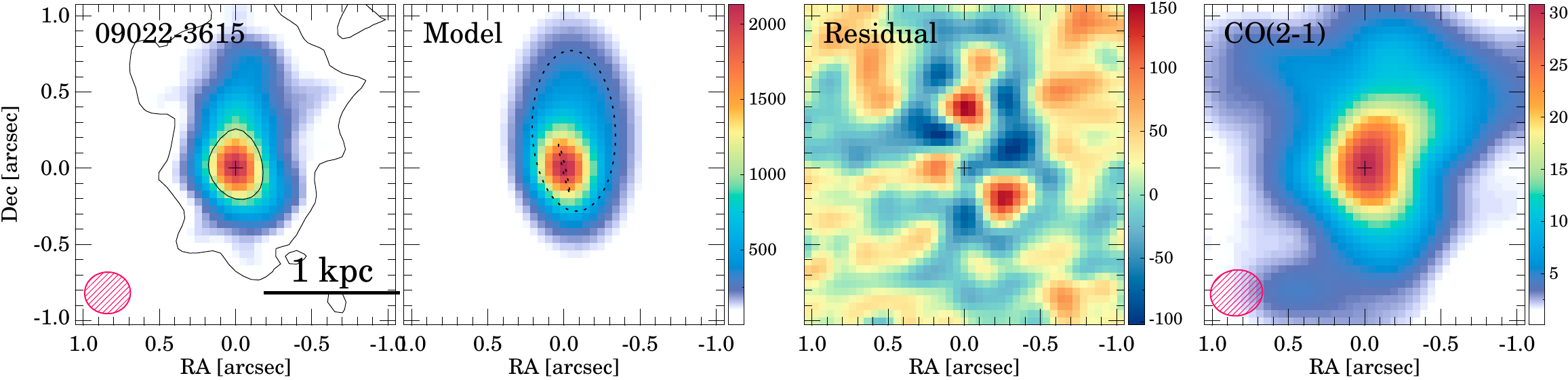}
\includegraphics[width=0.95\textwidth]{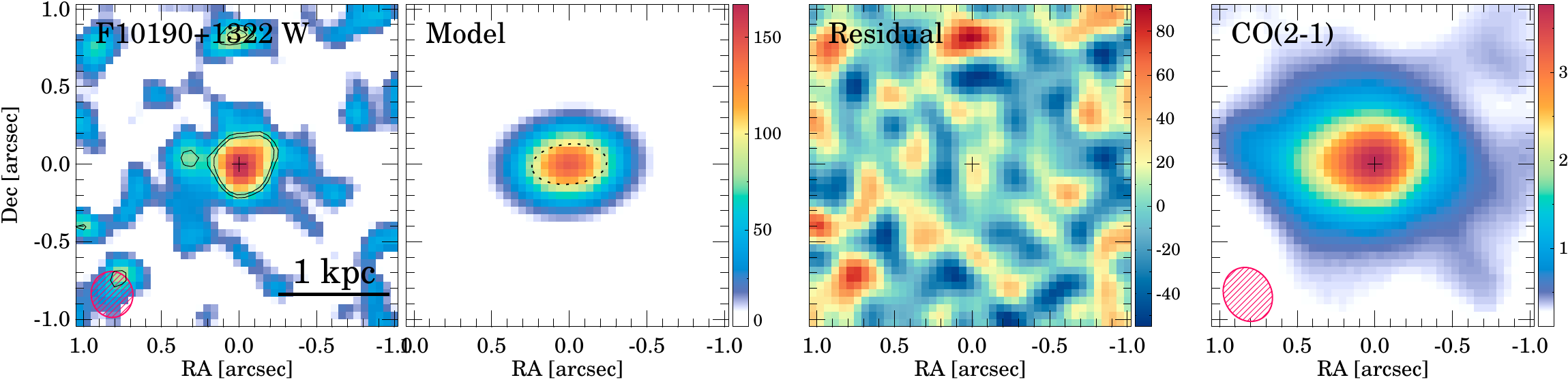}
\caption{(Continued)}
\end{figure}

\addtocounter{figure}{-1}
\begin{figure}[h]
\centering
\vspace{5mm}
\includegraphics[width=0.95\textwidth]{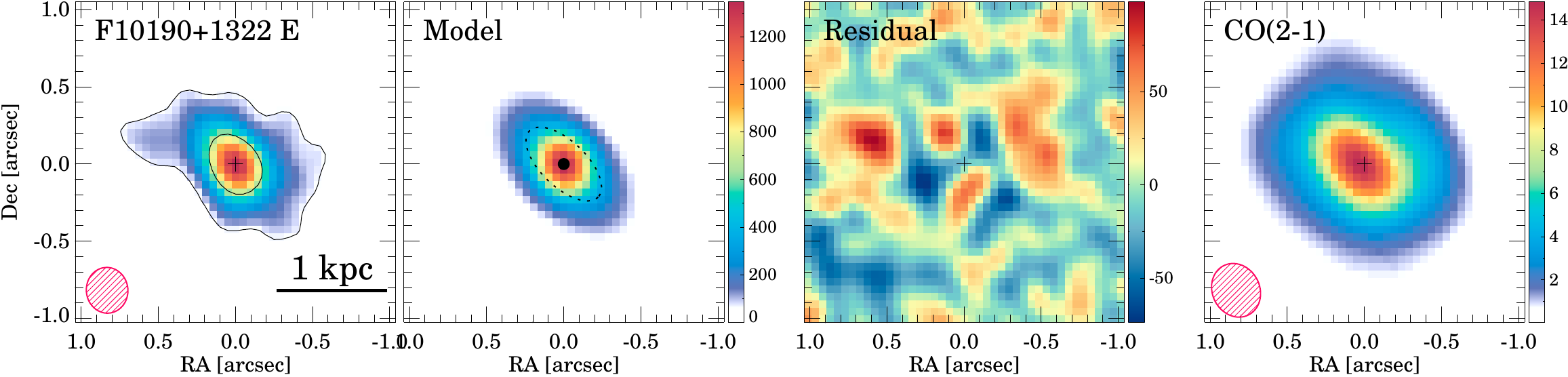}
\includegraphics[width=0.95\textwidth]{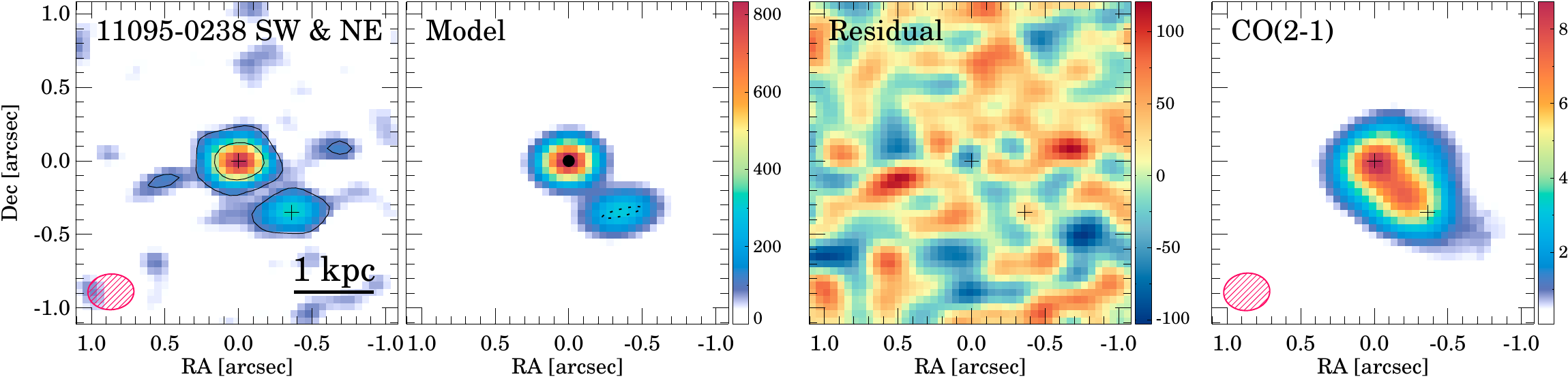}
\includegraphics[width=0.95\textwidth]{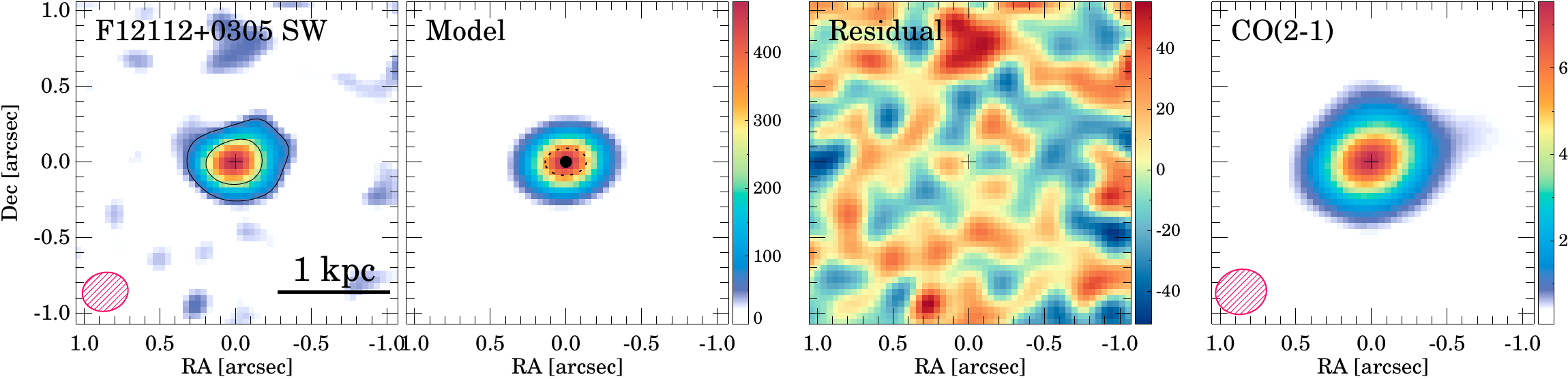}
\includegraphics[width=0.95\textwidth]{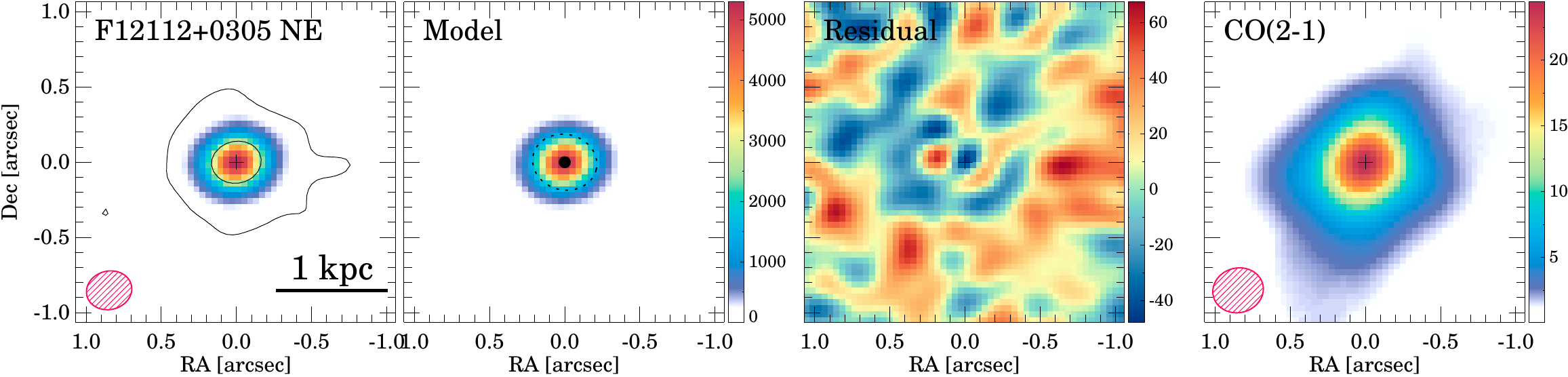}
\includegraphics[width=0.95\textwidth]{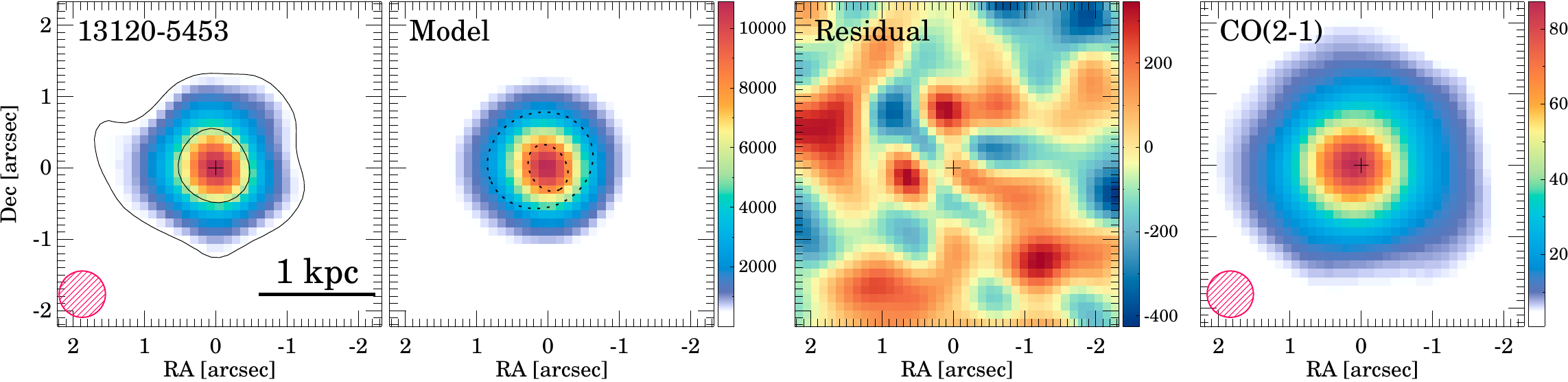}
\caption{(Continued)}
\end{figure}

\addtocounter{figure}{-1}
\begin{figure}[h]
\centering
\vspace{5mm}
\includegraphics[width=0.95\textwidth]{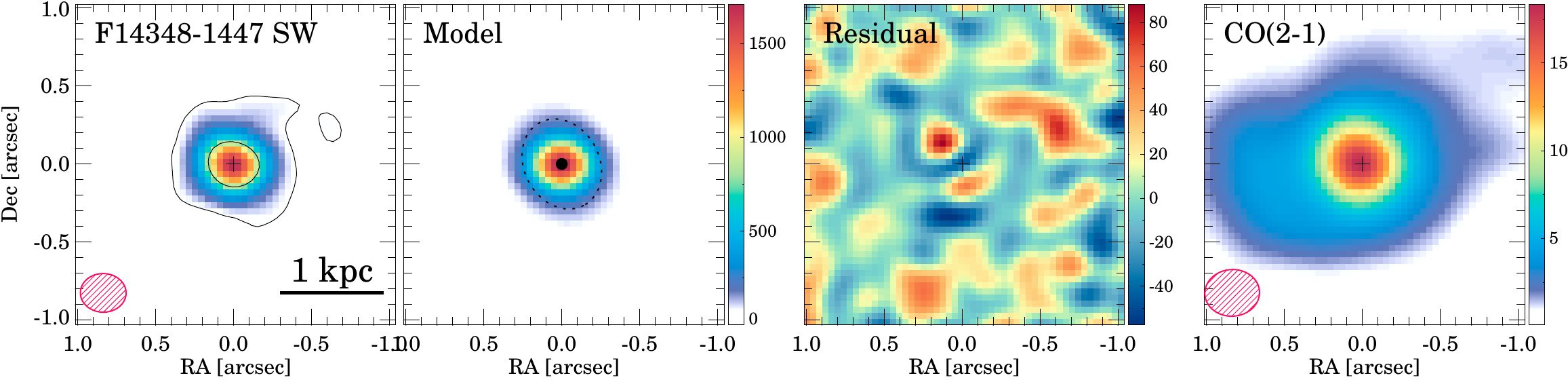}
\includegraphics[width=0.95\textwidth]{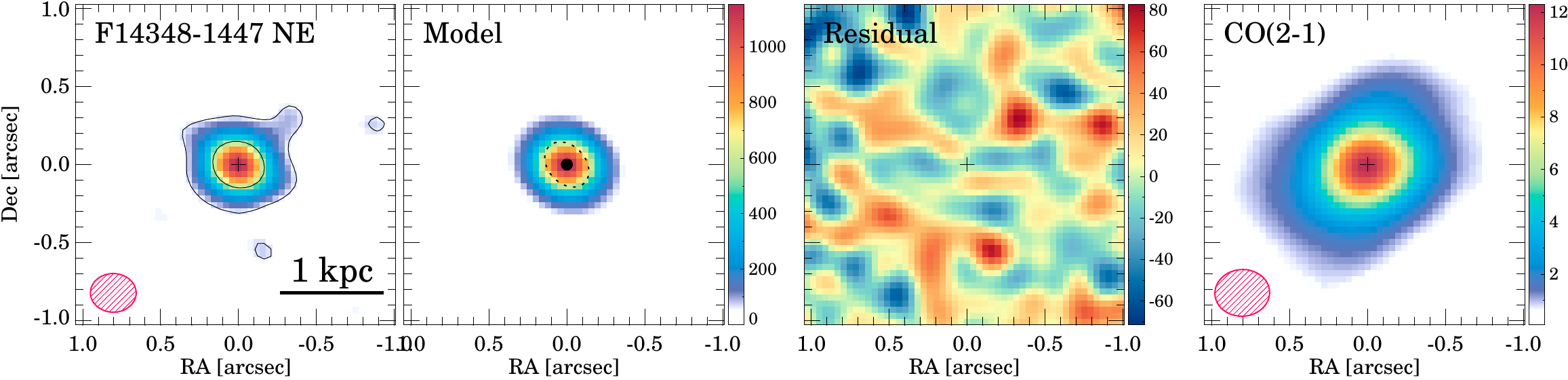}
\includegraphics[width=0.95\textwidth]{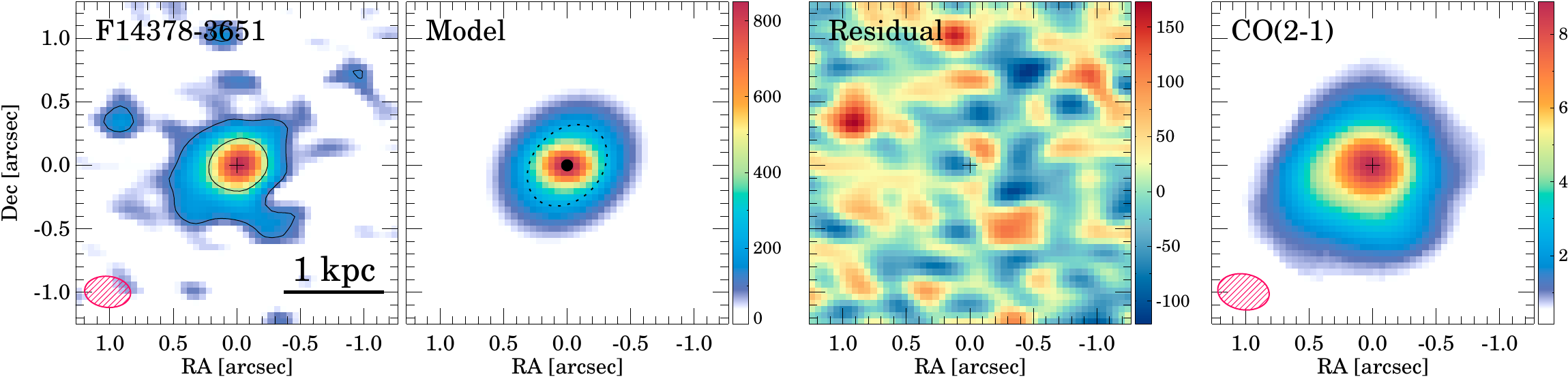}
\includegraphics[width=0.95\textwidth]{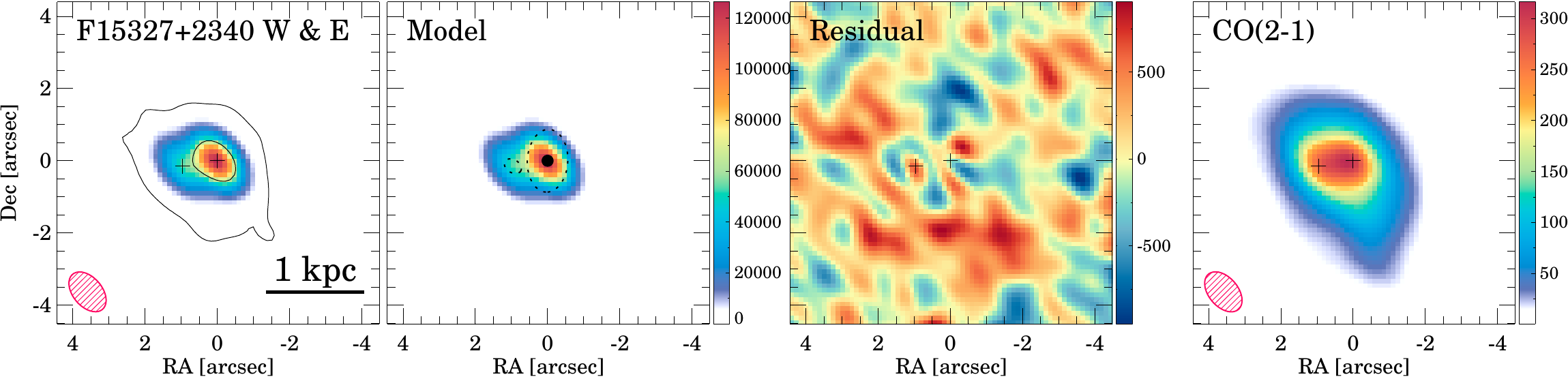}
\includegraphics[width=0.95\textwidth]{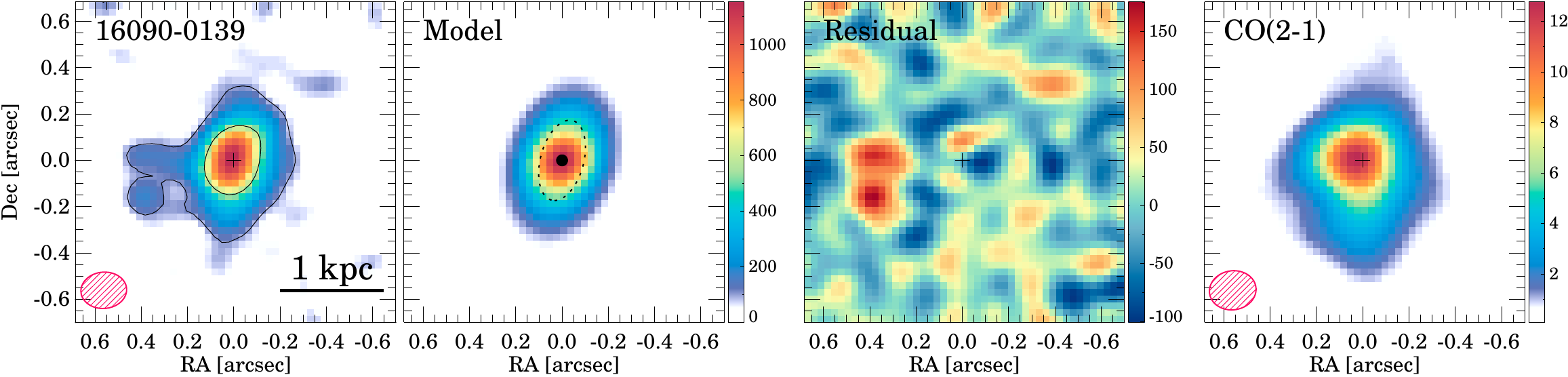}
\caption{(Continued)}
\end{figure}

\addtocounter{figure}{-1}
\begin{figure}[h]
\centering
\vspace{5mm}
\includegraphics[width=0.95\textwidth]{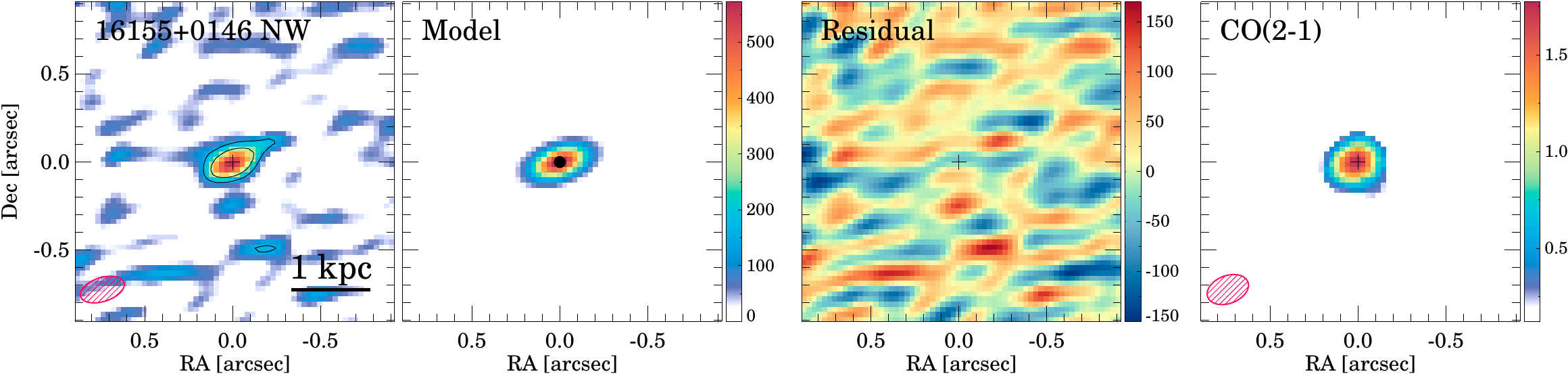}
\includegraphics[width=0.95\textwidth]{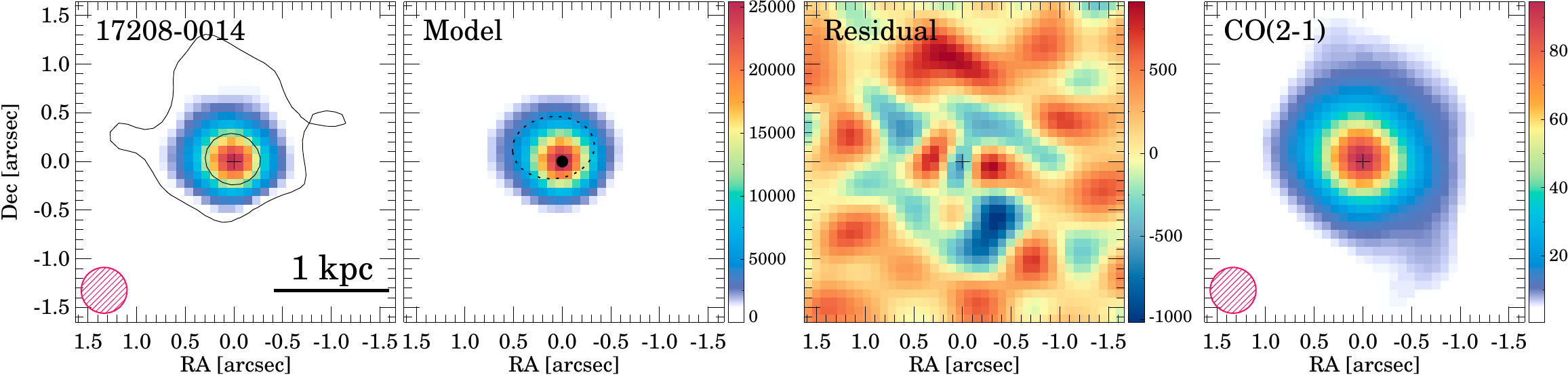}
\includegraphics[width=0.95\textwidth]{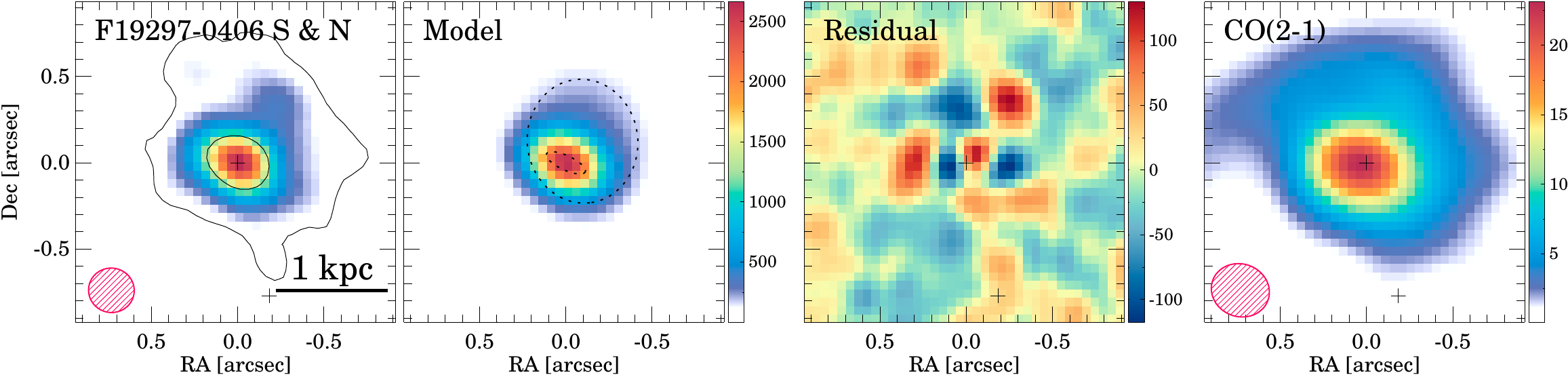}
\includegraphics[width=0.95\textwidth]{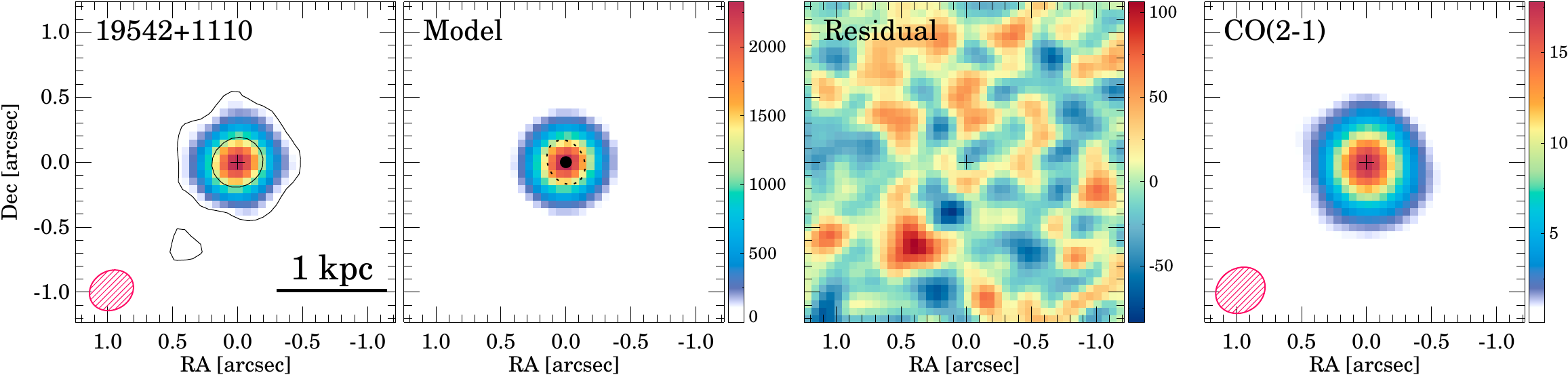}
\includegraphics[width=0.95\textwidth]{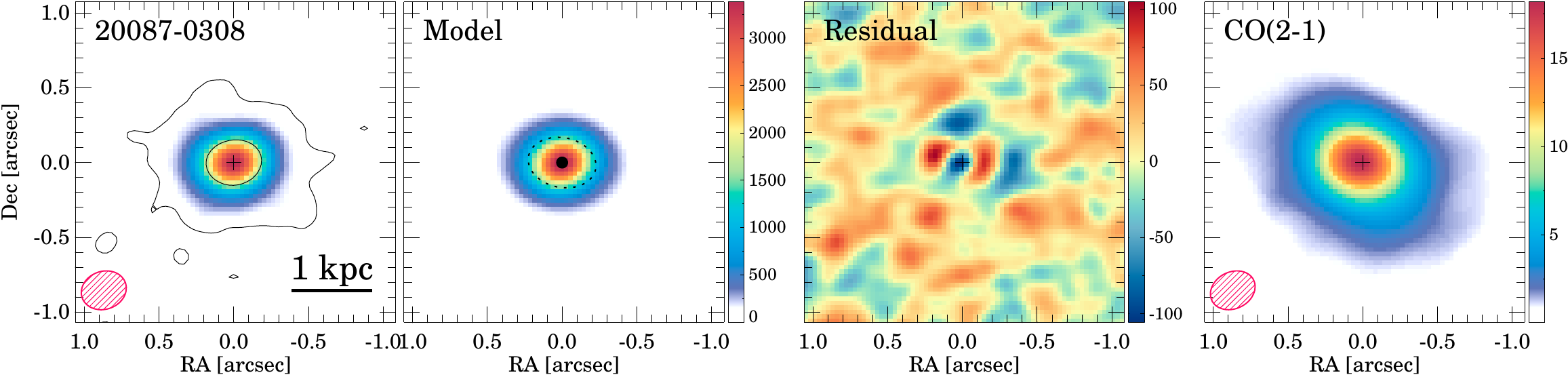}
\caption{(Continued)}
\end{figure}

\addtocounter{figure}{-1}
\begin{figure}[h]
\centering
\vspace{5mm}
\includegraphics[width=0.95\textwidth]{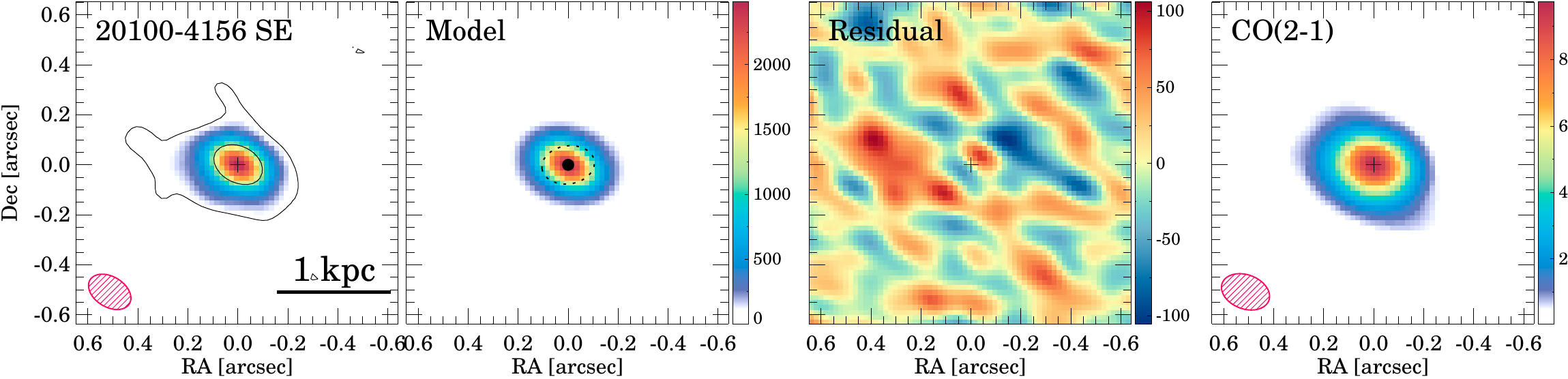}
\includegraphics[width=0.95\textwidth]{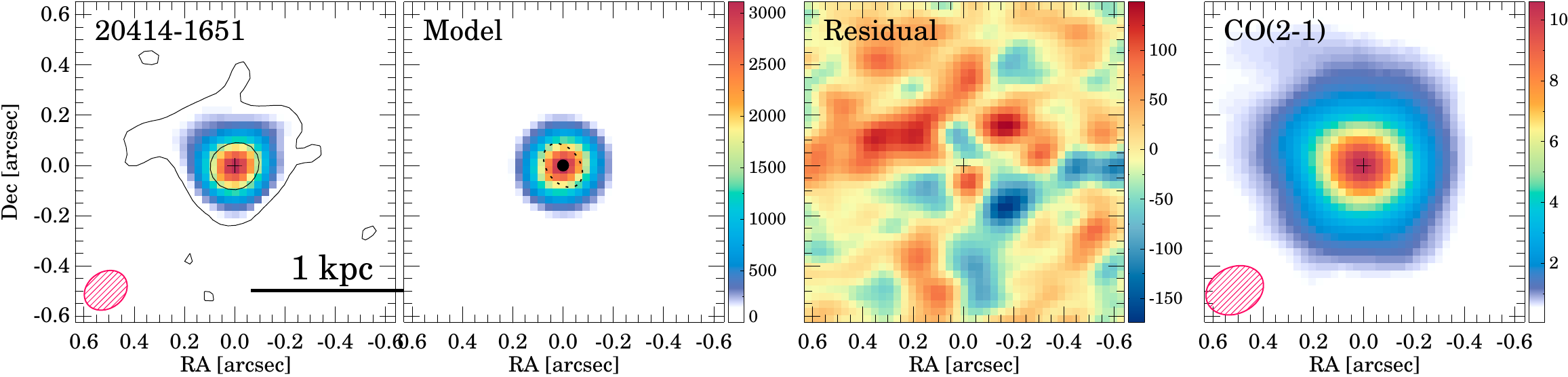}
\includegraphics[width=0.95\textwidth]{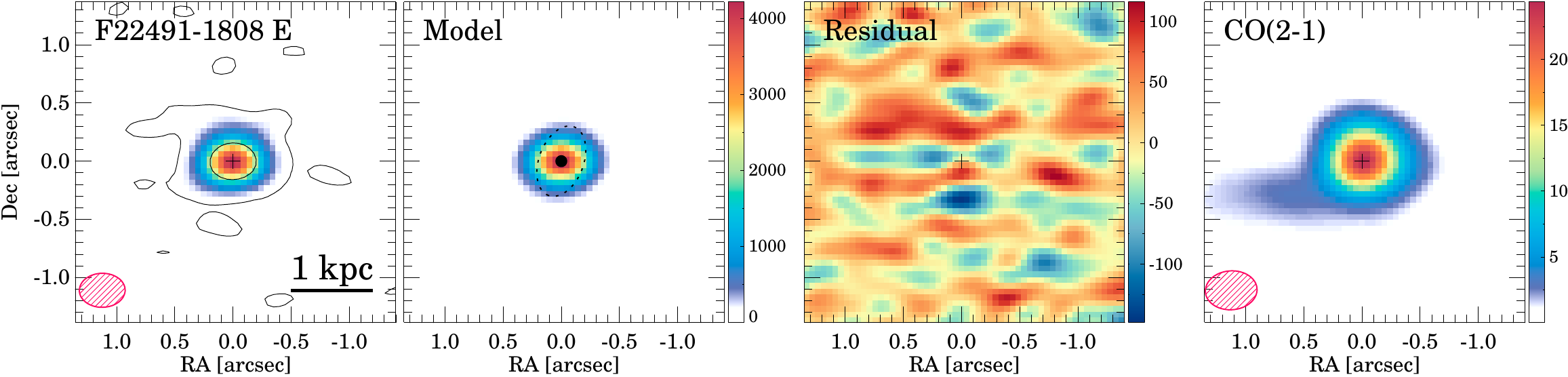}
\caption{(Continued)}
\end{figure}

\clearpage

\section{CO(2--1) emission models}\label{apx:co_model}
\begin{figure}[!h]
\centering
\vspace{5mm}
\includegraphics[trim=23mm 130mm 93mm 98mm, clip=true, width=0.32\textwidth]{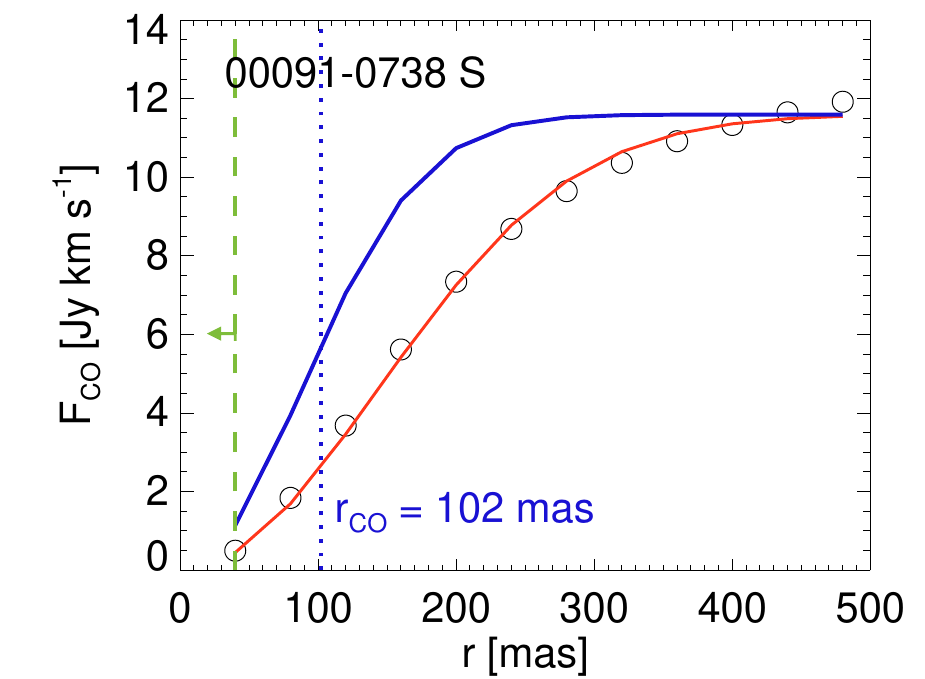}
\includegraphics[trim=23mm 130mm 93mm 98mm, clip=true, width=0.32\textwidth]{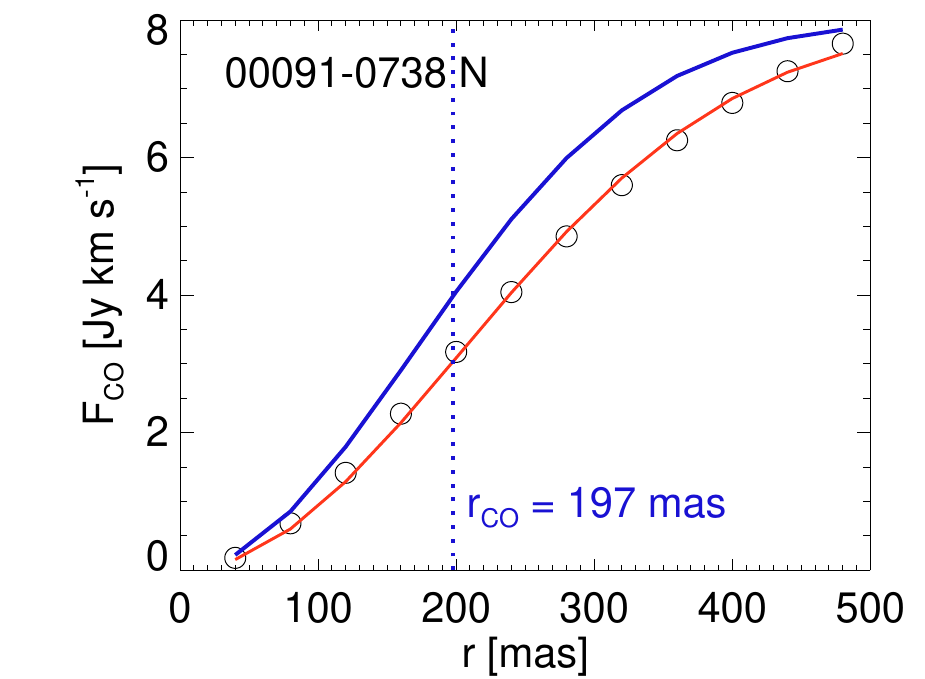}
\includegraphics[trim=23mm 130mm 93mm 98mm, clip=true, width=0.32\textwidth]{fig_profile_00188-0856.pdf}
\includegraphics[trim=23mm 130mm 93mm 98mm, clip=true, width=0.32\textwidth]{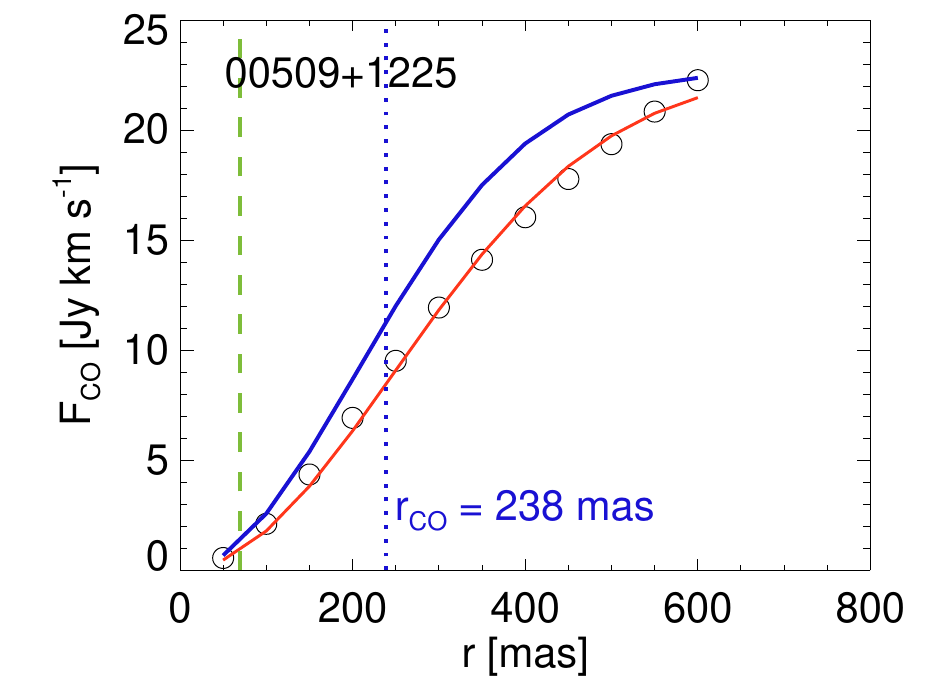}
\includegraphics[trim=23mm 130mm 93mm 98mm, clip=true, width=0.32\textwidth]{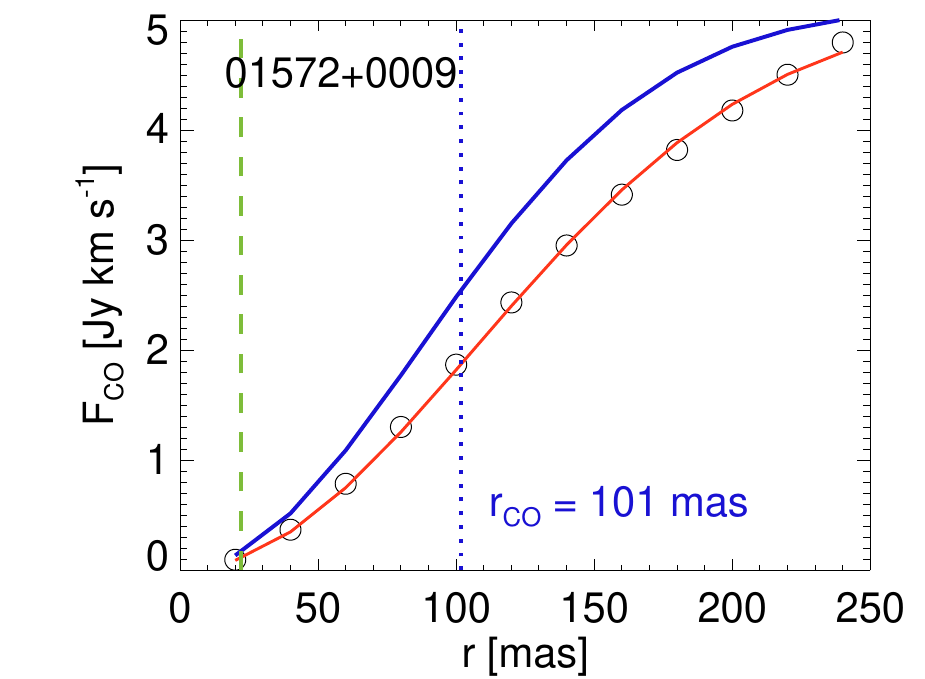}
\includegraphics[trim=23mm 130mm 93mm 98mm, clip=true, width=0.32\textwidth]{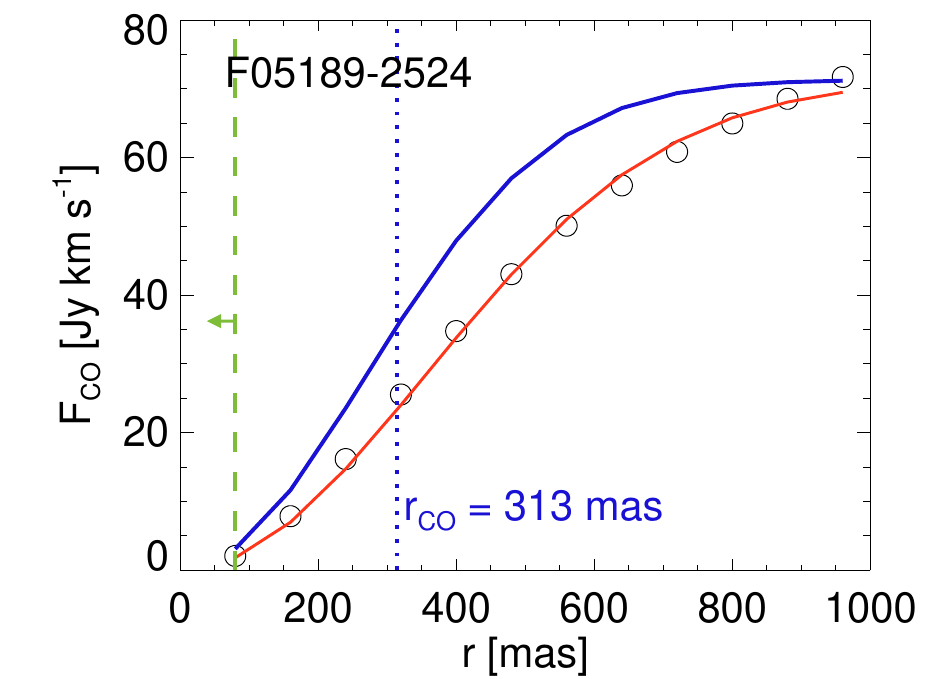}
\includegraphics[trim=23mm 130mm 93mm 98mm, clip=true, width=0.32\textwidth]{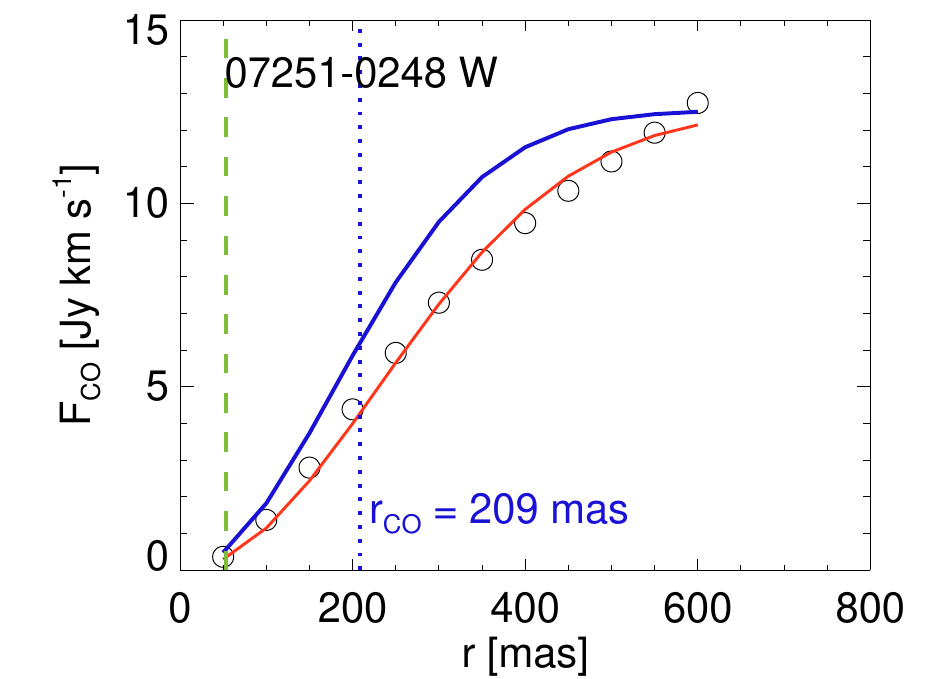}
\includegraphics[trim=23mm 130mm 93mm 98mm, clip=true, width=0.32\textwidth]{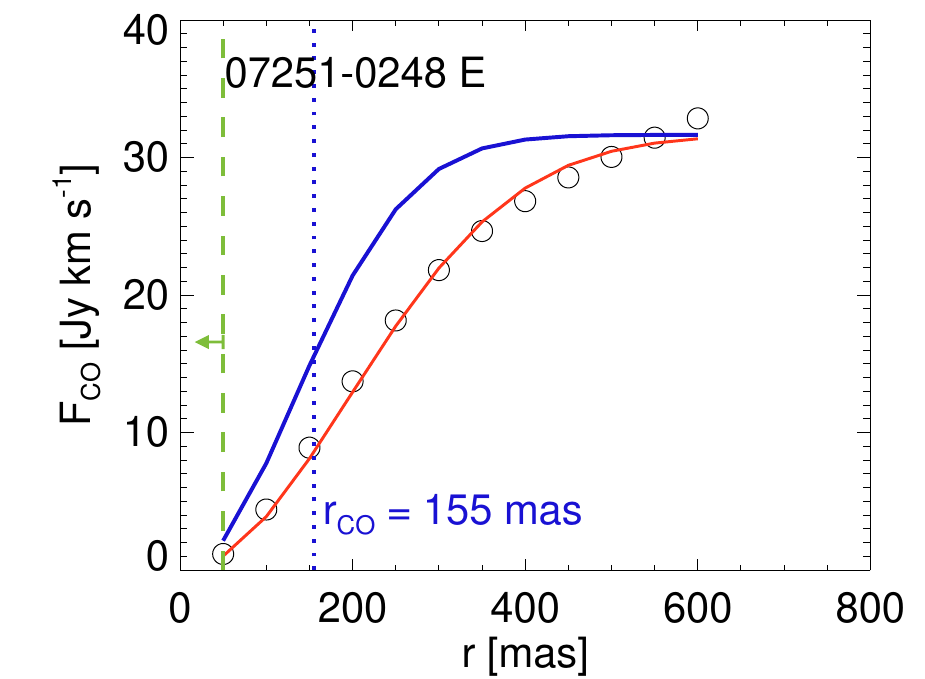}
\includegraphics[trim=23mm 130mm 93mm 98mm, clip=true, width=0.32\textwidth]{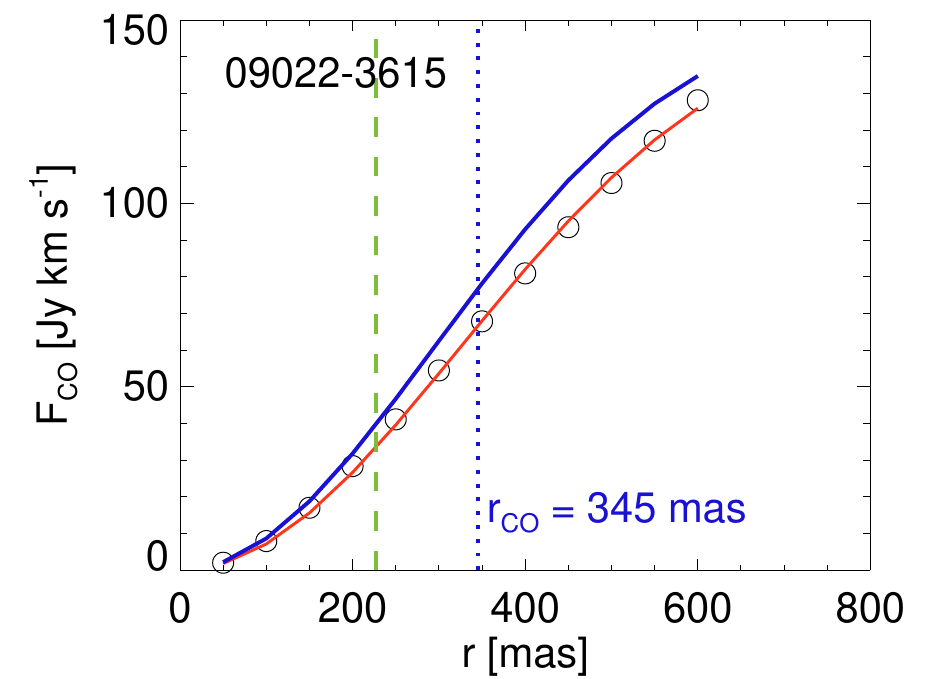}
\includegraphics[trim=23mm 130mm 93mm 98mm, clip=true, width=0.32\textwidth]{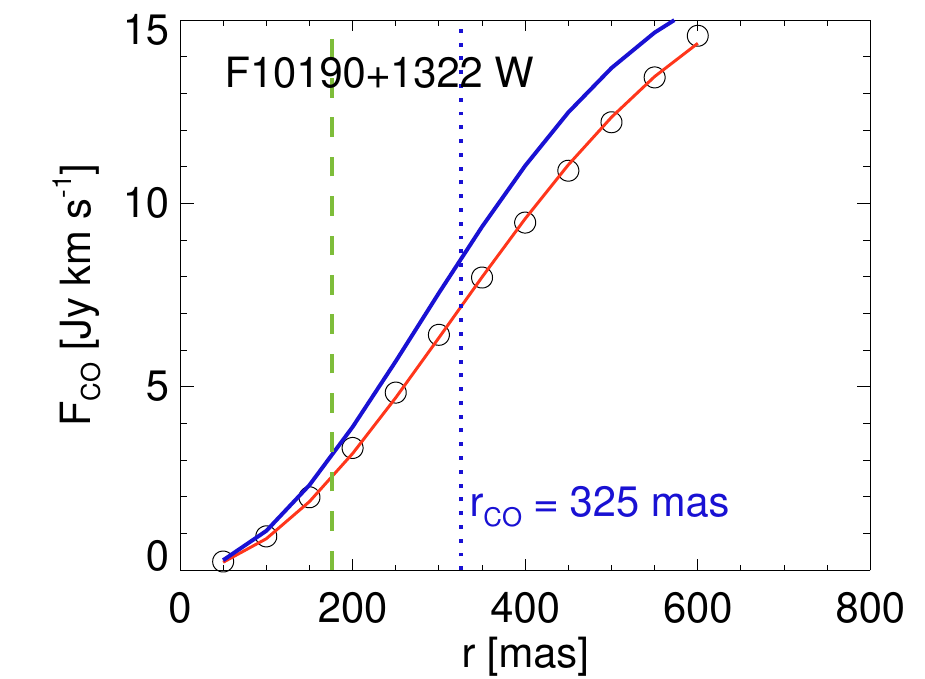}
\includegraphics[trim=23mm 130mm 93mm 98mm, clip=true, width=0.32\textwidth]{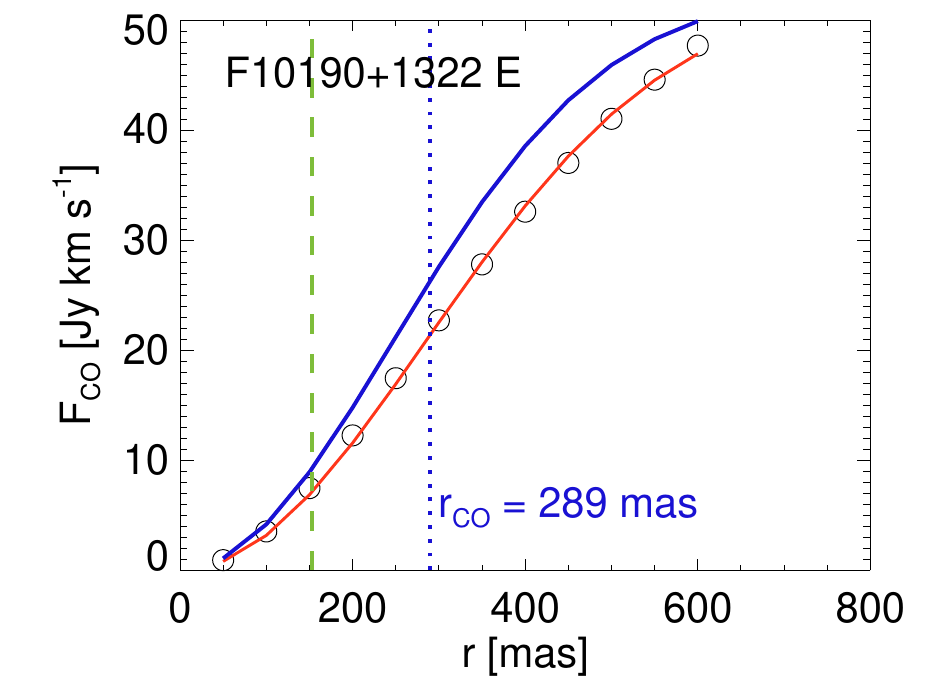}
\includegraphics[trim=23mm 130mm 93mm 98mm, clip=true, width=0.32\textwidth]{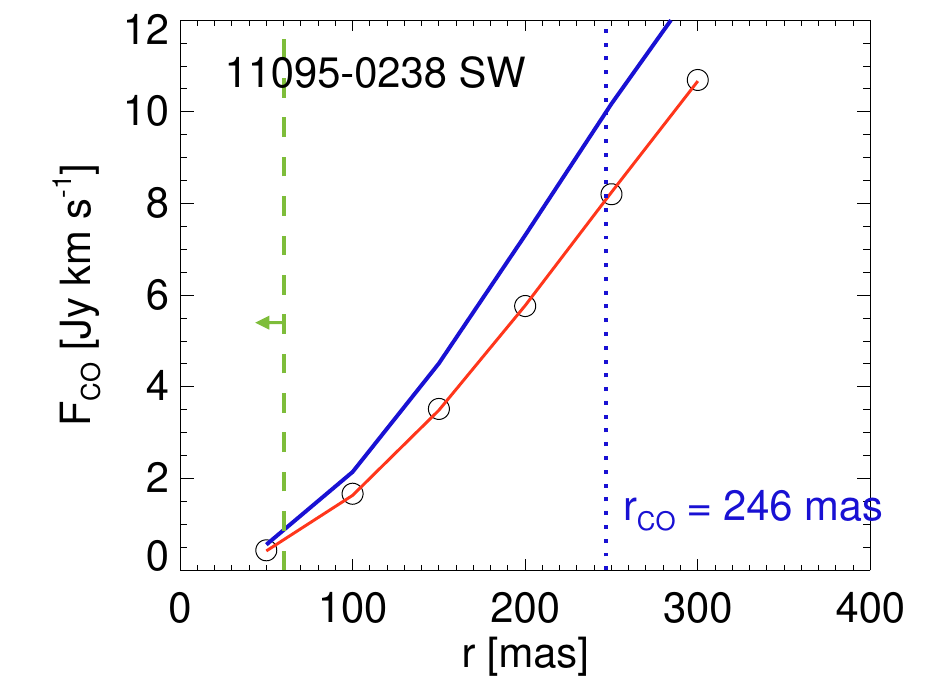}
\includegraphics[trim=23mm 130mm 93mm 98mm, clip=true, width=0.32\textwidth]{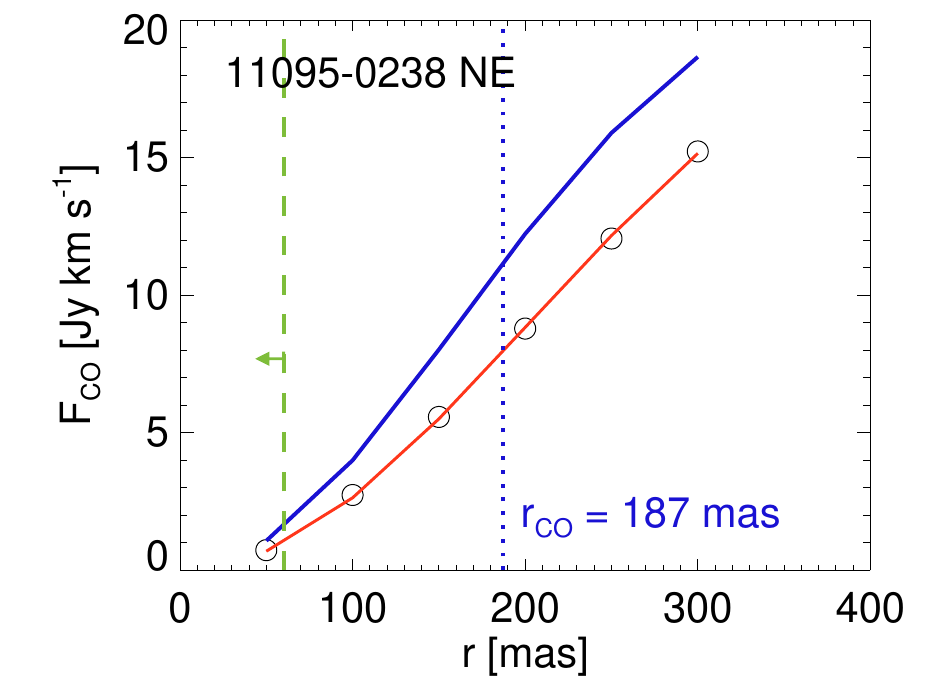}
\includegraphics[trim=23mm 130mm 93mm 98mm, clip=true, width=0.32\textwidth]{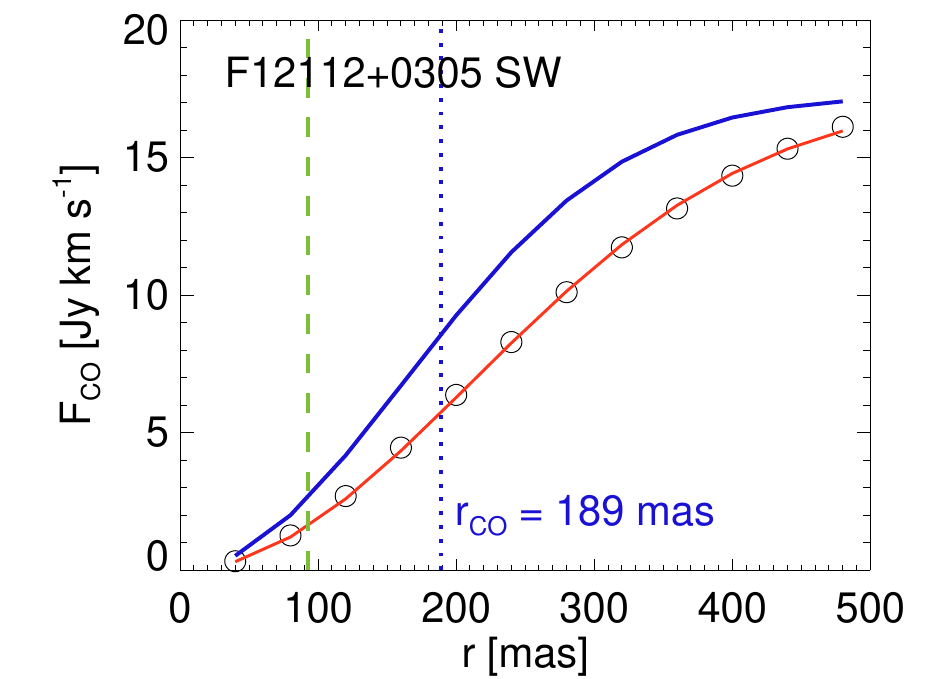}
\includegraphics[trim=23mm 130mm 93mm 98mm, clip=true, width=0.32\textwidth]{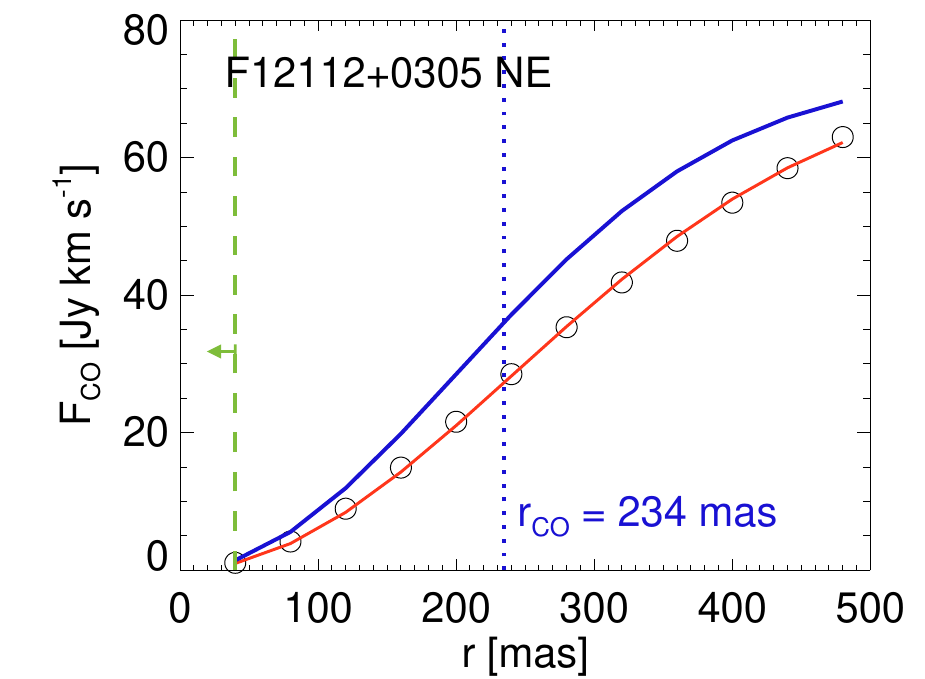}
\caption{Same as Fig.~\ref{fig:co_models}.\label{fig:apx_co_models}}
\end{figure}

\addtocounter{figure}{-1}
\begin{figure}[!h]
\centering
\vspace{5mm}
\includegraphics[trim=23mm 130mm 93mm 98mm, clip=true, width=0.32\textwidth]{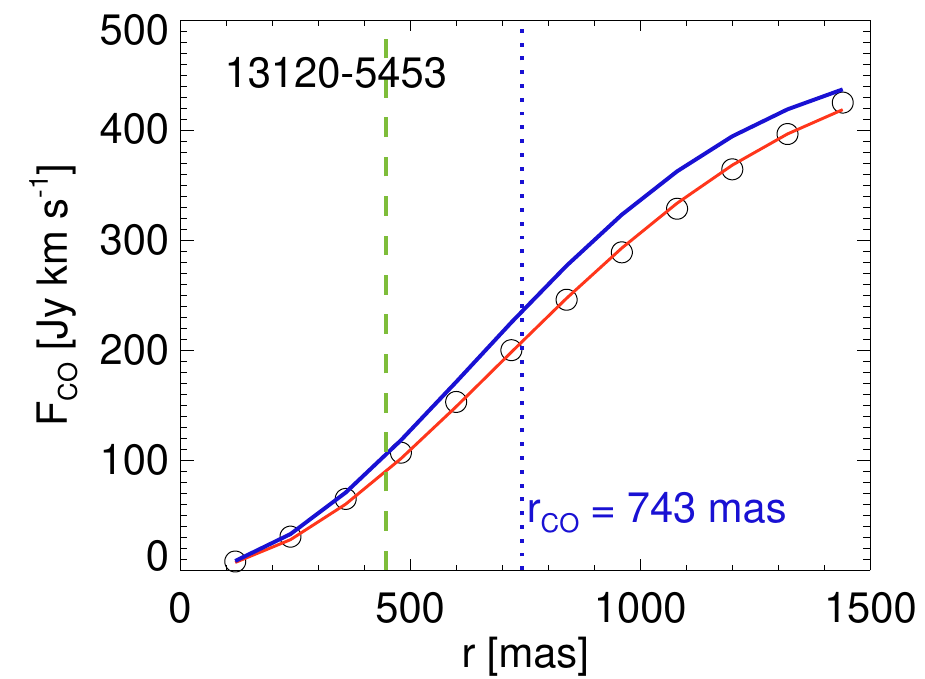}
\includegraphics[trim=23mm 130mm 93mm 98mm, clip=true, width=0.32\textwidth]{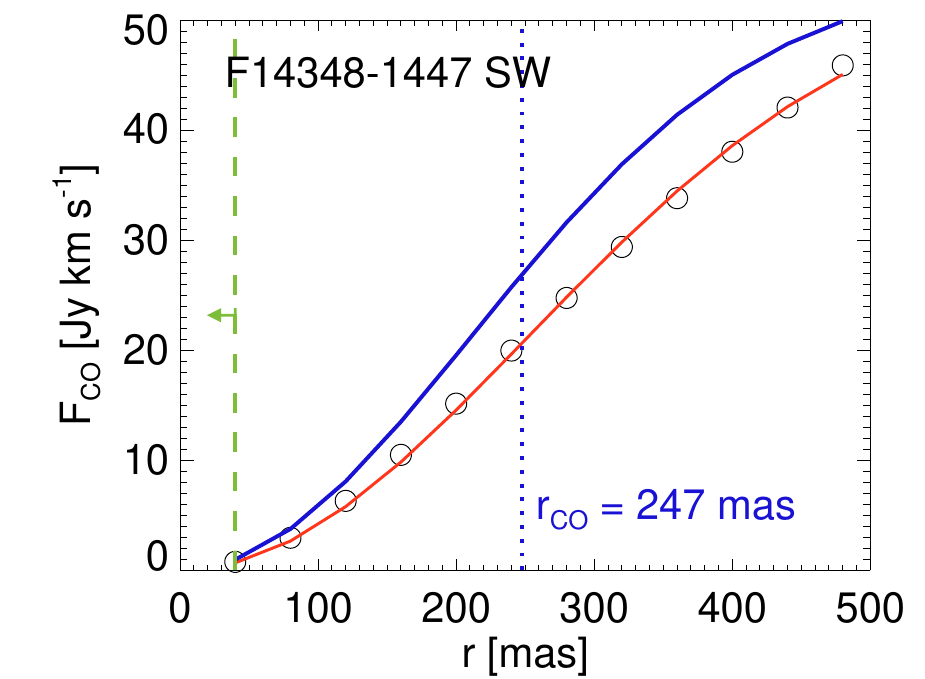}
\includegraphics[trim=23mm 130mm 93mm 98mm, clip=true, width=0.32\textwidth]{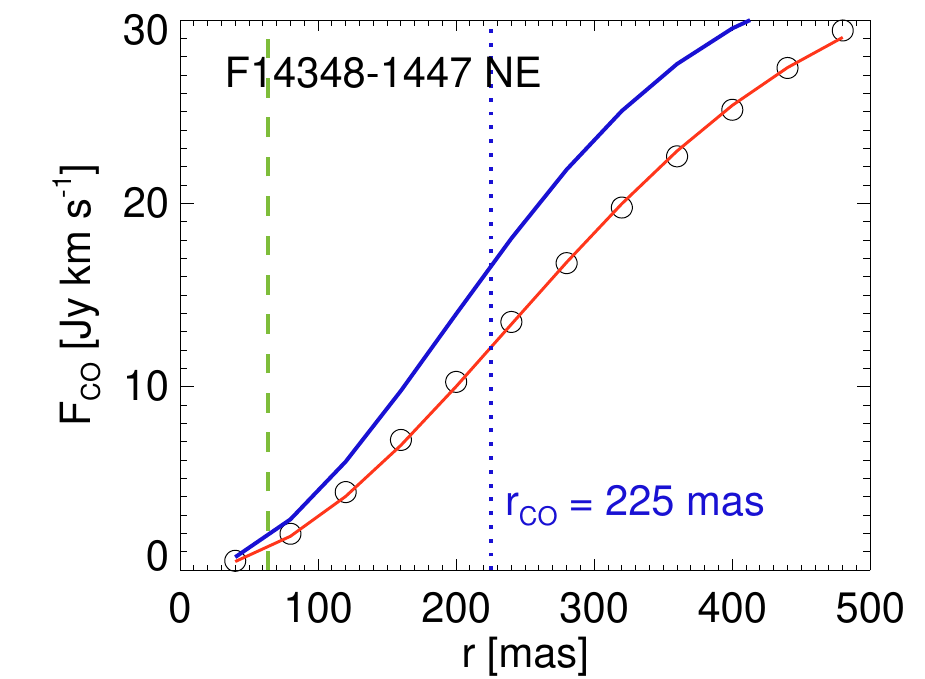}
\includegraphics[trim=23mm 130mm 93mm 98mm, clip=true, width=0.32\textwidth]{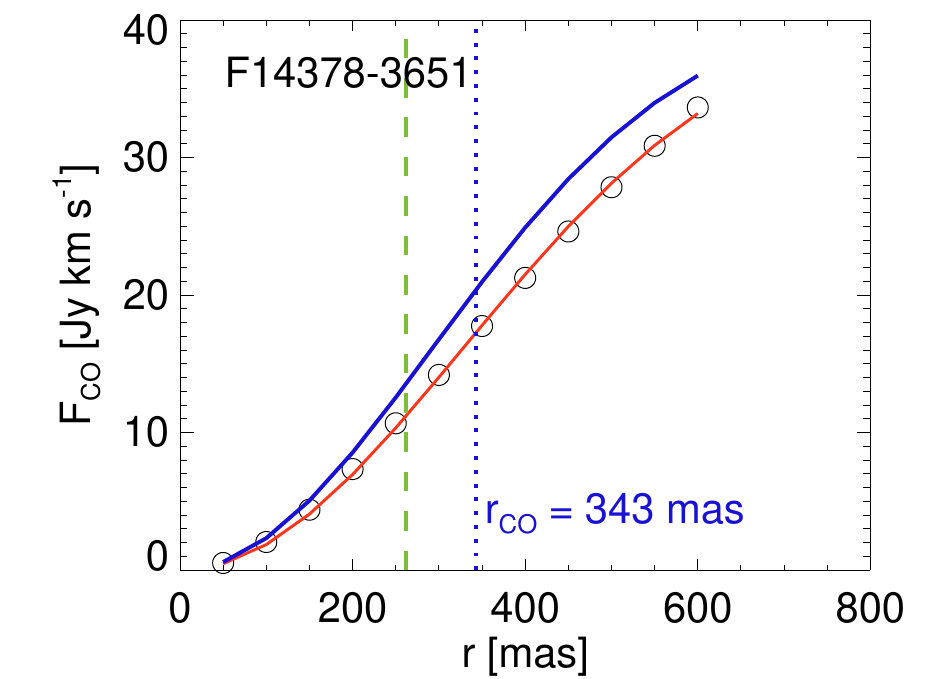}
\includegraphics[trim=23mm 130mm 93mm 98mm, clip=true, width=0.32\textwidth]{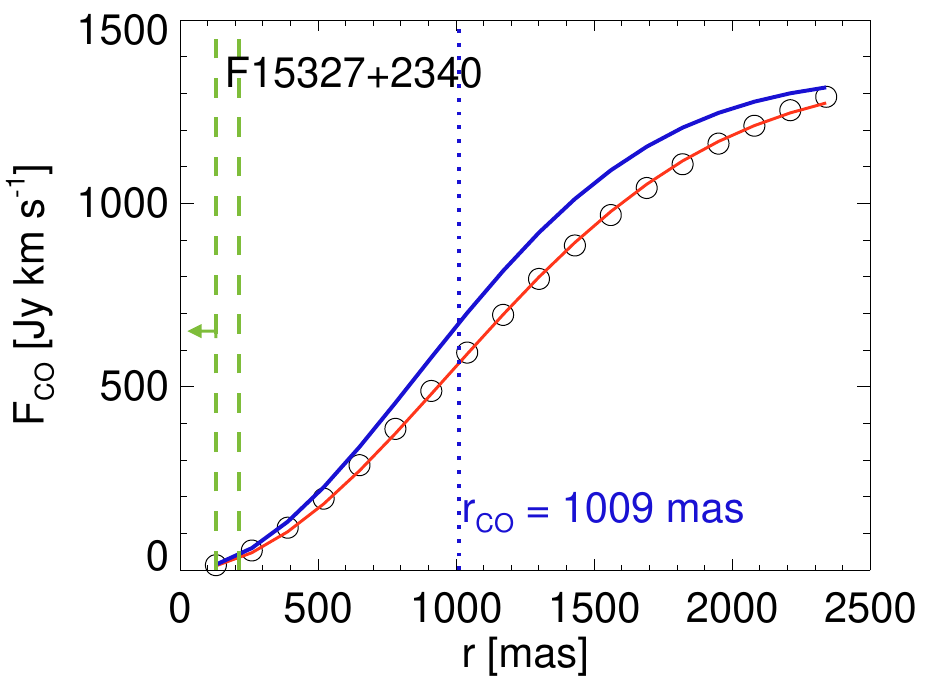}
\includegraphics[trim=23mm 130mm 93mm 98mm, clip=true, width=0.32\textwidth]{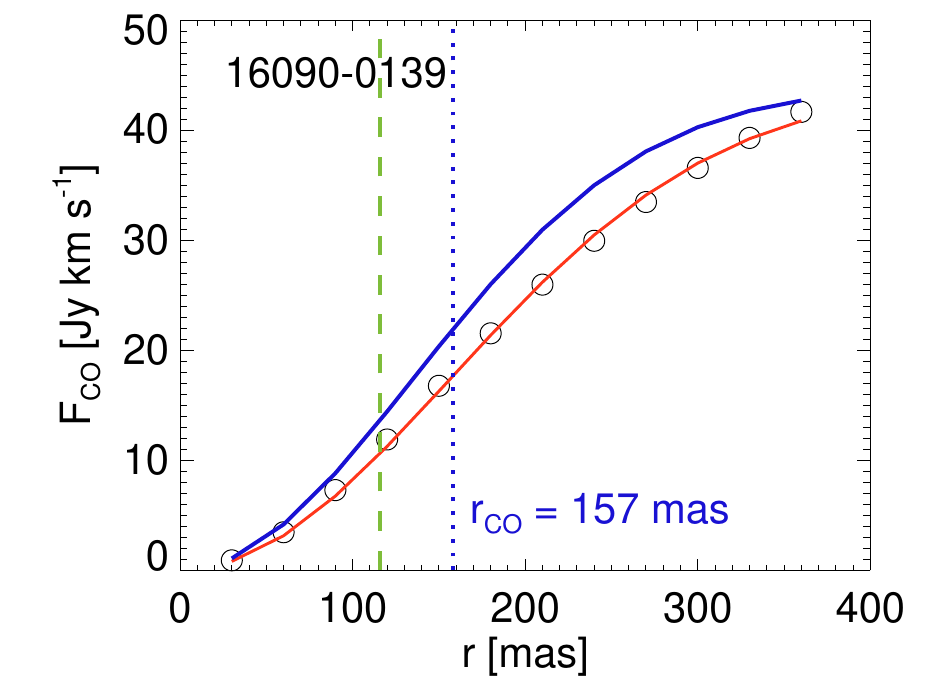}
\includegraphics[trim=23mm 130mm 93mm 98mm, clip=true, width=0.32\textwidth]{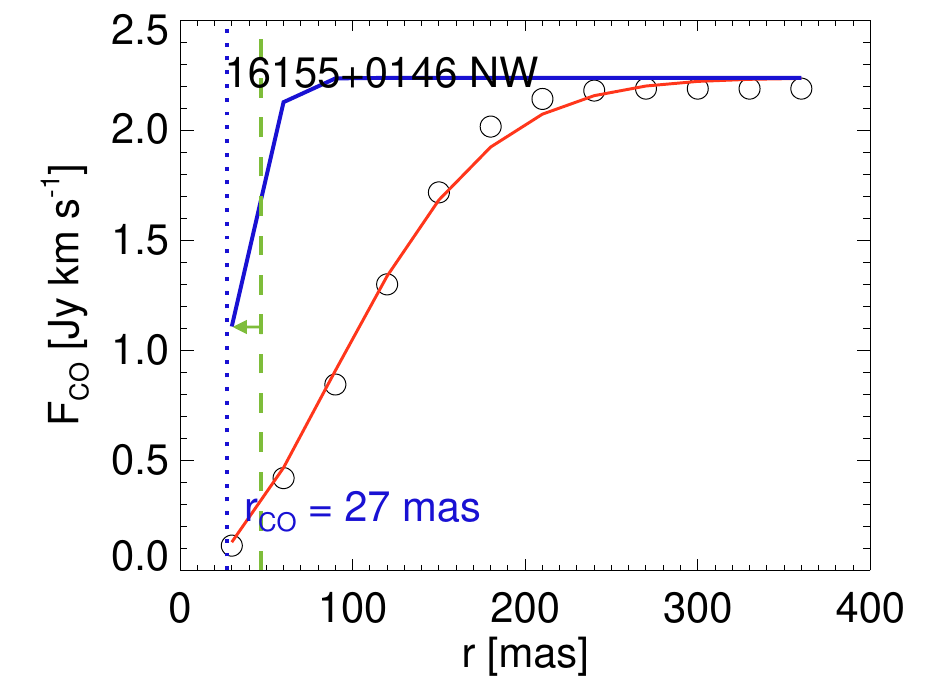}
\includegraphics[trim=23mm 130mm 93mm 98mm, clip=true, width=0.32\textwidth]{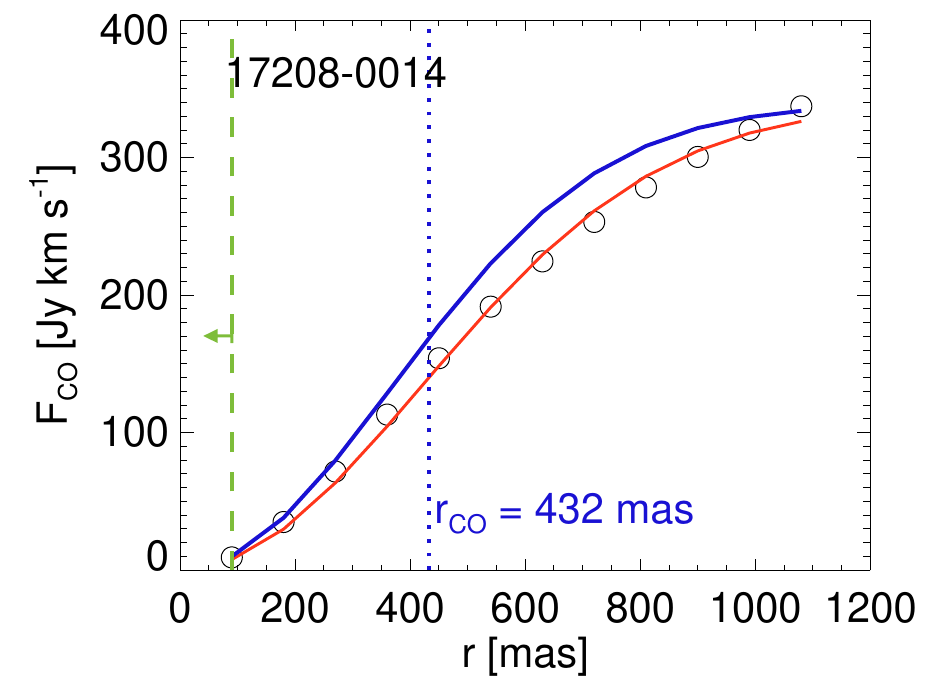}
\includegraphics[trim=23mm 130mm 93mm 98mm, clip=true, width=0.32\textwidth]{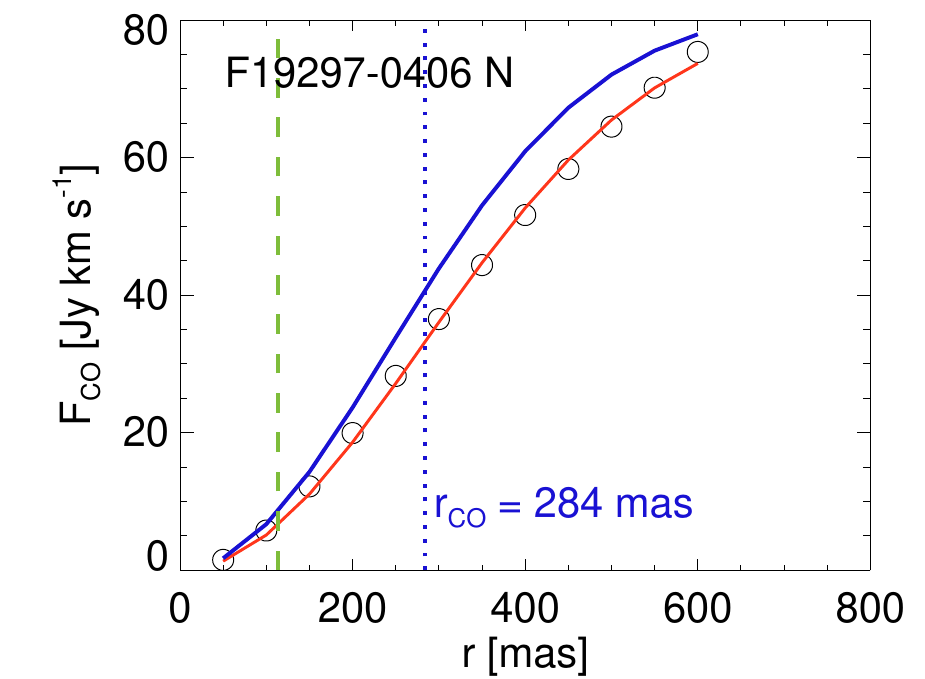}
\includegraphics[trim=23mm 130mm 93mm 98mm, clip=true, width=0.32\textwidth]{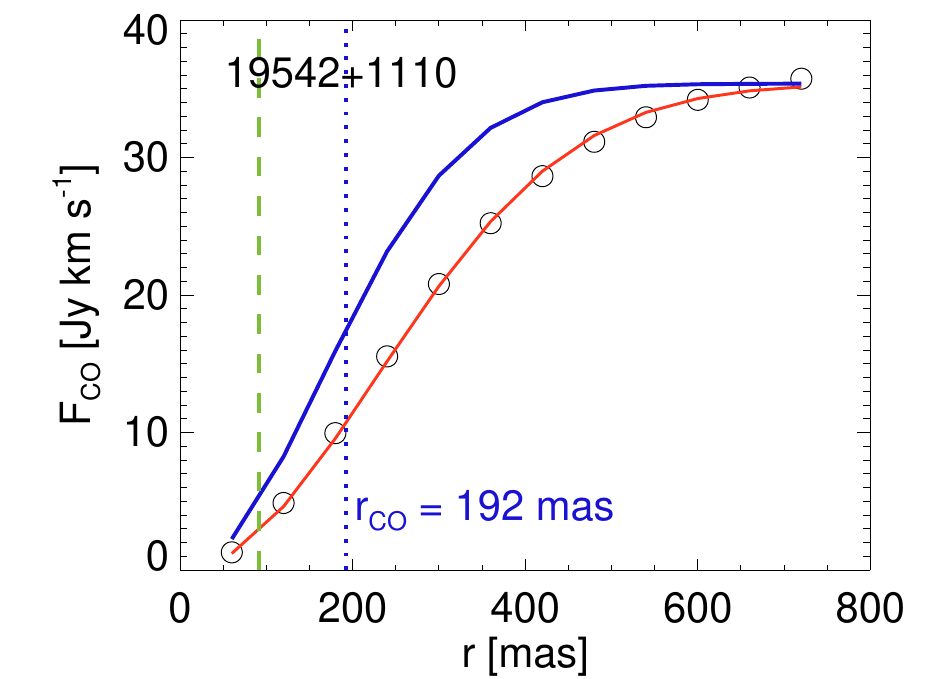}
\includegraphics[trim=23mm 130mm 93mm 98mm, clip=true, width=0.32\textwidth]{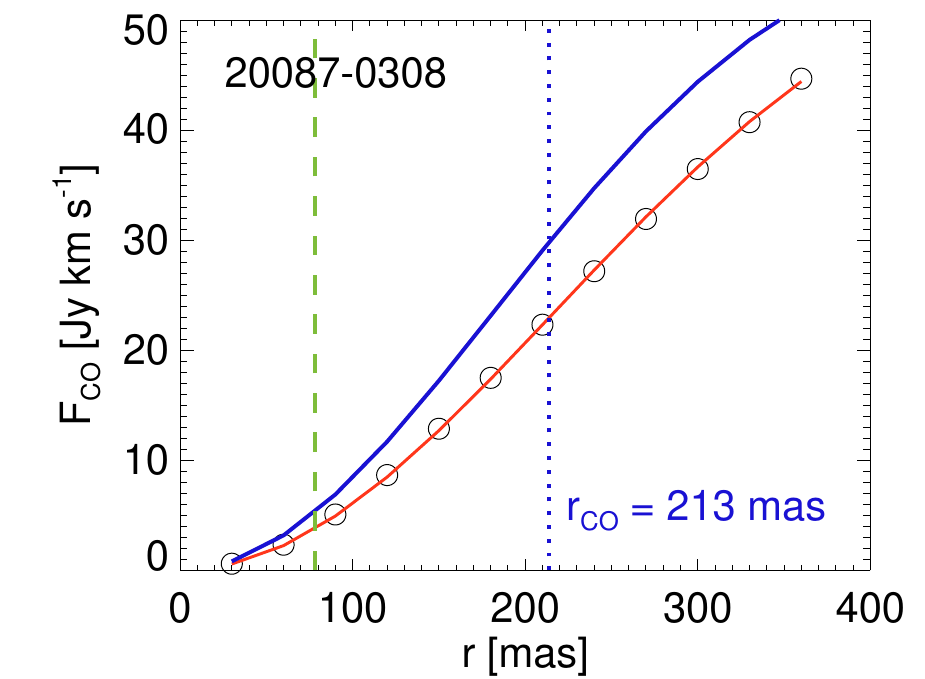}
\includegraphics[trim=23mm 130mm 93mm 98mm, clip=true, width=0.32\textwidth]{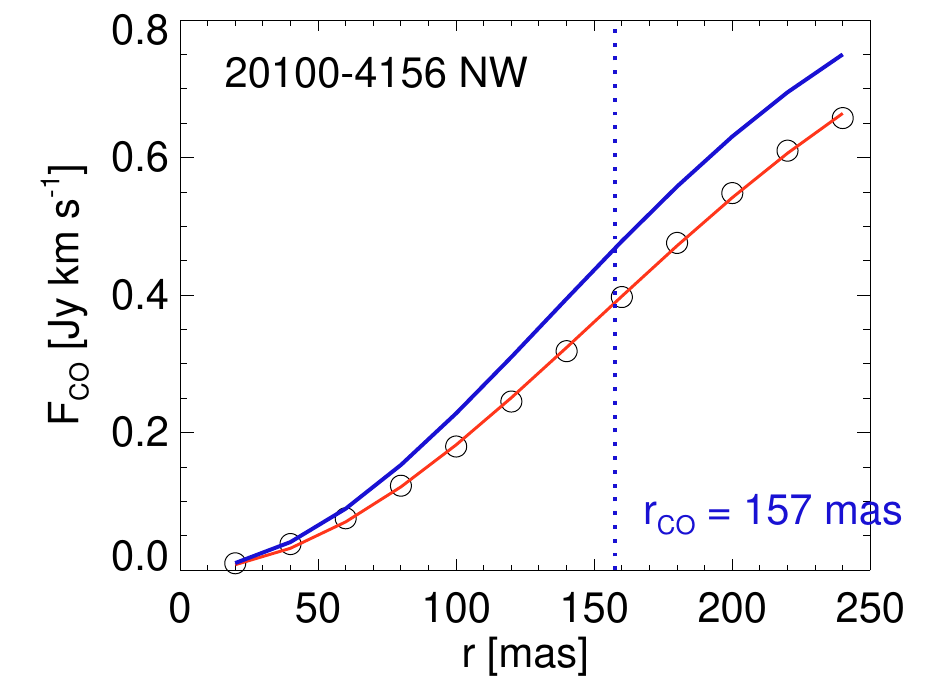}
\includegraphics[trim=23mm 130mm 93mm 98mm, clip=true, width=0.32\textwidth]{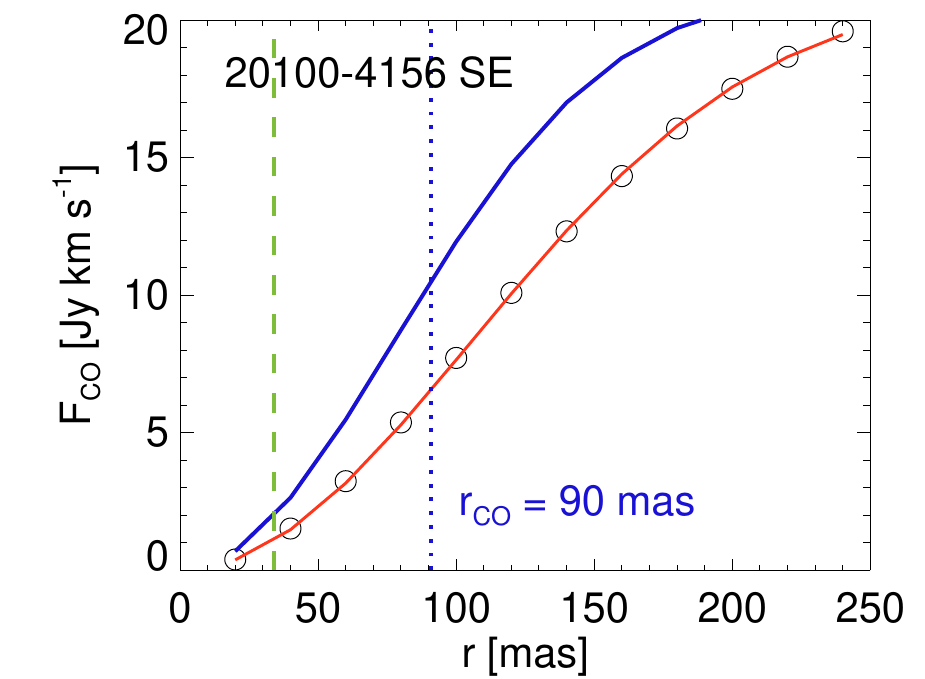}
\includegraphics[trim=23mm 130mm 93mm 98mm, clip=true, width=0.32\textwidth]{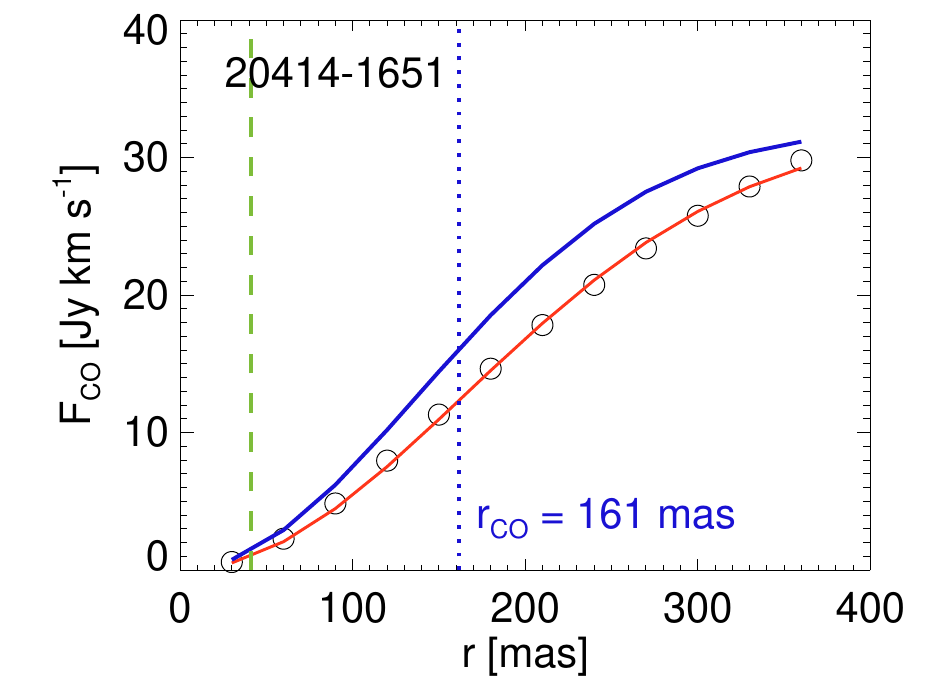}
\includegraphics[trim=23mm 130mm 93mm 98mm, clip=true, width=0.32\textwidth]{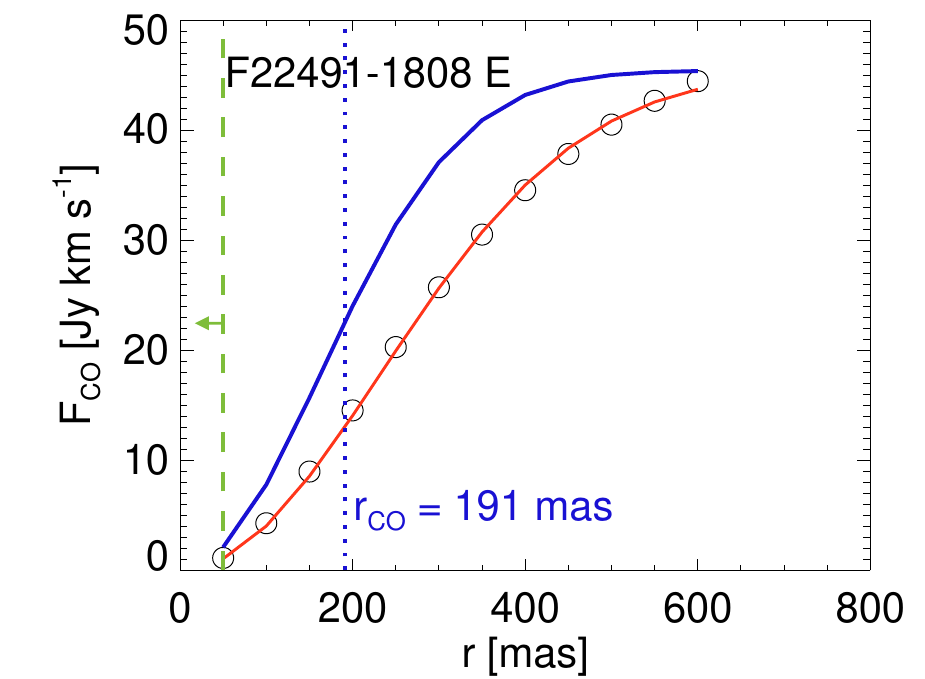}
\caption{(Continued)}
\end{figure}

\clearpage

\section{SED models}\label{apx:sed_models}
\begin{figure}[!h]
\centering
\vspace{5mm}
\includegraphics[width=0.32\textwidth]{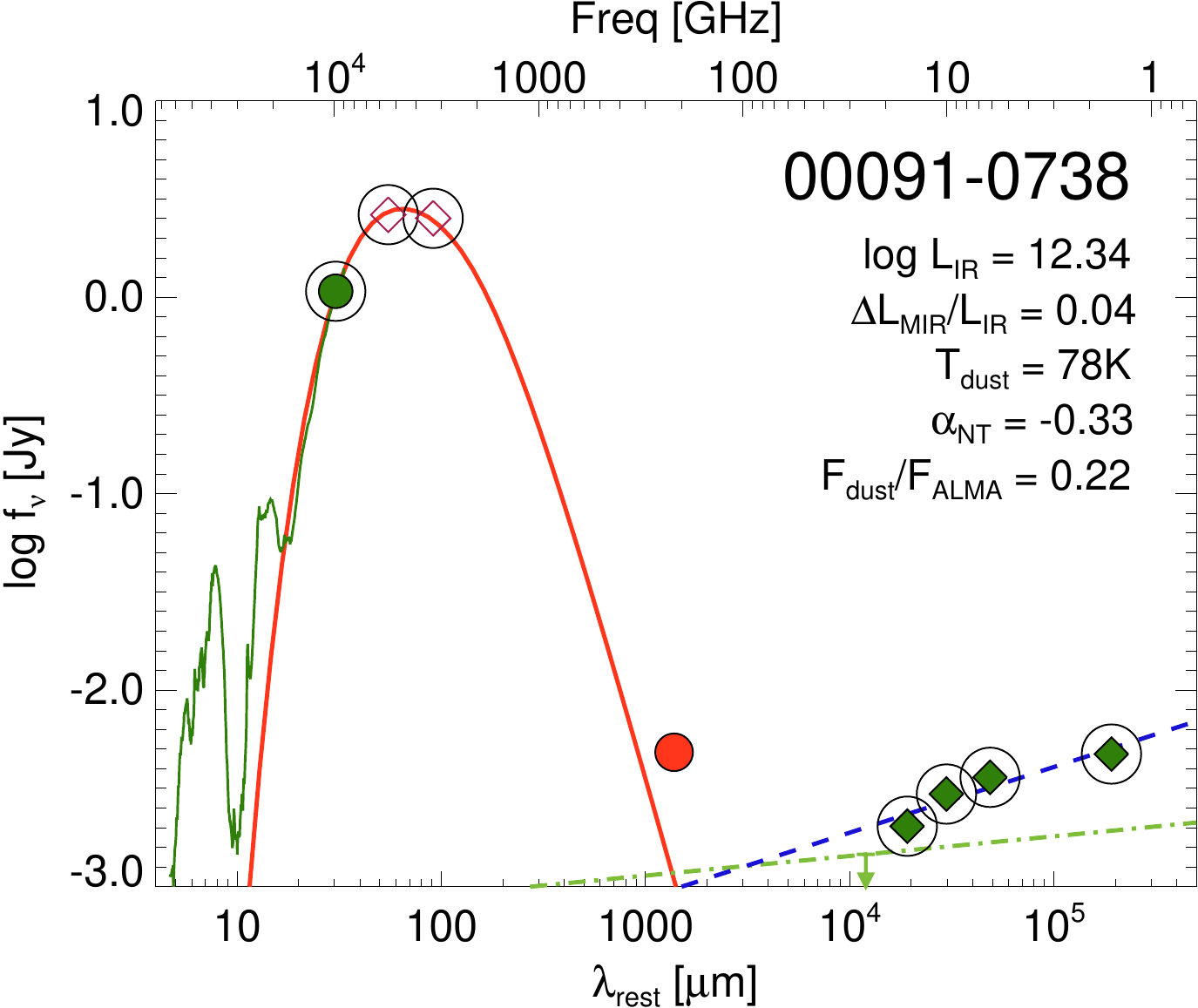}
\includegraphics[width=0.32\textwidth]{sed_00188-0856.pdf}
\includegraphics[width=0.32\textwidth]{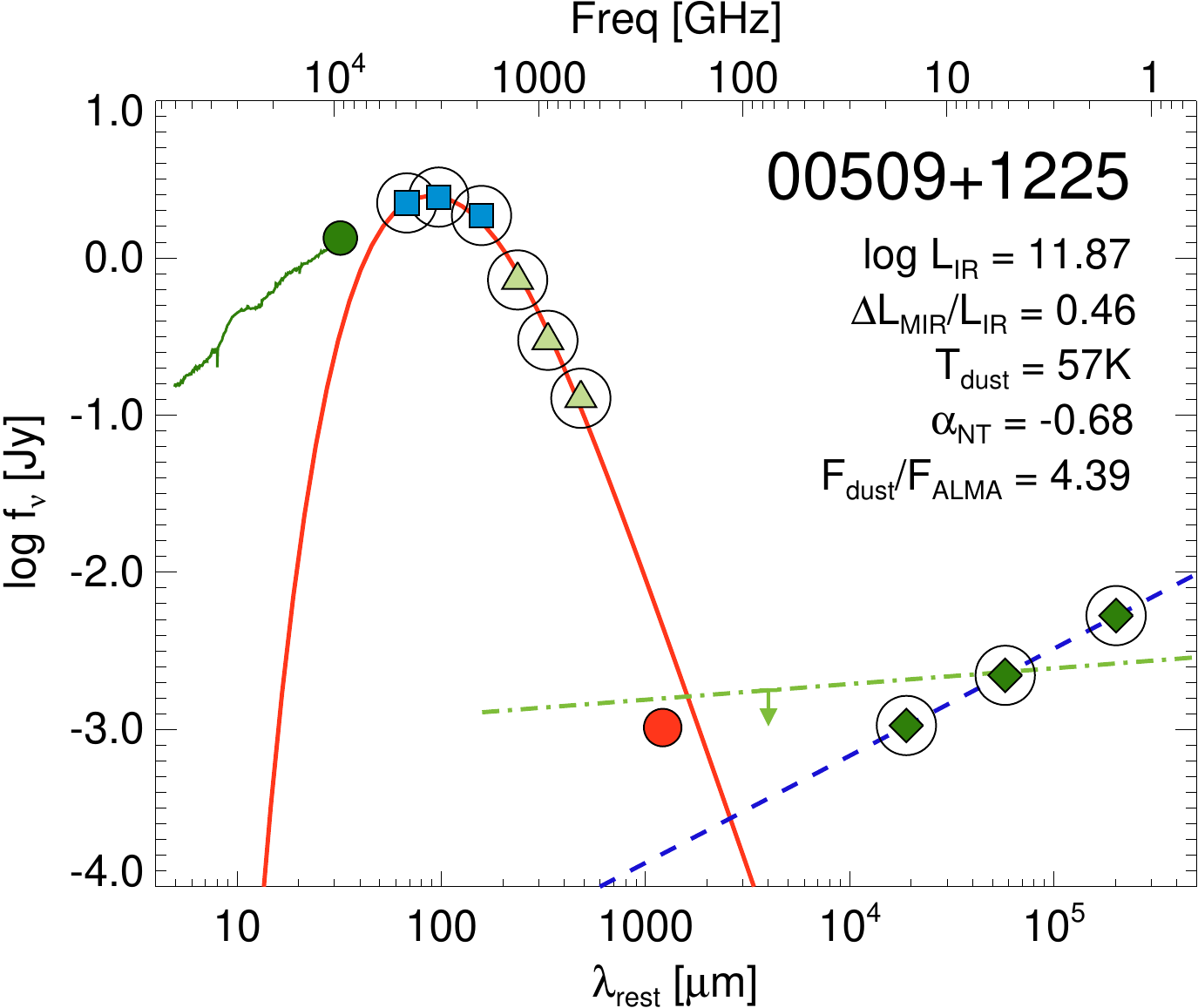}
\includegraphics[width=0.32\textwidth]{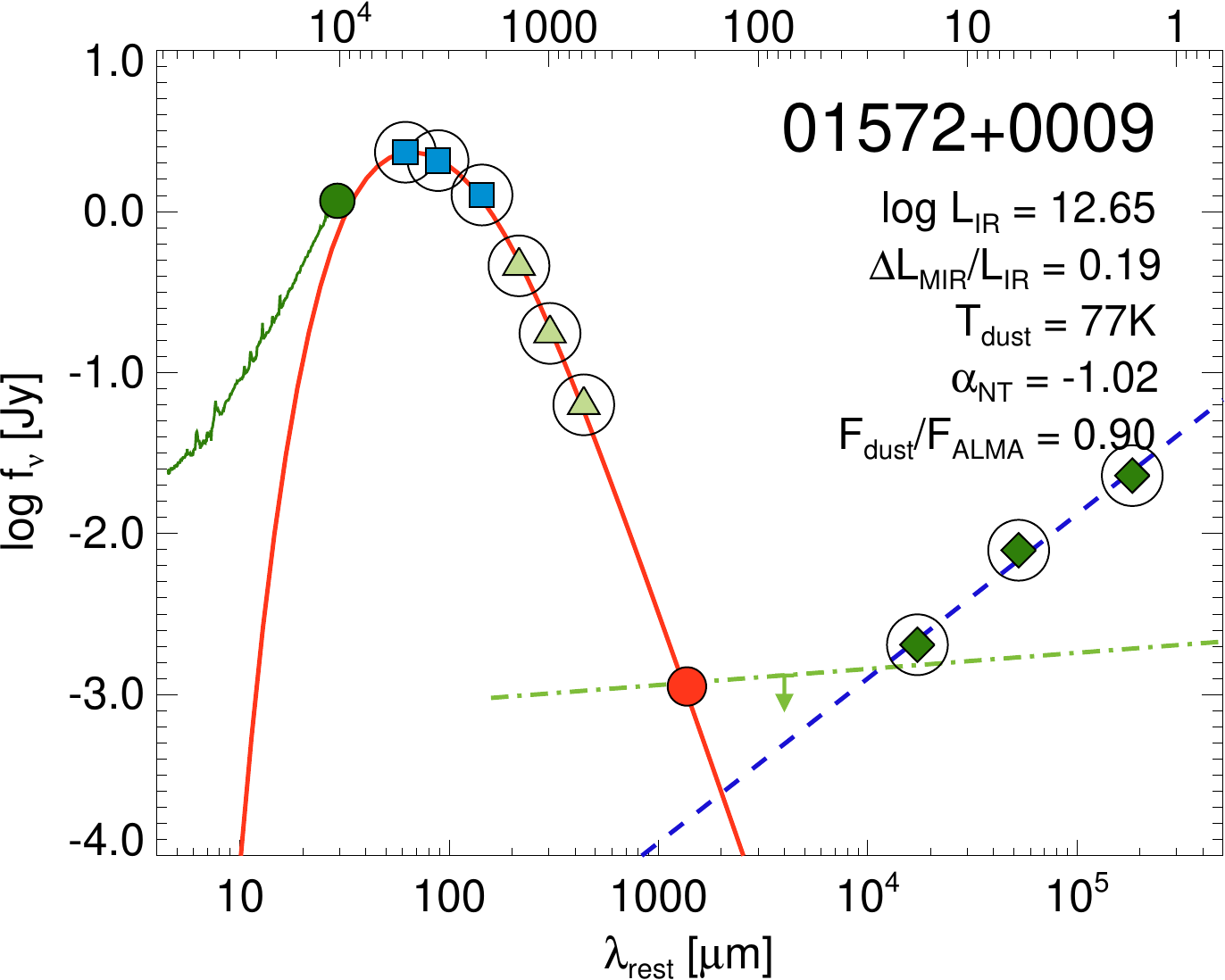}
\includegraphics[width=0.32\textwidth]{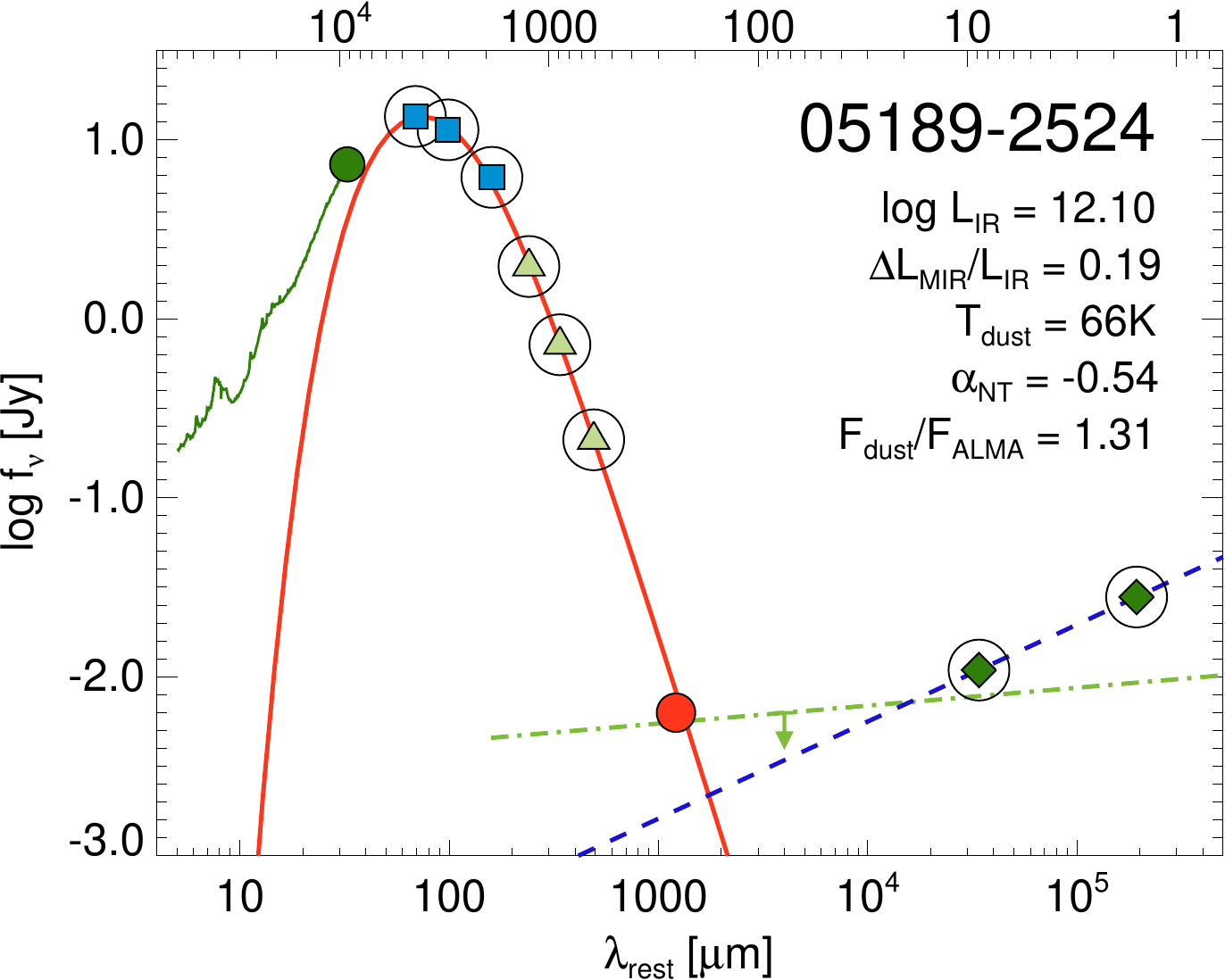}
\includegraphics[width=0.32\textwidth]{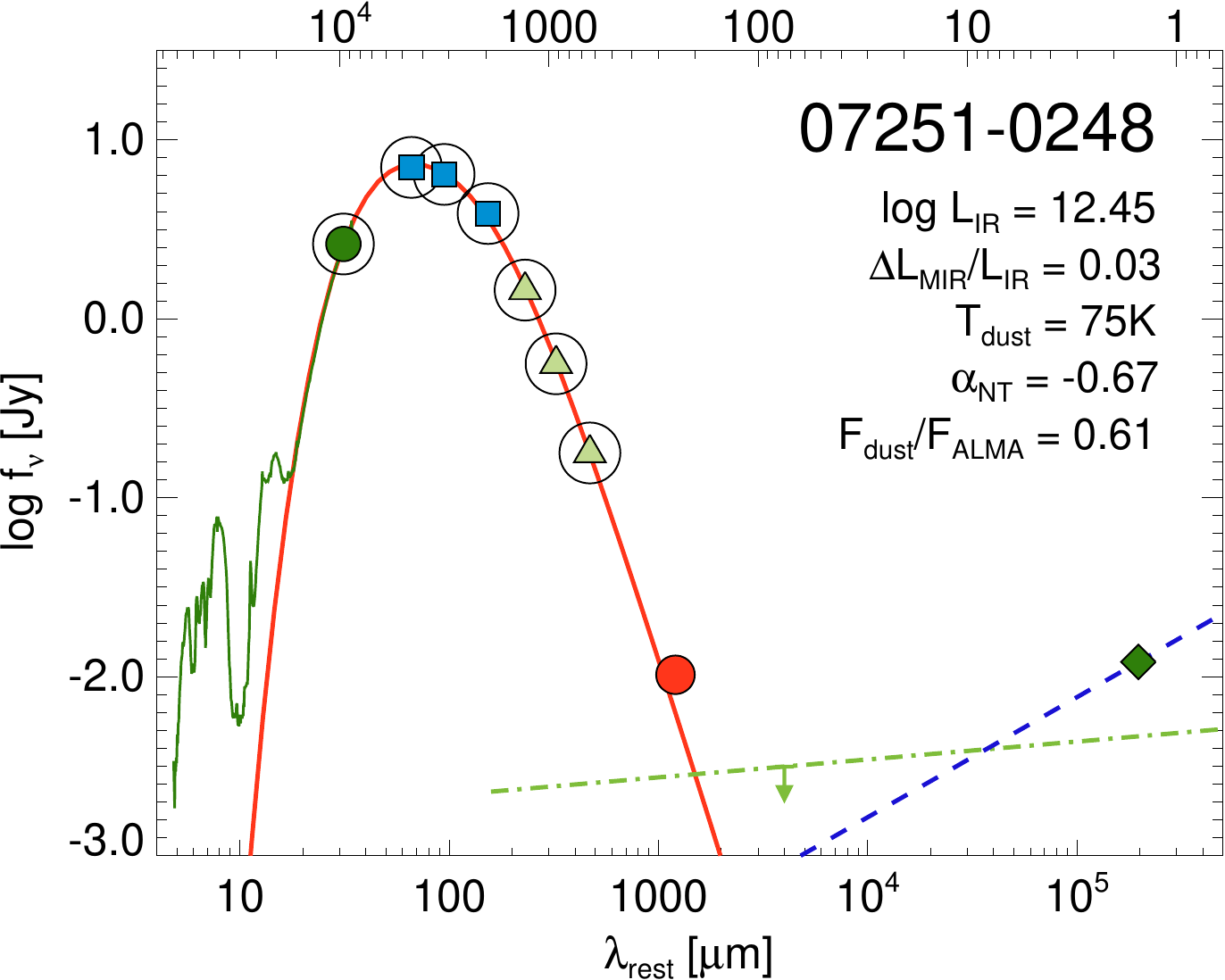}
\includegraphics[width=0.32\textwidth]{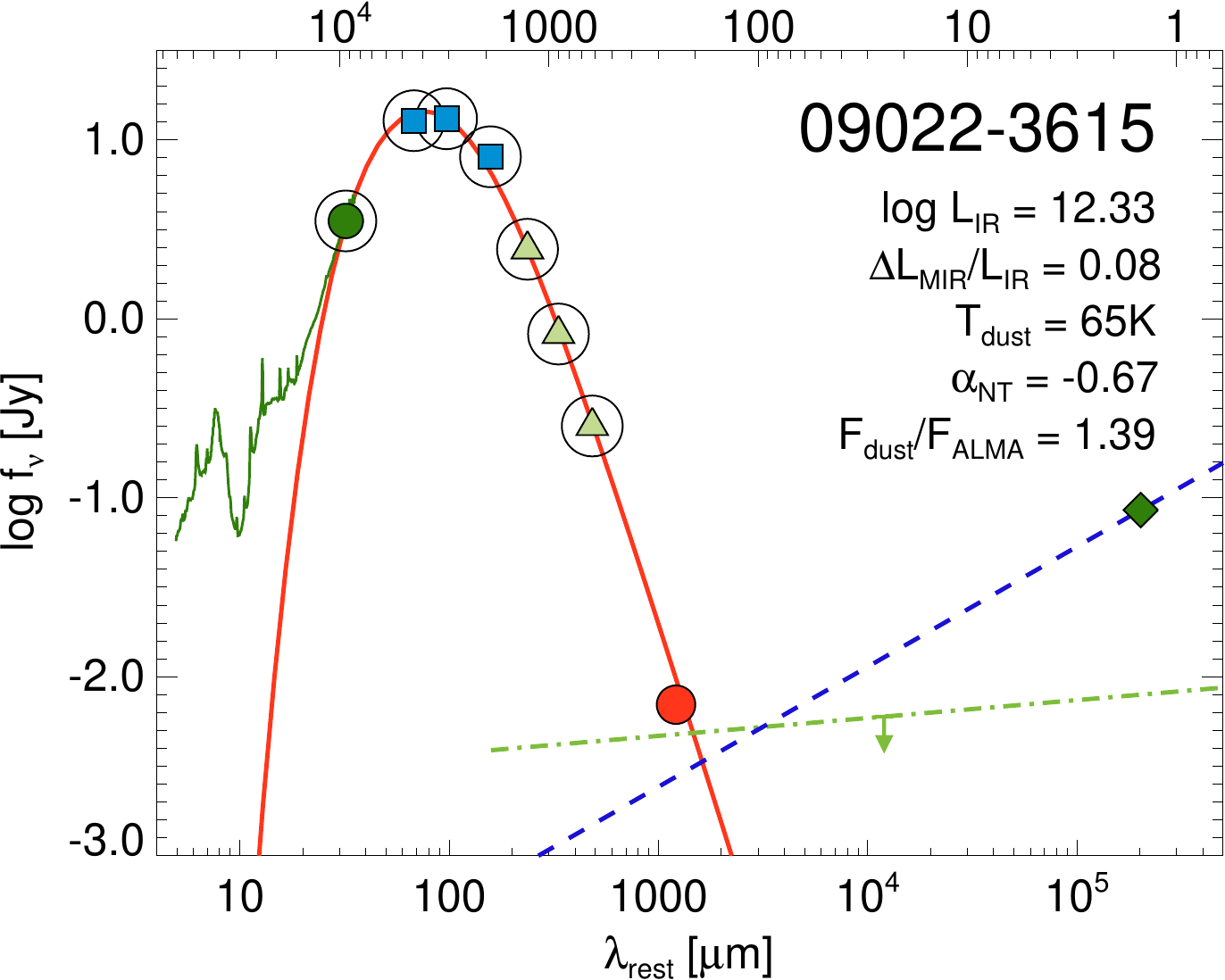}
\includegraphics[width=0.32\textwidth]{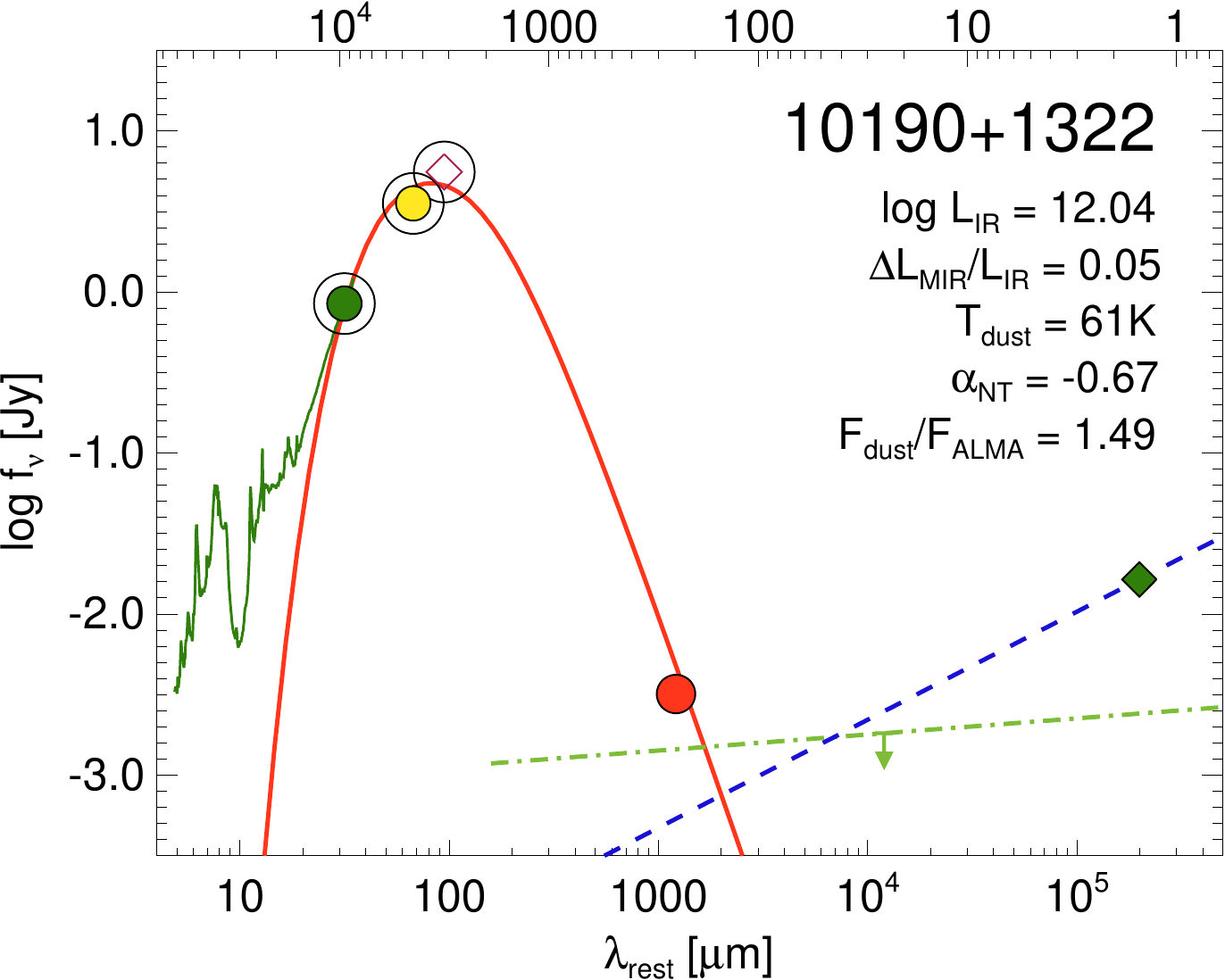}
\includegraphics[width=0.32\textwidth]{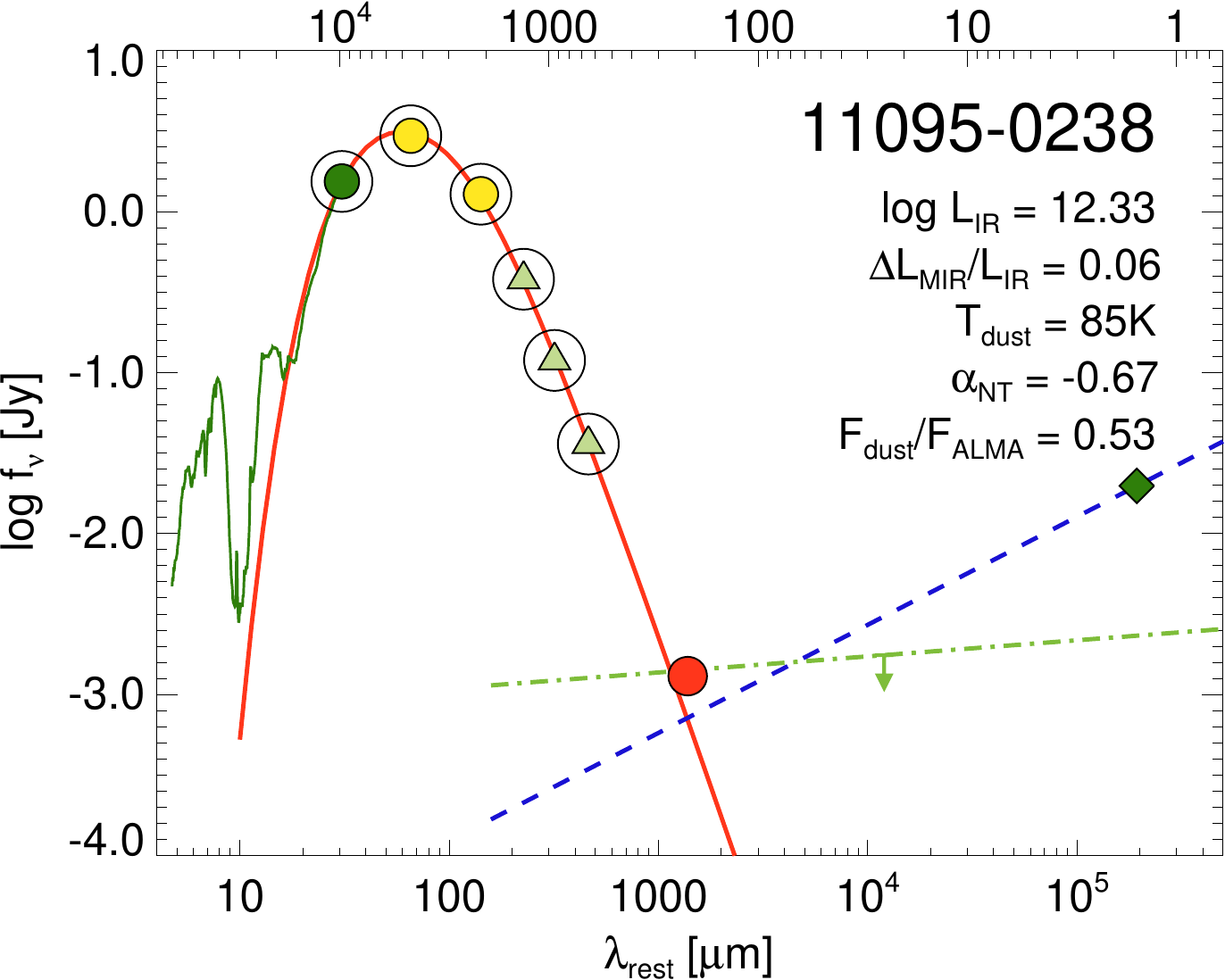}
\includegraphics[width=0.32\textwidth]{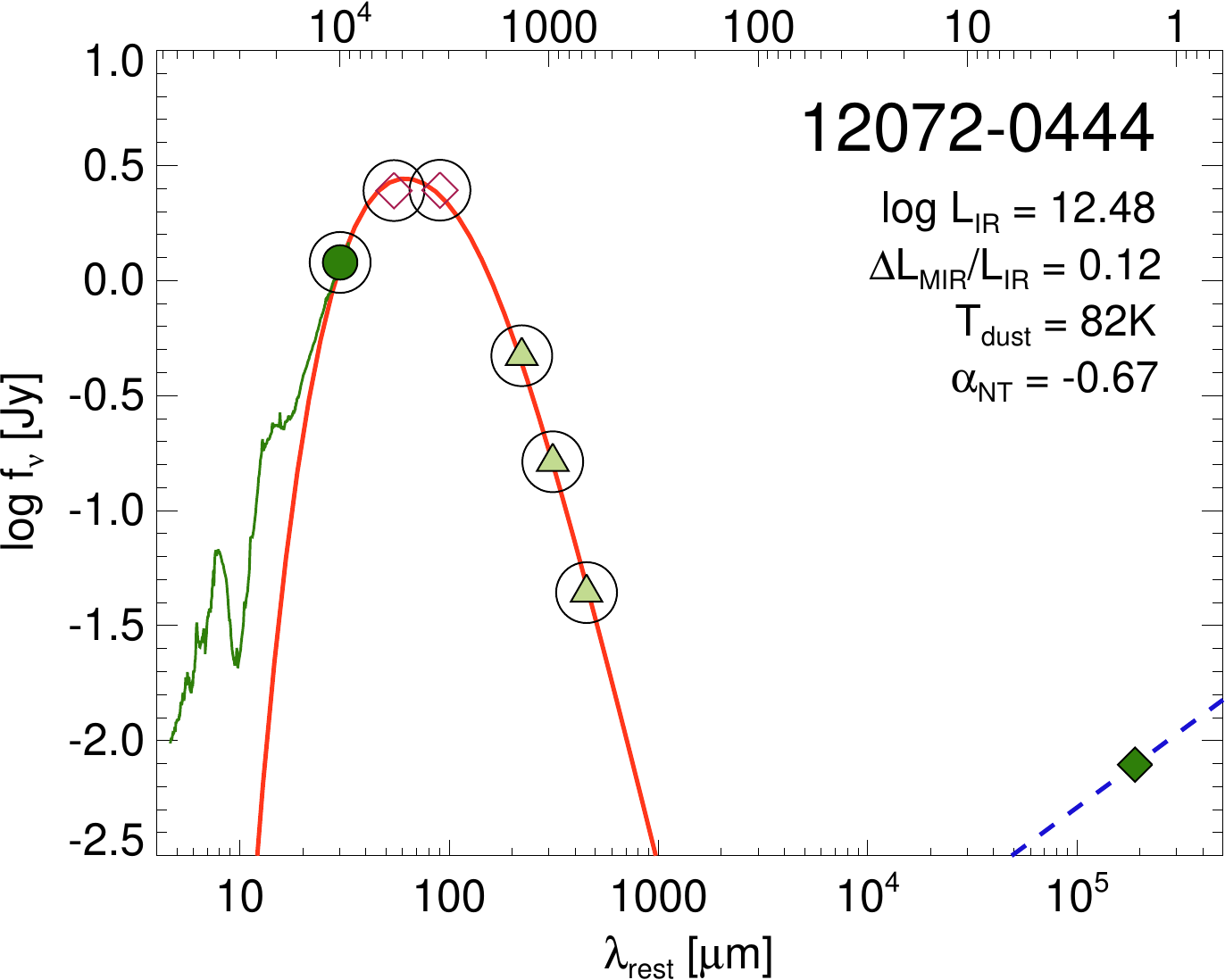}
\includegraphics[width=0.32\textwidth]{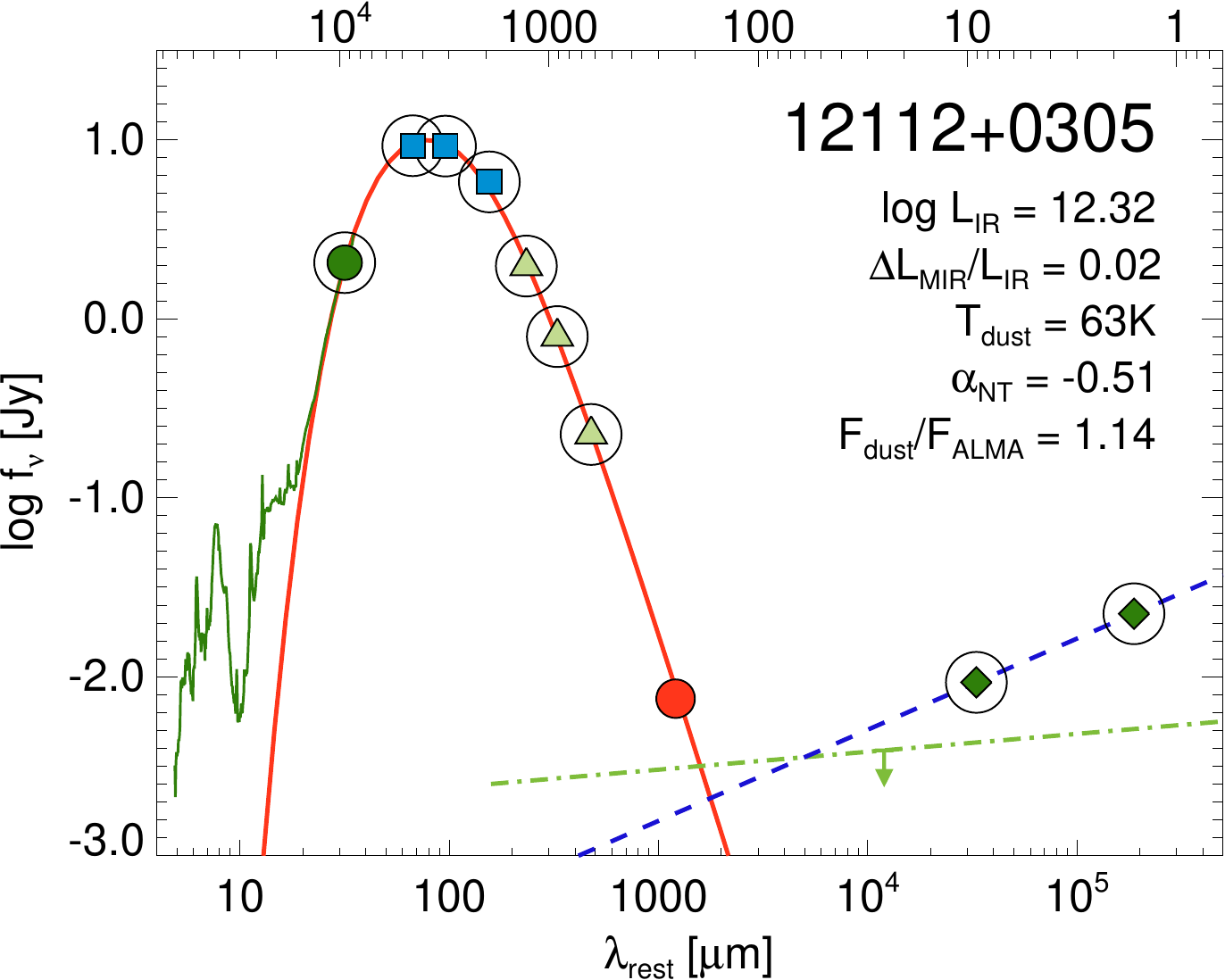}
\includegraphics[width=0.32\textwidth]{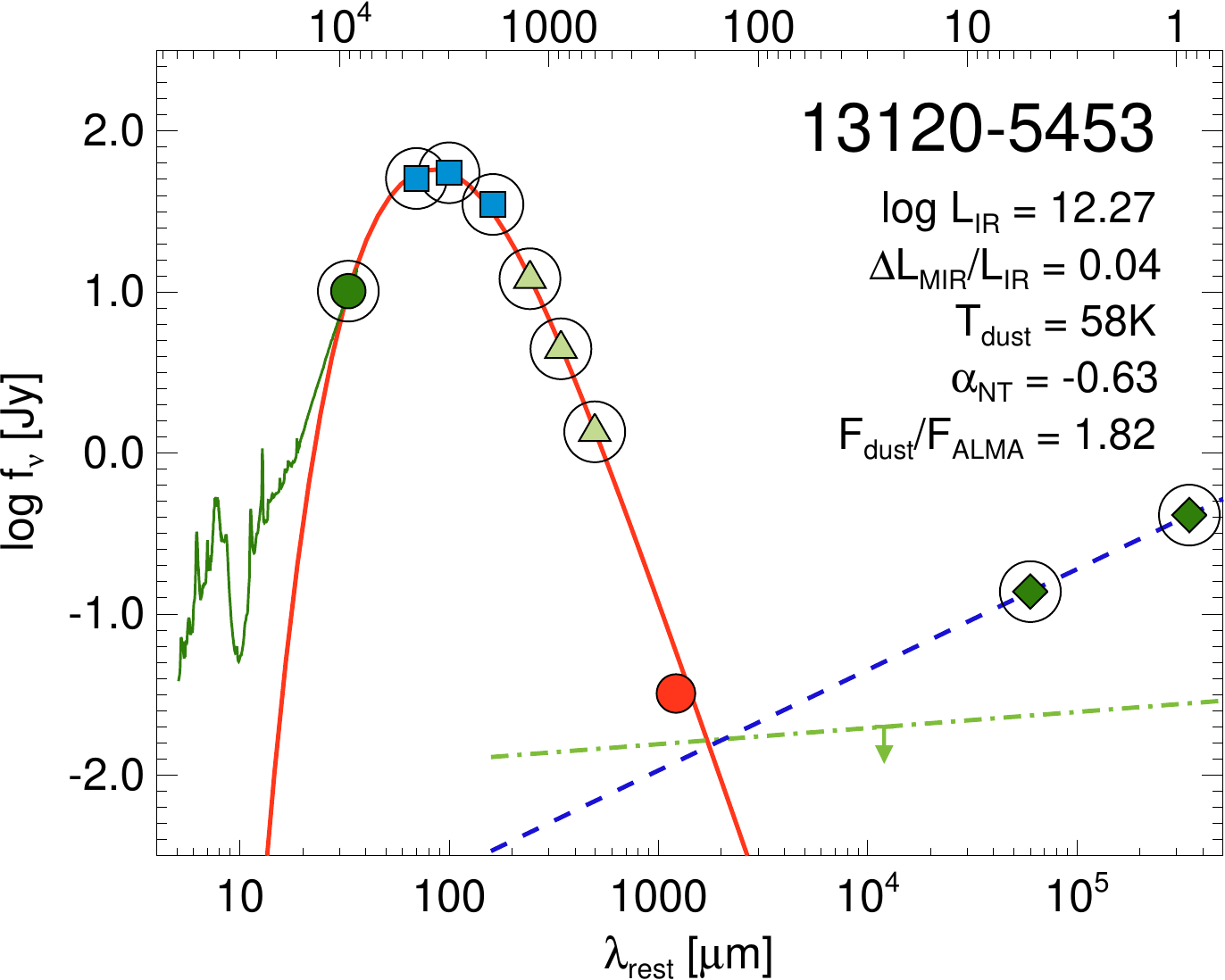}
\caption{Same as Fig.~\ref{fig:sed_models}. The IR observations are color coded as follows: {Spitzer}\slash IRS synthetic photometry at 34\micron\ (green circle); \textit{Spitzer}\slash MIPS (yellow circles); \textit{IRAS} (purple diamonds); \textit{Herschel}\slash PACS (blue squares); \textit{ISO}\slash ISOPHOT (blue circles); and  \textit{Herschel}\slash SPIRE (green triangles).
\label{fig:apx_sed_models}}
\end{figure}

\addtocounter{figure}{-1}
\begin{figure}[!h]
\centering
\vspace{5mm}
\includegraphics[width=0.32\textwidth]{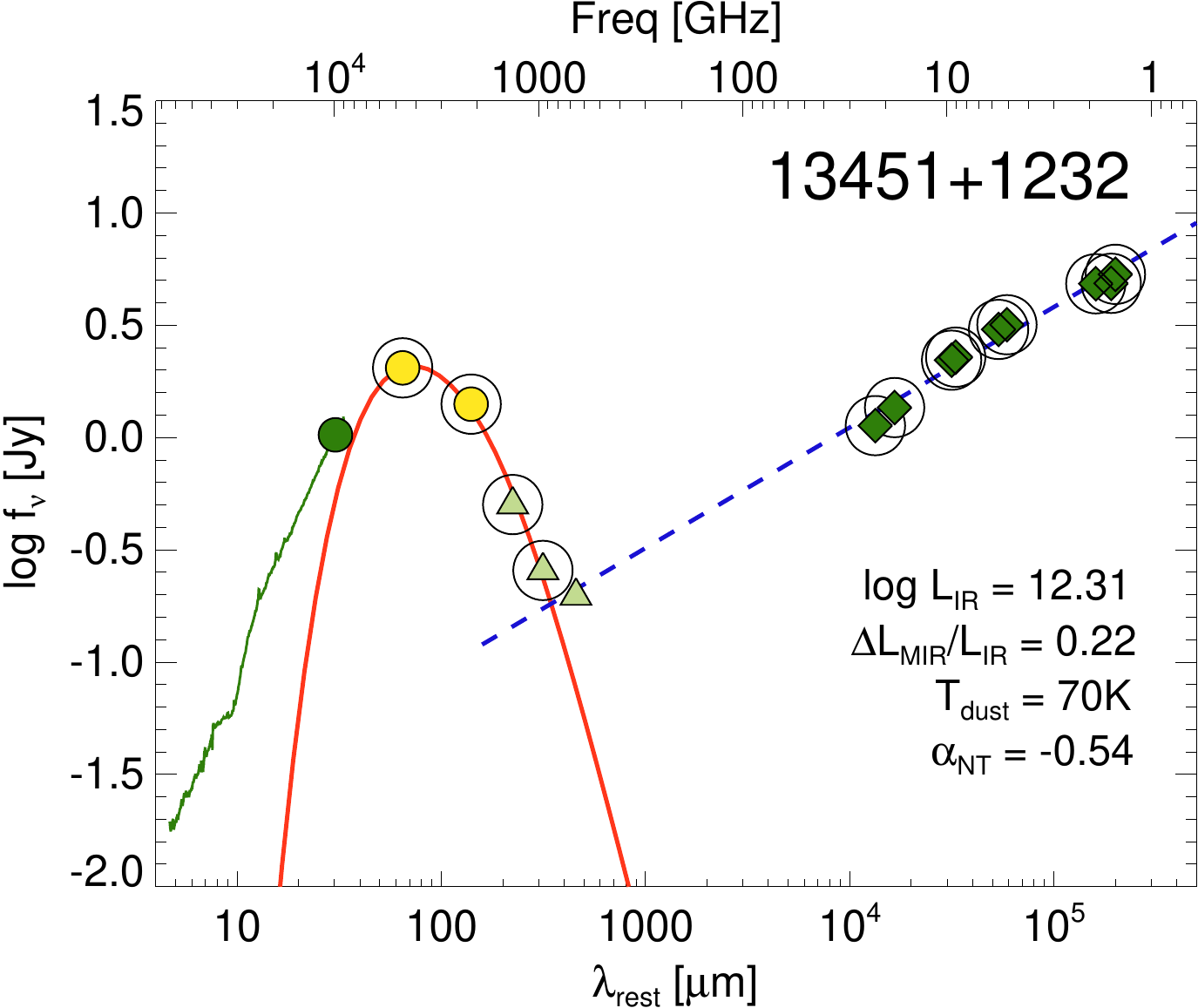}
\includegraphics[width=0.32\textwidth]{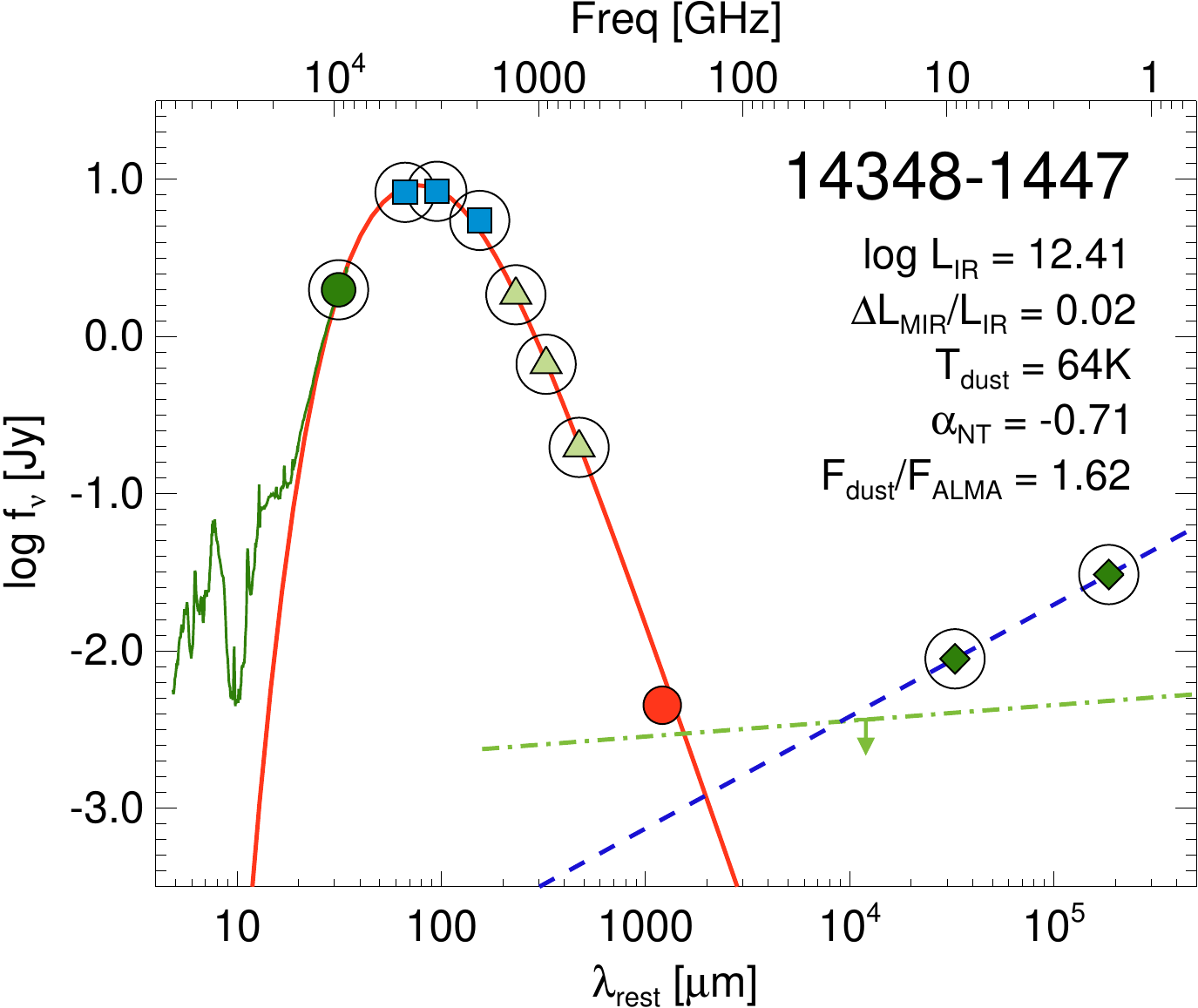}
\includegraphics[width=0.32\textwidth]{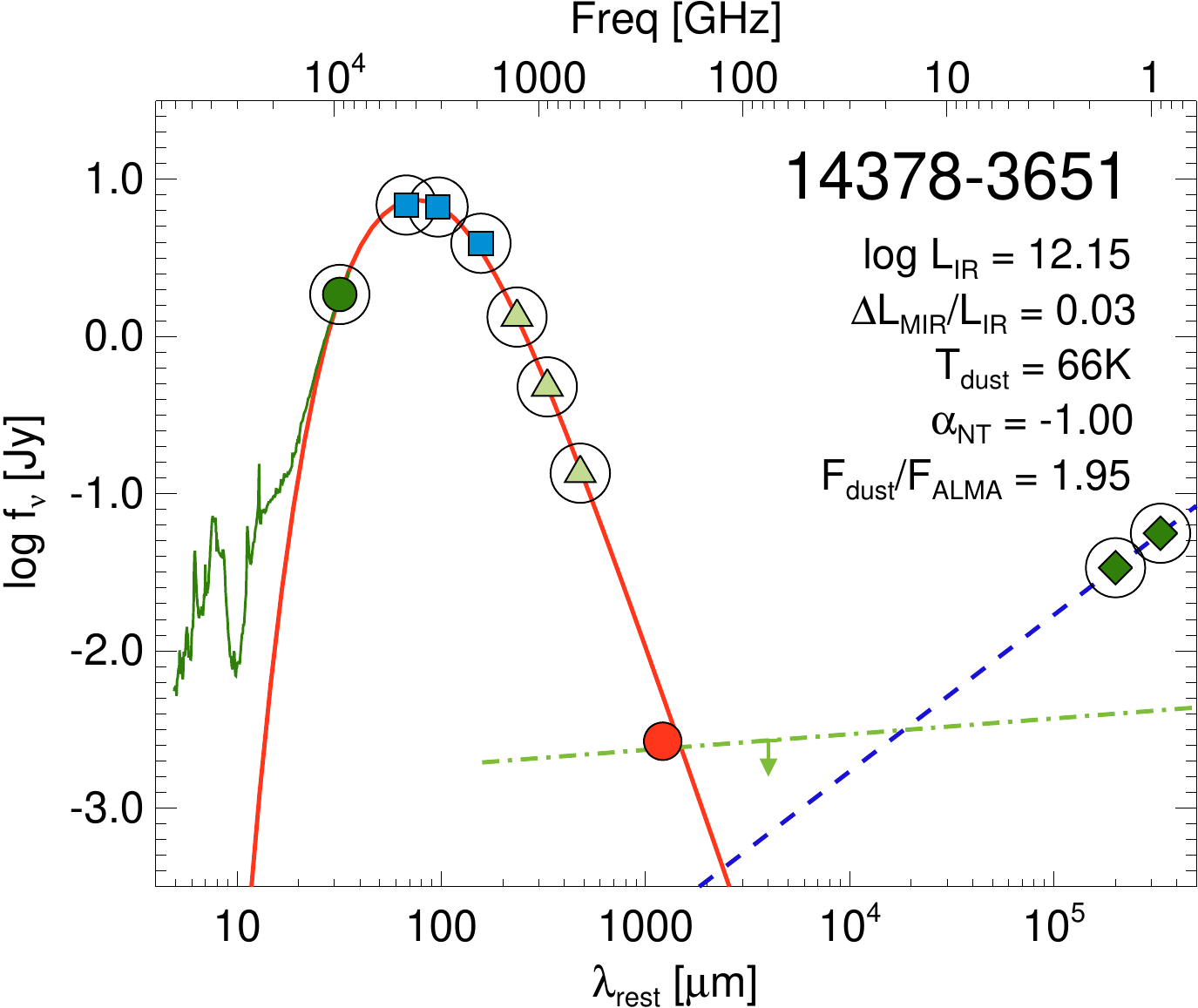}
\includegraphics[width=0.32\textwidth]{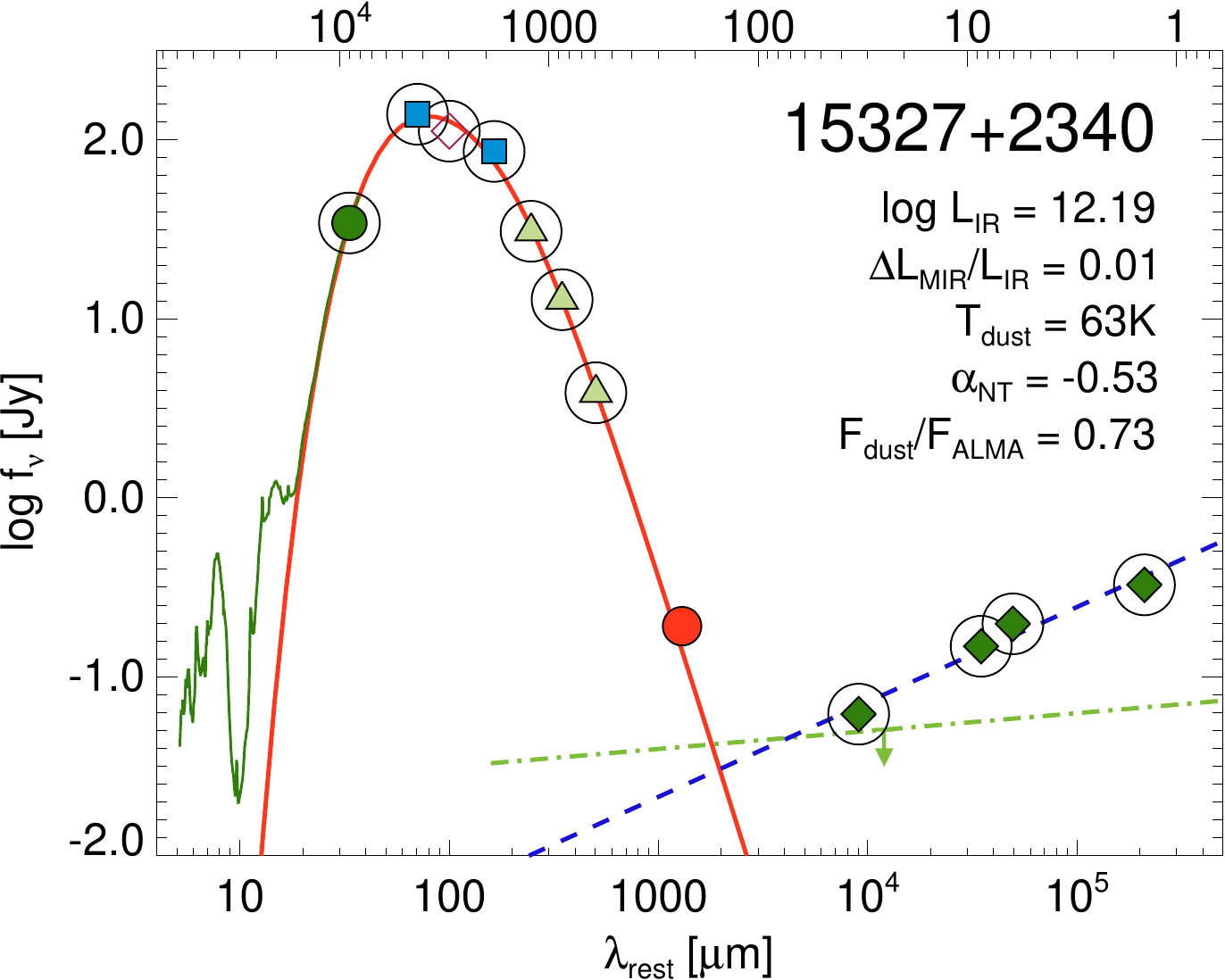}
\includegraphics[width=0.32\textwidth]{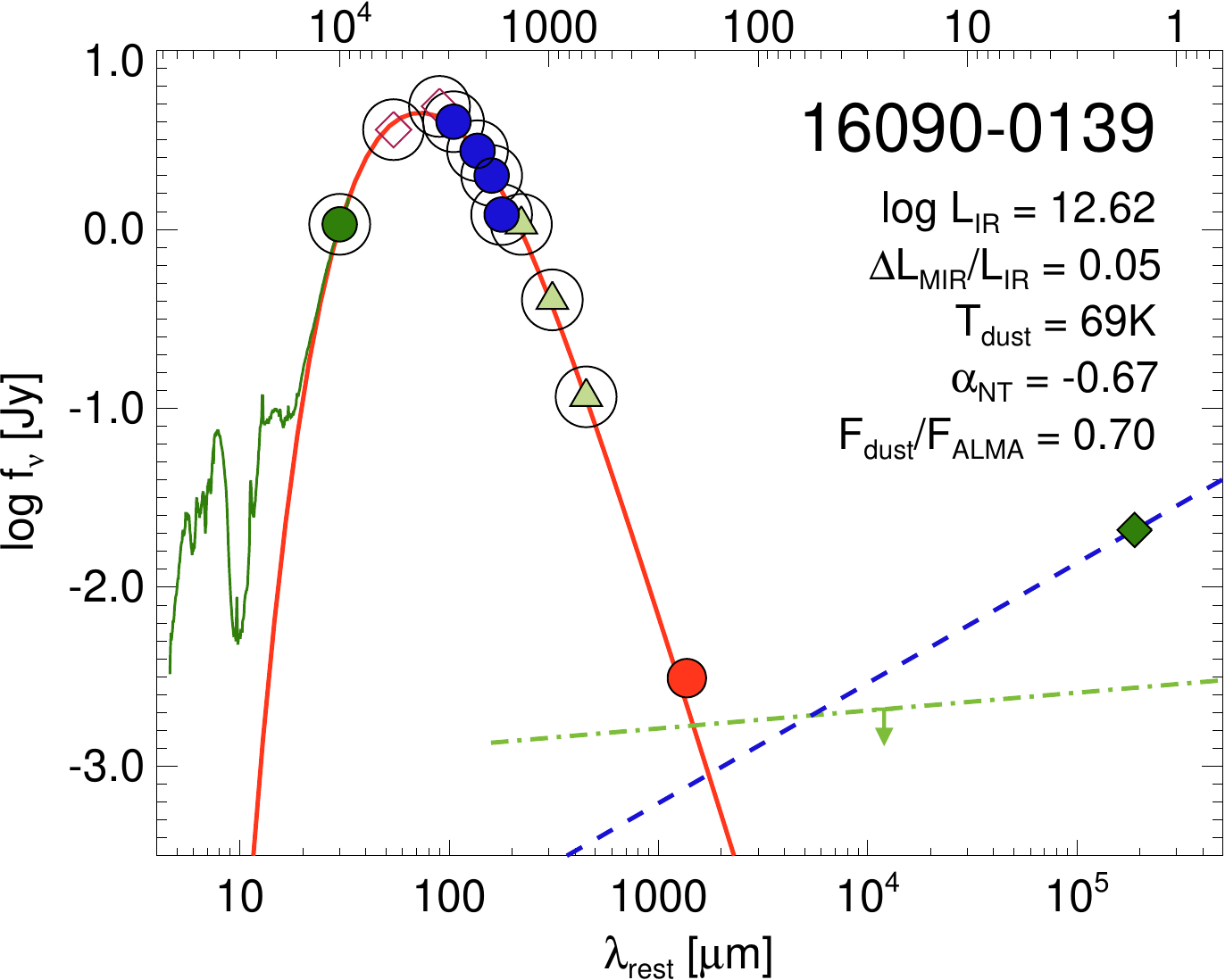}
\includegraphics[width=0.32\textwidth]{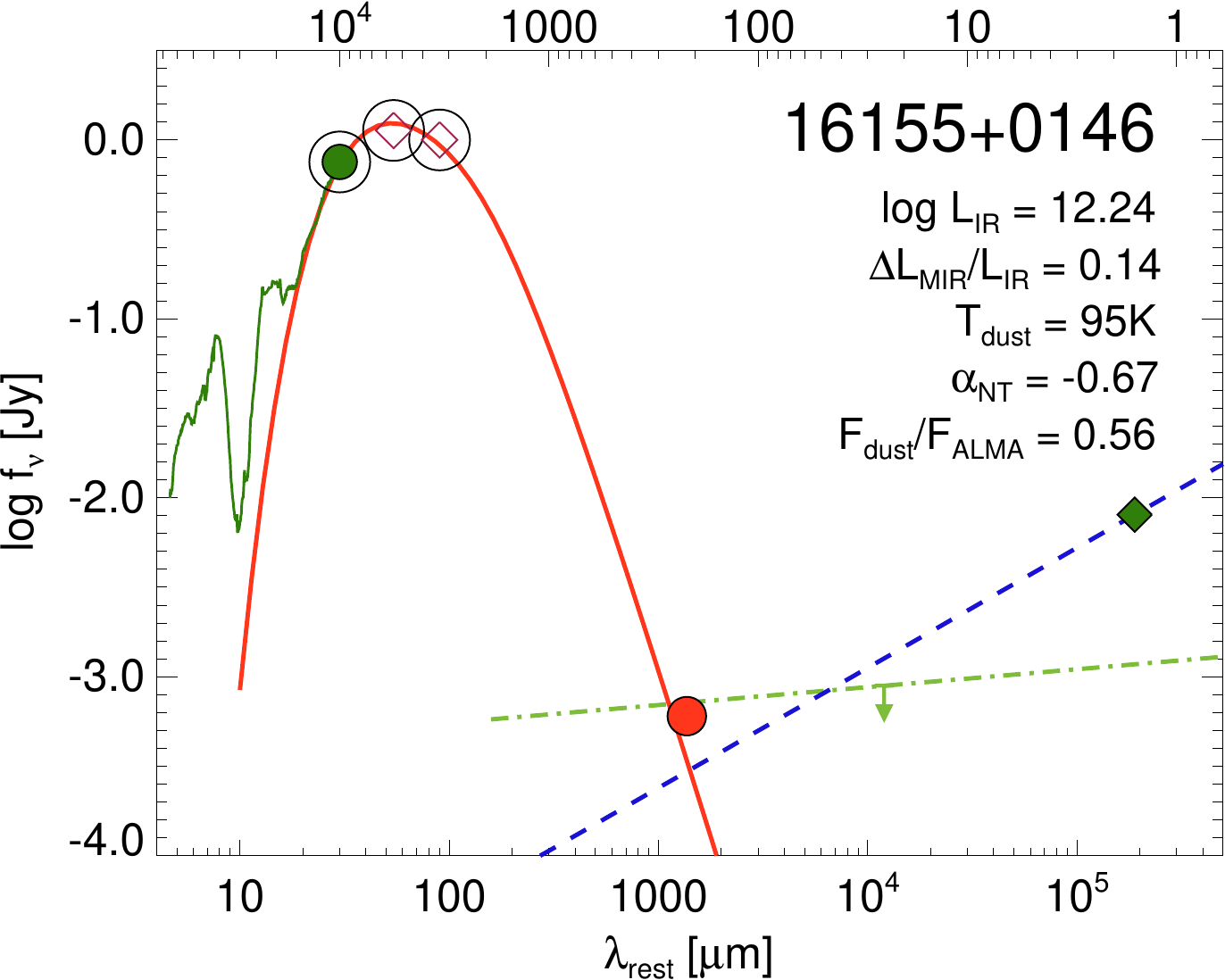}
\includegraphics[width=0.32\textwidth]{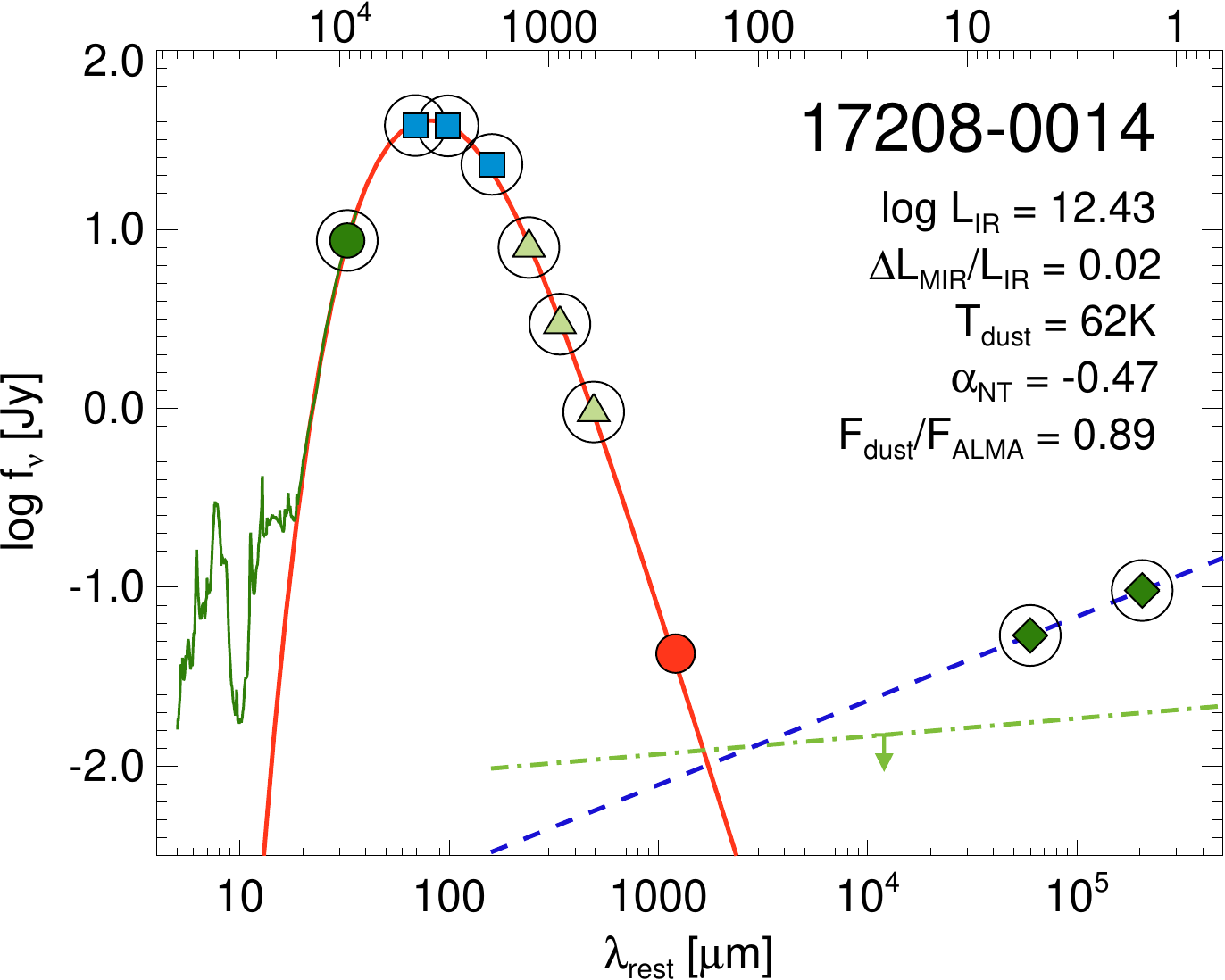}
\includegraphics[width=0.32\textwidth]{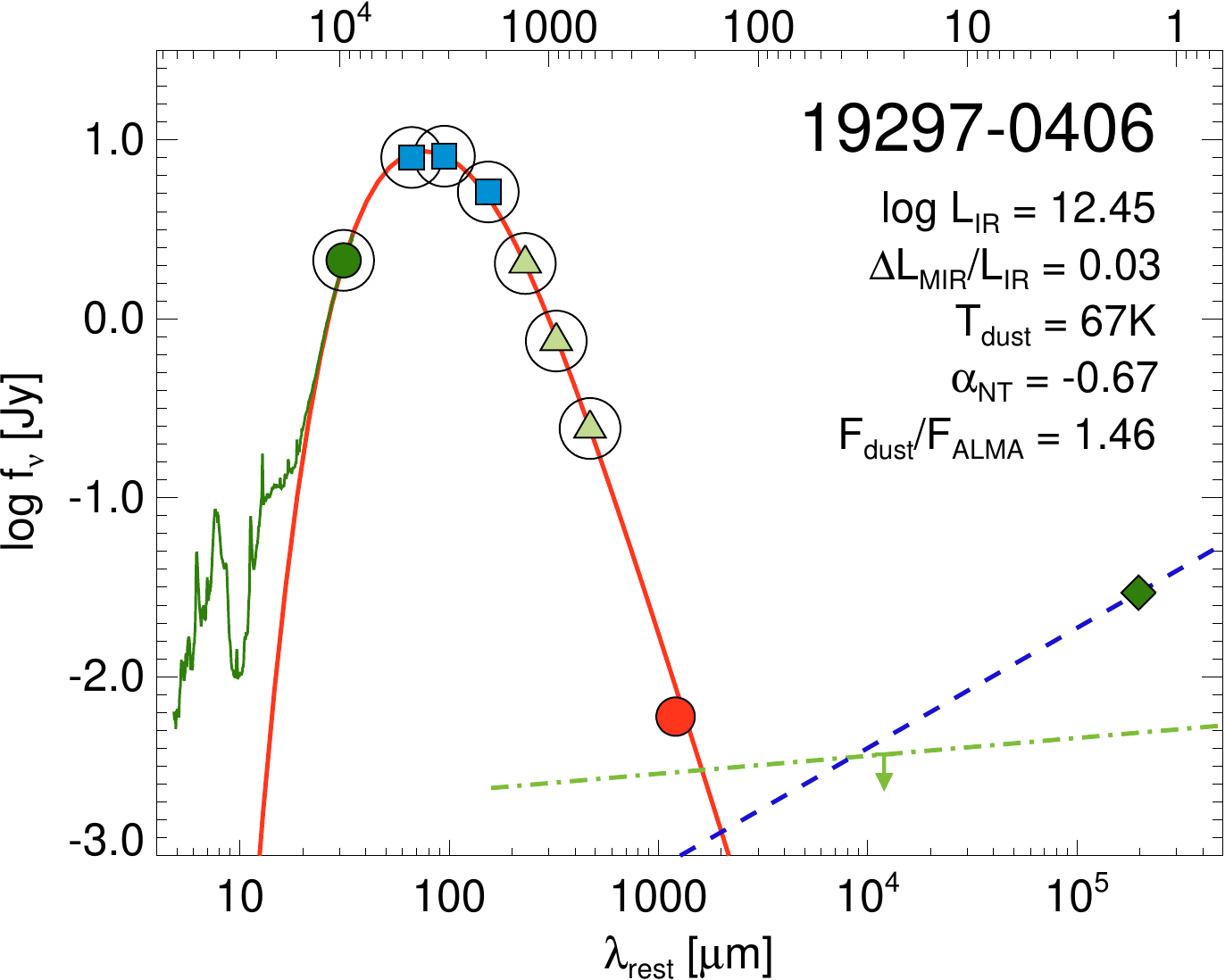}
\includegraphics[width=0.32\textwidth]{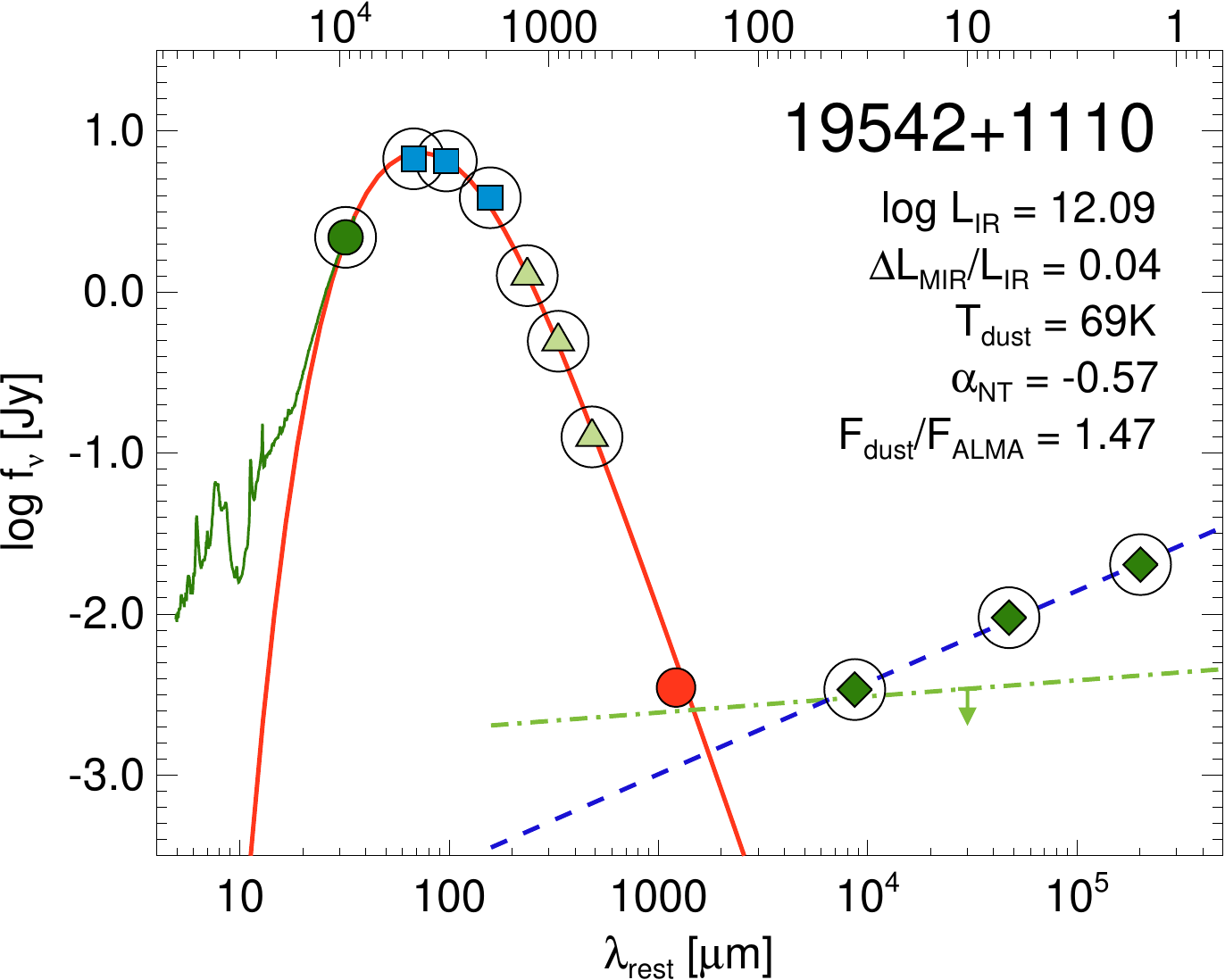}
\includegraphics[width=0.32\textwidth]{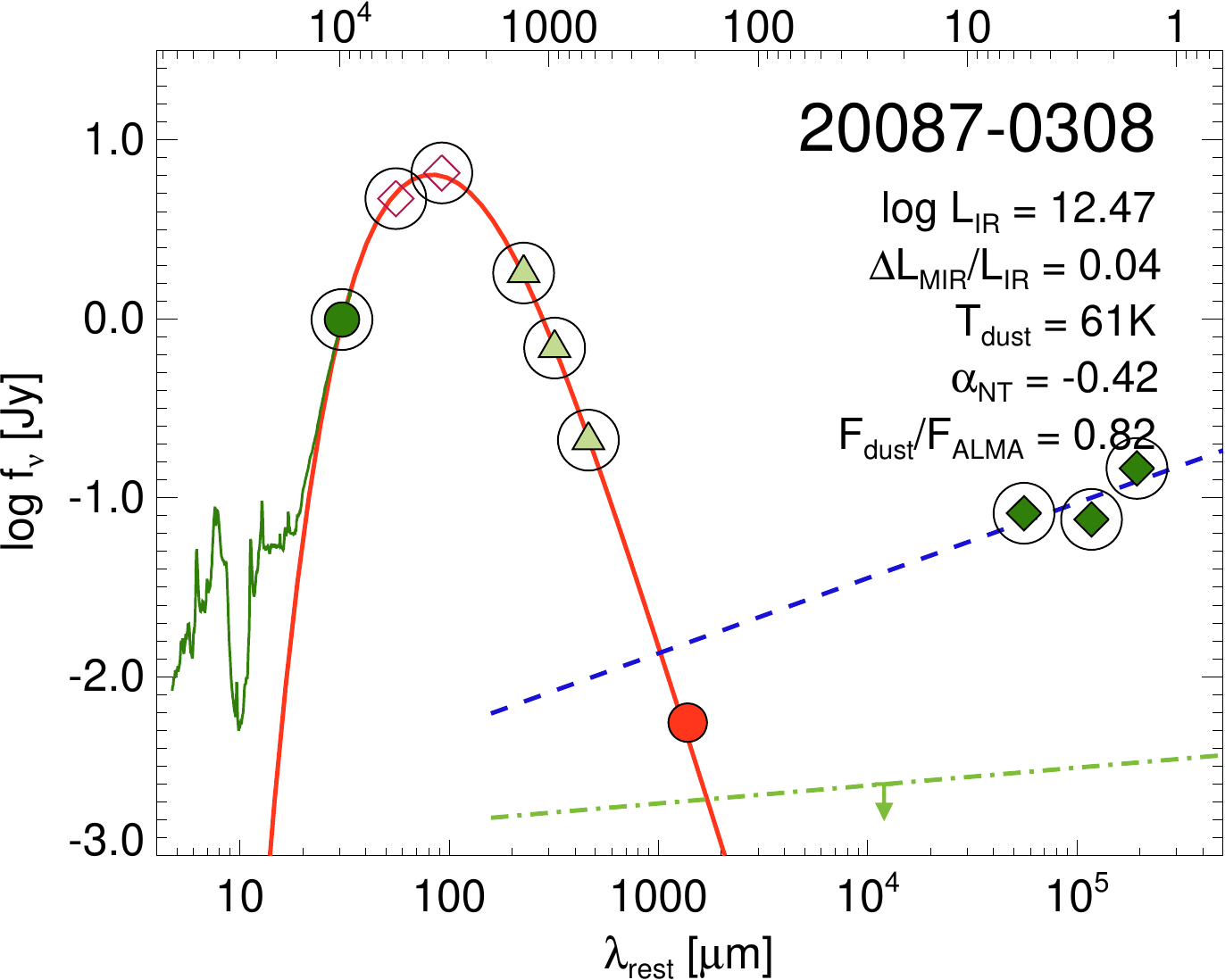}
\includegraphics[width=0.32\textwidth]{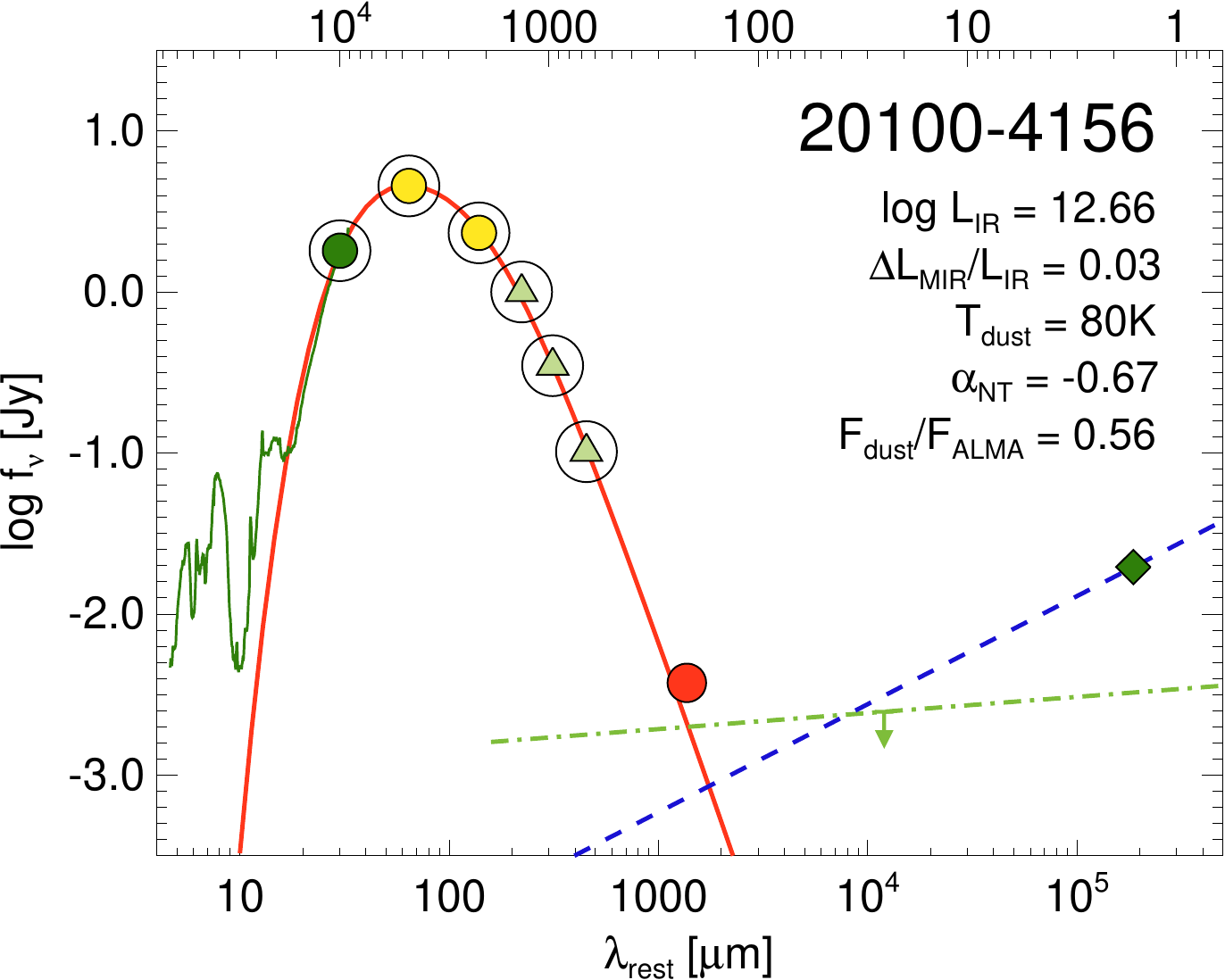}
\includegraphics[width=0.32\textwidth]{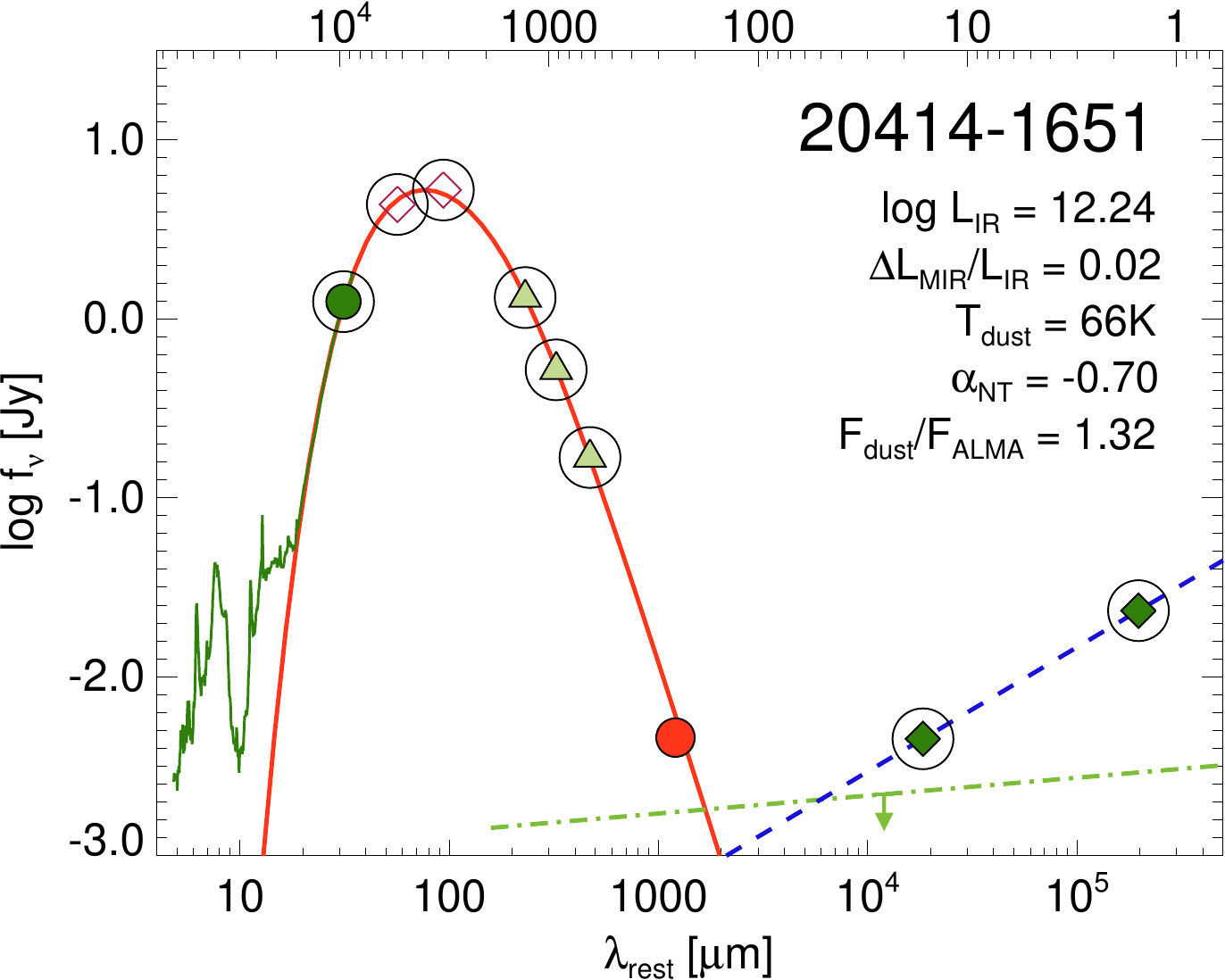}
\includegraphics[width=0.32\textwidth]{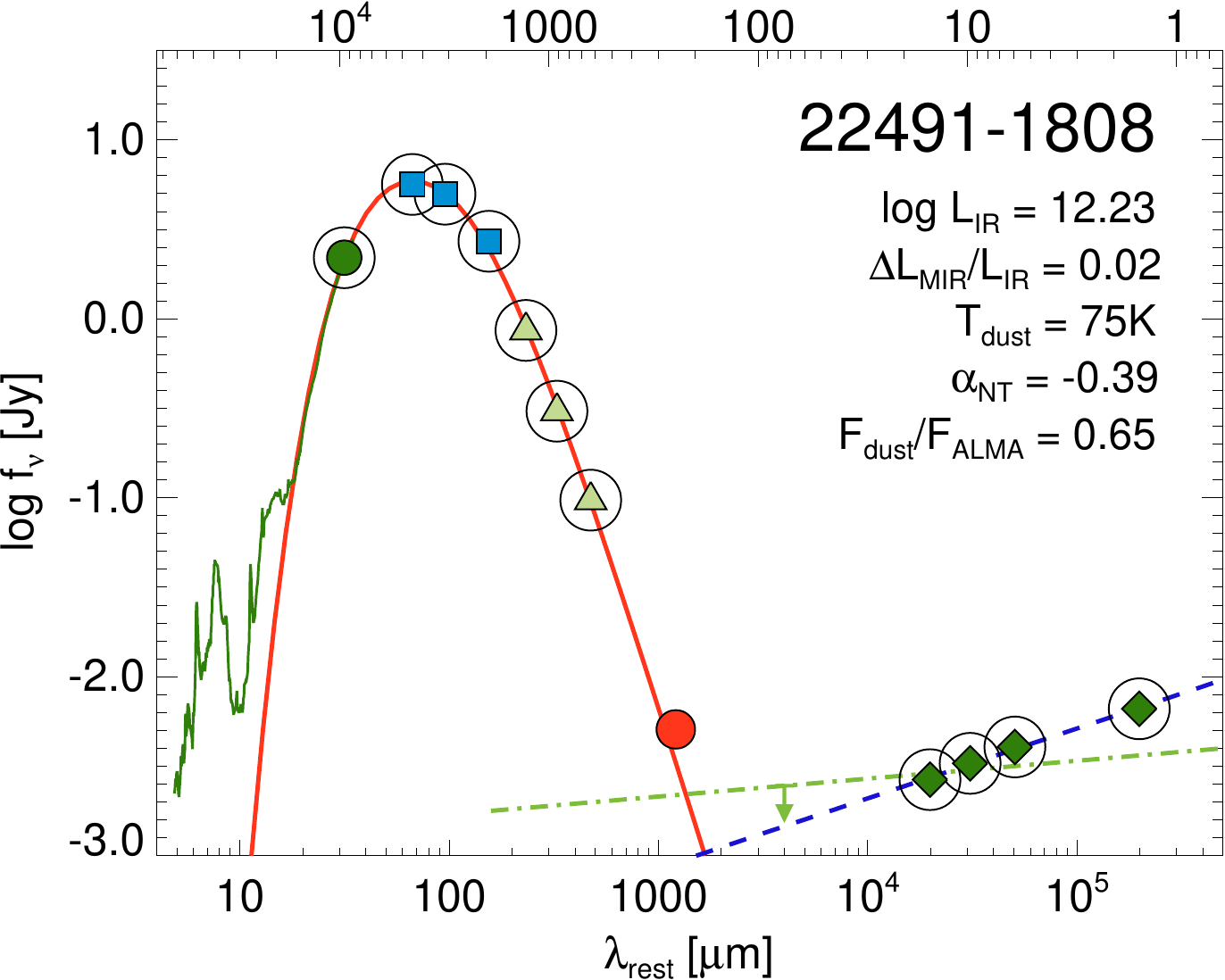}
\caption{(Continued)}
\end{figure}


\begin{thebibliography}{104}
\expandafter\ifx\csname natexlab\endcsname\relax\def\natexlab#1{#1}\fi

\bibitem[{{Abel} {et~al.}(2009){Abel}, {Dudley}, {Fischer}, {Satyapal}, \& {van
  Hoof}}]{Abel2009}
{Abel}, N.~P., {Dudley}, C., {Fischer}, J., {Satyapal}, S., \& {van Hoof},
  P.~A.~M. 2009, \apj, 701, 1147

\bibitem[{{Alonso-Herrero} {et~al.}(2016){Alonso-Herrero}, {Esquej}, {Roche},
  {Ramos Almeida}, {Gonz{\'a}lez-Mart{\'\i}n}, {Packham}, {Levenson}, {Mason},
  {Hern{\'a}n-Caballero}, {Pereira-Santaella}, {Alvarez}, {Aretxaga},
  {L{\'o}pez-Rodr{\'\i}guez}, {Colina}, {D{\'\i}az-Santos}, {Imanishi},
  {Rodr{\'\i}guez Espinosa}, \& {Perlman}}]{AAH2016}
{Alonso-Herrero}, A., {Esquej}, P., {Roche}, P.~F., {et~al.} 2016, \mnras, 455,
  563

\bibitem[{{Alonso-Herrero} {et~al.}(2014){Alonso-Herrero}, {Ramos Almeida},
  {Esquej}, {Roche}, {Hern{\'a}n-Caballero}, {H{\"o}nig},
  {Gonz{\'a}lez-Mart{\'{\i}}n}, {Aretxaga}, {Mason}, {Packham}, {Levenson},
  {Rodr{\'{\i}}guez Espinosa}, {Siebenmorgen}, {Pereira-Santaella},
  {D{\'{\i}}az-Santos}, {Colina}, {Alvarez}, \& {Telesco}}]{AAH2014}
{Alonso-Herrero}, A., {Ramos Almeida}, C., {Esquej}, P., {et~al.} 2014, \mnras,
  443, 2766

\bibitem[{{Andrews} \& {Thompson}(2011)}]{Andrews2011}
{Andrews}, B.~H. \& {Thompson}, T.~A. 2011, \apj, 727, 97

\bibitem[{{Arribas} {et~al.}(2012){Arribas}, {Colina}, {Alonso-Herrero},
  {Rosales-Ortega}, {Monreal-Ibero}, {Garc{\'{\i}}a-Mar{\'{\i}}n},
  {Garc{\'{\i}}a-Burillo}, \& {Rodr{\'{\i}}guez-Zaur{\'{\i}}n}}]{Arribas2012}
{Arribas}, S., {Colina}, L., {Alonso-Herrero}, A., {et~al.} 2012, \aap, 541,
  A20

\bibitem[{{Baan} \& {Kl{\"o}ckner}(2006)}]{Baan2006}
{Baan}, W.~A. \& {Kl{\"o}ckner}, H.-R. 2006, \aap, 449, 559

\bibitem[{{Barcos-Mu{\~n}oz} {et~al.}(2017){Barcos-Mu{\~n}oz}, {Leroy},
  {Evans}, {Condon}, {Privon}, {Thompson}, {Armus}, {D{\'\i}az-Santos},
  {Mazzarella}, {Meier}, {Momjian}, {Murphy}, {Ott}, {Sanders}, {Schinnerer},
  {Stierwalt}, {Surace}, \& {Walter}}]{BarcosMunoz2017}
{Barcos-Mu{\~n}oz}, L., {Leroy}, A.~K., {Evans}, A.~S., {et~al.} 2017, \apj,
  843, 117

\bibitem[{{Barcos-Mu{\~n}oz} {et~al.}(2015){Barcos-Mu{\~n}oz}, {Leroy},
  {Evans}, {Privon}, {Armus}, {Condon}, {Mazzarella}, {Meier}, {Momjian},
  {Murphy}, {Ott}, {Reichardt}, {Sakamoto}, {Sanders}, {Schinnerer},
  {Stierwalt}, {Surace}, {Thompson}, \& {Walter}}]{BarcosMunoz2015}
{Barcos-Mu{\~n}oz}, L., {Leroy}, A.~K., {Evans}, A.~S., {et~al.} 2015, \apj,
  799, 10

\bibitem[{{Barvainis} \& {Antonucci}(1989)}]{Barvainis1989}
{Barvainis}, R. \& {Antonucci}, R. 1989, \apjs, 70, 257

\bibitem[{{Bolatto} {et~al.}(2013){Bolatto}, {Wolfire}, \&
  {Leroy}}]{Bolatto2013}
{Bolatto}, A.~D., {Wolfire}, M., \& {Leroy}, A.~K. 2013, \araa, 51, 207

\bibitem[{{Brown} \& {Wilson}(2019)}]{Brown2019}
{Brown}, T. \& {Wilson}, C.~D. 2019, \apj, 879, 17

\bibitem[{{Casey} {et~al.}(2014){Casey}, {Narayanan}, \& {Cooray}}]{Casey2014}
{Casey}, C.~M., {Narayanan}, D., \& {Cooray}, A. 2014, \physrep, 541, 45

\bibitem[{{Chu} {et~al.}(2017){Chu}, {Sanders}, {Larson}, {Mazzarella},
  {Howell}, {D{\'\i}az-Santos}, {Xu}, {Paladini}, {Schulz}, {Shupe},
  {Appleton}, {Armus}, {Billot}, {Chan}, {Evans}, {Fadda}, {Frayer}, {Haan},
  {Ishida}, {Iwasawa}, {Kim}, {Lord}, {Murphy}, {Petric}, {Privon}, {Surace},
  \& {Treister}}]{Chu2017}
{Chu}, J.~K., {Sanders}, D.~B., {Larson}, K.~L., {et~al.} 2017, \apjs, 229, 25

\bibitem[{{Cicone} {et~al.}(2014){Cicone}, {Maiolino}, {Sturm},
  {Graci{\'a}-Carpio}, {Feruglio}, {Neri}, {Aalto}, {Davies}, {Fiore},
  {Fischer}, {Garc{\'{\i}}a-Burillo}, {Gonz{\'a}lez-Alfonso},
  {Hailey-Dunsheath}, {Piconcelli}, \& {Veilleux}}]{Cicone2014}
{Cicone}, C., {Maiolino}, R., {Sturm}, E., {et~al.} 2014, \aap, 562, A21

\bibitem[{{Clemens} {et~al.}(2008){Clemens}, {Vega}, {Bressan}, {Granato},
  {Silva}, \& {Panuzzo}}]{Clemens2008}
{Clemens}, M.~S., {Vega}, O., {Bressan}, A., {et~al.} 2008, \aap, 477, 95

\bibitem[{{Condon} {et~al.}(1998){Condon}, {Cotton}, {Greisen}, {Yin},
  {Perley}, {Taylor}, \& {Broderick}}]{Condon1998}
{Condon}, J.~J., {Cotton}, W.~D., {Greisen}, E.~W., {et~al.} 1998, \aj, 115,
  1693

\bibitem[{{Condon} {et~al.}(1990){Condon}, {Helou}, {Sanders}, \&
  {Soifer}}]{Condon1990}
{Condon}, J.~J., {Helou}, G., {Sanders}, D.~B., \& {Soifer}, B.~T. 1990, \apjs,
  73, 359

\bibitem[{{Condon} {et~al.}(1996){Condon}, {Helou}, {Sanders}, \&
  {Soifer}}]{Condon1996}
{Condon}, J.~J., {Helou}, G., {Sanders}, D.~B., \& {Soifer}, B.~T. 1996, \apjs,
  103, 81

\bibitem[{{Condon} {et~al.}(1991){Condon}, {Huang}, {Yin}, \&
  {Thuan}}]{Condon1991}
{Condon}, J.~J., {Huang}, Z.~P., {Yin}, Q.~F., \& {Thuan}, T.~X. 1991, \apj,
  378, 65

\bibitem[{{Condon} \& {Ransom}(2016)}]{Condon2016}
{Condon}, J.~J. \& {Ransom}, S.~M. 2016, {Essential Radio Astronomy}

\bibitem[{{D{\'\i}az-Santos} {et~al.}(2010){D{\'\i}az-Santos}, {Charmandaris},
  {Armus}, {Petric}, {Howell}, {Murphy}, {Mazzarella}, {Veilleux}, {Bothun},
  {Inami}, {Appleton}, {Evans}, {Haan}, {Marshall}, {Sanders}, {Stierwalt}, \&
  {Surace}}]{DiazSantos2010}
{D{\'\i}az-Santos}, T., {Charmandaris}, V., {Armus}, L., {et~al.} 2010, \apj,
  723, 993

\bibitem[{{Downes} \& {Eckart}(2007)}]{Downes2007}
{Downes}, D. \& {Eckart}, A. 2007, \aap, 468, L57

\bibitem[{{Duc} {et~al.}(1997){Duc}, {Mirabel}, \& {Maza}}]{Duc1997}
{Duc}, P.~A., {Mirabel}, I.~F., \& {Maza}, J. 1997, \aaps, 124, 533

\bibitem[{{Falstad} {et~al.}(2021){Falstad}, {Aalto}, {K{\"o}nig}, {Onishi},
  {Muller}, {Gorski}, {Sato}, {Stanley}, {Combes}, {Gonz{\'a}lez-Alfonso},
  {Mangum}, {Evans}, {Barcos-Mu{\~n}oz}, {Privon}, {Linden},
  {D{\'\i}az-Santos}, {Mart{\'\i}n}, {Sakamoto}, {Harada}, {Fuller},
  {Gallagher}, {van der Werf}, {Viti}, {Greve}, {Garc{\'\i}a-Burillo},
  {Henkel}, {Imanishi}, {Izumi}, {Nishimura}, {Ricci}, \&
  {M{\"u}hle}}]{Falstad2021}
{Falstad}, N., {Aalto}, S., {K{\"o}nig}, S., {et~al.} 2021, arXiv e-prints,
  arXiv:2102.13563

\bibitem[{{Garcia-Burillo} {et~al.}(2021){Garcia-Burillo}, {Alonso-Herrero},
  {Ramos Almeida}, {Gonzalez-Martin}, {Combes}, {Usero}, {Hoenig}, {Querejeta},
  {Hicks}, {Hunt}, {Rosario}, {Davies}, {Boorman}, {Bunker}, {Burstcher},
  {Colina}, {D{\'\i}az-Santos}, {Gandhi}, {Garcia-Bernete}, {Garcia-Lorenzo},
  {Ichikawa}, {Imanishi}, {Izumi}, {Labiano}, {Levenson}, {Lopez-Rodriguez},
  {Packham}, {Pereira-Santaella}, {Ricci}, {Rigopoulou}, {Rouan}, {Stalevsk},
  {Wada}, \& {Williamson}}]{GarciaBurillo2021}
{Garcia-Burillo}, S., {Alonso-Herrero}, A., {Ramos Almeida}, C., {et~al.} 2021,
  arXiv e-prints, arXiv:2104.10227

\bibitem[{{Genzel} {et~al.}(1998){Genzel}, {Lutz}, {Sturm}, {Egami}, {Kunze},
  {Moorwood}, {Rigopoulou}, {Spoon}, {Sternberg}, {Tacconi-Garman}, {Tacconi},
  \& {Thatte}}]{Genzel1998}
{Genzel}, R., {Lutz}, D., {Sturm}, E., {et~al.} 1998, \apj, 498, 579

\bibitem[{{G{\'o}mez-Guijarro} {et~al.}(2018){G{\'o}mez-Guijarro}, {Toft},
  {Karim}, {Magnelli}, {Magdis}, {Jim{\'e}nez-Andrade}, {Capak}, {Fraternali},
  {Fujimoto}, {Riechers}, {Schinnerer}, {Smol{\v{c}}i{\'c}}, {Aravena},
  {Bertoldi}, {Cortzen}, {Hasinger}, {Hu}, {Jones}, {Koekemoer}, {Lee},
  {McCracken}, {Micha{\l}owski}, {Navarrete}, {Povi{\'c}}, {Puglisi},
  {Romano-D{\'\i}az}, {Sheth}, {Silverman}, {Staguhn}, {Steinhardt},
  {Stockmann}, {Tanaka}, {Valentino}, {van Kampen}, \&
  {Zirm}}]{GomezGuijarro2018}
{G{\'o}mez-Guijarro}, C., {Toft}, S., {Karim}, A., {et~al.} 2018, \apj, 856,
  121

\bibitem[{{Gonz{\'a}lez-Alfonso} {et~al.}(2017){Gonz{\'a}lez-Alfonso},
  {Fischer}, {Spoon}, {Stewart}, {Ashby}, {Veilleux}, {Smith}, {Sturm},
  {Farrah}, {Falstad}, {Mel{\'e}ndez}, {Graci{\'a}-Carpio}, {Janssen}, \&
  {Lebouteiller}}]{GonzalezAlfonso2017}
{Gonz{\'a}lez-Alfonso}, E., {Fischer}, J., {Spoon}, H.~W.~W., {et~al.} 2017,
  \apj, 836, 11

\bibitem[{{Gonz{\'a}lez-Alfonso} {et~al.}(2015){Gonz{\'a}lez-Alfonso},
  {Fischer}, {Sturm}, {Graci{\'a}-Carpio}, {Veilleux}, {Mel{\'e}ndez}, {Lutz},
  {Poglitsch}, {Aalto}, {Falstad}, {Spoon}, {Farrah}, {Blasco}, {Henkel},
  {Contursi}, {Verma}, {Spaans}, {Smith}, {Ashby}, {Hailey-Dunsheath},
  {Garc{\'\i}a-Burillo}, {Mart{\'\i}n-Pintado}, {van der Werf}, {Meijerink}, \&
  {Genzel}}]{GonzalezAlfonso2015}
{Gonz{\'a}lez-Alfonso}, E., {Fischer}, J., {Sturm}, E., {et~al.} 2015, \apj,
  800, 69

\bibitem[{{Gonz{\'a}lez-Alfonso} {et~al.}(2021){Gonz{\'a}lez-Alfonso},
  {Pereira-Santaella}, {Fischer}, {Garc{\'\i}a-Burillo}, {Yang},
  {Alonso-Herrero}, {Colina}, {Ashby}, {Smith}, {Rico-Villas},
  {Mart{\'\i}n-Pintado}, {Cazzoli}, \& {Stewart}}]{GonzalezAlfonso2021}
{Gonz{\'a}lez-Alfonso}, E., {Pereira-Santaella}, M., {Fischer}, J., {et~al.}
  2021, \aap, 645, A49

\bibitem[{{Gonz{\'a}lez-Alfonso} \& {Sakamoto}(2019)}]{GonzalezAlfonso2019}
{Gonz{\'a}lez-Alfonso}, E. \& {Sakamoto}, K. 2019, \apj, 882, 153

\bibitem[{{Gullberg} {et~al.}(2018){Gullberg}, {Swinbank}, {Smail}, {Biggs},
  {Bertoldi}, {De Breuck}, {Chapman}, {Chen}, {Cooke}, {Coppin}, {Cox},
  {Dannerbauer}, {Dunlop}, {Edge}, {Farrah}, {Geach}, {Greve}, {Hodge}, {Ibar},
  {Ivison}, {Karim}, {Schinnerer}, {Scott}, {Simpson}, {Stach}, {Thomson}, {van
  der Werf}, {Walter}, {Wardlow}, \& {Weiss}}]{Gullberg2018}
{Gullberg}, B., {Swinbank}, A.~M., {Smail}, I., {et~al.} 2018, \apj, 859, 12

\bibitem[{{Hardcastle} \& {Croston}(2020)}]{Hardcastle2020}
{Hardcastle}, M.~J. \& {Croston}, J.~H. 2020, \nar, 88, 101539

\bibitem[{{Hayashi} {et~al.}(2021){Hayashi}, {Hagiwara}, \&
  {Imanishi}}]{Hayashi2021}
{Hayashi}, T.~J., {Hagiwara}, Y., \& {Imanishi}, M. 2021, arXiv e-prints,
  arXiv:2101.12058

\bibitem[{{Helfand} {et~al.}(2015){Helfand}, {White}, \&
  {Becker}}]{Helfand2015}
{Helfand}, D.~J., {White}, R.~L., \& {Becker}, R.~H. 2015, \apj, 801, 26

\bibitem[{{Hern{\'a}n-Caballero} {et~al.}(2020){Hern{\'a}n-Caballero}, {Spoon},
  {Alonso-Herrero}, {Hatziminaoglou}, {Magdis}, {P{\'e}rez-Gonz{\'a}lez},
  {Pereira-Santaella}, {Arribas}, {Cortzen}, {Labiano}, {Piqueras}, \&
  {Rigopoulou}}]{HernanCaballero2020}
{Hern{\'a}n-Caballero}, A., {Spoon}, H. W.~W., {Alonso-Herrero}, A., {et~al.}
  2020, \mnras, 497, 4614

\bibitem[{{Hopkins} {et~al.}(2013){Hopkins}, {Cox}, {Hernquist}, {Narayanan},
  {Hayward}, \& {Murray}}]{Hopkins2013}
{Hopkins}, P.~F., {Cox}, T.~J., {Hernquist}, L., {et~al.} 2013, \mnras, 430,
  1901

\bibitem[{{Houck} {et~al.}(2004){Houck}, {Roellig}, {van Cleve}, {Forrest},
  {Herter}, {Lawrence}, {Matthews}, {Reitsema}, {Soifer}, {Watson}, {Weedman},
  {Huisjen}, {Troeltzsch}, {Barry}, {Bernard-Salas}, {Blacken}, {Brandl},
  {Charmandaris}, {Devost}, {Gull}, {Hall}, {Henderson}, {Higdon}, {Pirger},
  {Schoenwald}, {Sloan}, {Uchida}, {Appleton}, {Armus}, {Burgdorf},
  {Fajardo-Acosta}, {Grillmair}, {Ingalls}, {Morris}, \& {Teplitz}}]{HouckIRS}
{Houck}, J.~R., {Roellig}, T.~L., {van Cleve}, J., {et~al.} 2004, \apjs, 154,
  18

\bibitem[{{Imanishi} {et~al.}(2011){Imanishi}, {Imase}, {Oi}, \&
  {Ichikawa}}]{Imanishi2011}
{Imanishi}, M., {Imase}, K., {Oi}, N., \& {Ichikawa}, K. 2011, \aj, 141, 156

\bibitem[{{Imanishi} {et~al.}(2020){Imanishi}, {Kawamuro}, {Kikuta}, {Nakano},
  \& {Saito}}]{Imanishi2020}
{Imanishi}, M., {Kawamuro}, T., {Kikuta}, S., {Nakano}, S., \& {Saito}, Y.
  2020, \apj, 891, 140

\bibitem[{{Imanishi} {et~al.}(2016){Imanishi}, {Nakanishi}, \&
  {Izumi}}]{Imanishi2016}
{Imanishi}, M., {Nakanishi}, K., \& {Izumi}, T. 2016, \aj, 152, 218

\bibitem[{{Imanishi} {et~al.}(2019){Imanishi}, {Nakanishi}, \&
  {Izumi}}]{Imanishi2019}
{Imanishi}, M., {Nakanishi}, K., \& {Izumi}, T. 2019, \apjs, 241, 19

\bibitem[{{Imanishi} \& {Terashima}(2004)}]{Imanishi2004c}
{Imanishi}, M. \& {Terashima}, Y. 2004, \aj, 127, 758

\bibitem[{{Iwasawa} {et~al.}(2011){Iwasawa}, {Sanders}, {Teng}, {U}, {Armus},
  {Evans}, {Howell}, {Komossa}, {Mazzarella}, {Petric}, {Surace}, {Vavilkin},
  {Veilleux}, \& {Trentham}}]{Iwasawa2011}
{Iwasawa}, K., {Sanders}, D.~B., {Teng}, S.~H., {et~al.} 2011, \aap, 529, A106+

\bibitem[{{Jiang} {et~al.}(2006){Jiang}, {Gao}, {Omont}, {Schuller}, \&
  {Simon}}]{Jiang2006}
{Jiang}, B.~W., {Gao}, J., {Omont}, A., {Schuller}, F., \& {Simon}, G. 2006,
  \aap, 446, 551

\bibitem[{{Kennicutt} \& {Evans}(2012)}]{Kennicutt2012}
{Kennicutt}, R.~C. \& {Evans}, N.~J. 2012, \araa, 50, 531

\bibitem[{{Kim} {et~al.}(1998){Kim}, {Veilleux}, \& {Sanders}}]{Kim1998}
{Kim}, D., {Veilleux}, S., \& {Sanders}, D.~B. 1998, \apj, 508, 627

\bibitem[{{Klaas} {et~al.}(2001){Klaas}, {Haas}, {M{\"u}ller}, {Chini},
  {Schulz}, {Coulson}, {Hippelein}, {Wilke}, {Albrecht}, \&
  {Lemke}}]{Klaas2001}
{Klaas}, U., {Haas}, M., {M{\"u}ller}, S.~A.~H., {et~al.} 2001, \aap, 379, 823

\bibitem[{{Kov{\'a}cs} {et~al.}(2010){Kov{\'a}cs}, {Omont}, {Beelen},
  {Lonsdale}, {Polletta}, {Fiolet}, {Greve}, {Borys}, {Cox}, {De Breuck},
  {Dole}, {Dowell}, {Farrah}, {Lagache}, {Menten}, {Bell}, \&
  {Owen}}]{Kovacs2010}
{Kov{\'a}cs}, A., {Omont}, A., {Beelen}, A., {et~al.} 2010, \apj, 717, 29

\bibitem[{{Kroupa}(2001)}]{Kroupa2001}
{Kroupa}, P. 2001, \mnras, 322, 231

\bibitem[{{Lahuis} {et~al.}(2007){Lahuis}, {Spoon}, {Tielens}, {Doty}, {Armus},
  {Charmandaris}, {Houck}, {St{\"a}uber}, \& {van Dishoeck}}]{Lahuis2007}
{Lahuis}, F., {Spoon}, H.~W.~W., {Tielens}, A.~G.~G.~M., {et~al.} 2007, \apj,
  659, 296

\bibitem[{{Lebouteiller} {et~al.}(2011){Lebouteiller}, {Barry}, {Spoon},
  {Bernard-Salas}, {Sloan}, {Houck}, \& {Weedman}}]{Lebouteiller2011}
{Lebouteiller}, V., {Barry}, D.~J., {Spoon}, H.~W.~W., {et~al.} 2011, \apjs,
  196, 8

\bibitem[{{Leipski} {et~al.}(2006){Leipski}, {Falcke}, {Bennert}, \&
  {H{\"u}ttemeister}}]{Leipski2006}
{Leipski}, C., {Falcke}, H., {Bennert}, N., \& {H{\"u}ttemeister}, S. 2006,
  \aap, 455, 161

\bibitem[{{Leitherer} {et~al.}(1999){Leitherer}, {Schaerer}, {Goldader},
  {Delgado}, {Robert}, {Kune}, {de Mello}, {Devost}, \&
  {Heckman}}]{Leitherer1999}
{Leitherer}, C., {Schaerer}, D., {Goldader}, J.~D., {et~al.} 1999, \apjs, 123,
  3

\bibitem[{{Leroy} {et~al.}(2011){Leroy}, {Evans}, {Momjian}, {Murphy}, {Ott},
  {Armus}, {Condon}, {Haan}, {Mazzarella}, {Meier}, {Privon}, {Schinnerer},
  {Surace}, \& {Walter}}]{Leroy2011radio}
{Leroy}, A.~K., {Evans}, A.~S., {Momjian}, E., {et~al.} 2011, \apjl, 739, L25

\bibitem[{{Lonsdale} {et~al.}(2006){Lonsdale}, {Farrah}, \&
  {Smith}}]{Lonsdale2006}
{Lonsdale}, C.~J., {Farrah}, D., \& {Smith}, H.~E. 2006, {Ultraluminous
  Infrared Galaxies}, ed. {Mason, J.~W.} (Springer Verlag), 285--+

\bibitem[{{Lutz} {et~al.}(2020){Lutz}, {Sturm}, {Janssen}, {Veilleux}, {Aalto},
  {Cicone}, {Contursi}, {Davies}, {Feruglio}, {Fischer}, {Fluetsch},
  {Garcia-Burillo}, {Genzel}, {Gonz{\'a}lez-Alfonso}, {Graci{\'a}-Carpio},
  {Herrera-Camus}, {Maiolino}, {Schruba}, {Shimizu}, {Sternberg}, {Tacconi}, \&
  {Wei{\ss}}}]{Lutz2020}
{Lutz}, D., {Sturm}, E., {Janssen}, A., {et~al.} 2020, \aap, 633, A134

\bibitem[{{Marconi} {et~al.}(2004){Marconi}, {Risaliti}, {Gilli}, {Hunt},
  {Maiolino}, \& {Salvati}}]{Marconi2004}
{Marconi}, A., {Risaliti}, G., {Gilli}, R., {et~al.} 2004, \mnras, 351, 169

\bibitem[{{Mauch} {et~al.}(2003){Mauch}, {Murphy}, {Buttery}, {Curran},
  {Hunstead}, {Piestrzynski}, {Robertson}, \& {Sadler}}]{Mauch2003}
{Mauch}, T., {Murphy}, T., {Buttery}, H.~J., {et~al.} 2003, \mnras, 342, 1117

\bibitem[{{McMullin} {et~al.}(2007){McMullin}, {Waters}, {Schiebel}, {Young},
  \& {Golap}}]{McMullin2007}
{McMullin}, J.~P., {Waters}, B., {Schiebel}, D., {Young}, W., \& {Golap}, K.
  2007, in Astronomical Society of the Pacific Conference Series, Vol. 376,
  Astronomical Data Analysis Software and Systems XVI, ed. R.~A. {Shaw},
  F.~{Hill}, \& D.~J. {Bell}, 127

\bibitem[{{Meyers} {et~al.}(2017){Meyers}, {Hurley-Walker}, {Hancock},
  {Franzen}, {Carretti}, {Staveley-Smith}, {Gaensler}, {Haverkorn}, \&
  {Poppi}}]{Meyers2017}
{Meyers}, B.~W., {Hurley-Walker}, N., {Hancock}, P.~J., {et~al.} 2017, \pasa,
  34, e013

\bibitem[{{Michiyama} {et~al.}(2020){Michiyama}, {Iono}, {Nakanishi}, {Ueda},
  {Saito}, {Yamashita}, {Bolatto}, \& {Yun}}]{Michiyama2020}
{Michiyama}, T., {Iono}, D., {Nakanishi}, K., {et~al.} 2020, \apj, 895, 85

\bibitem[{{Moshir} \& {et al.}(1990)}]{Moshir1990}
{Moshir}, M. \& {et al.} 1990, IRAS Faint Source Catalogue, 0

\bibitem[{{Murphy} {et~al.}(2011){Murphy}, {Condon}, {Schinnerer}, {Kennicutt},
  {Calzetti}, {Armus}, {Helou}, {Turner}, {Aniano}, {Beir{\~a}o}, {Bolatto},
  {Brandl}, {Croxall}, {Dale}, {Donovan Meyer}, {Draine}, {Engelbracht},
  {Hunt}, {Hao}, {Koda}, {Roussel}, {Skibba}, \& {Smith}}]{Murphy2011}
{Murphy}, E.~J., {Condon}, J.~J., {Schinnerer}, E., {et~al.} 2011, \apj, 737,
  67

\bibitem[{{Murphy} {et~al.}(2007){Murphy}, {Mauch}, {Green}, {Hunstead},
  {Piestrzynska}, {Kels}, \& {Sztajer}}]{Murphy2007}
{Murphy}, T., {Mauch}, T., {Green}, A., {et~al.} 2007, \mnras, 382, 382

\bibitem[{{Nagar} {et~al.}(2003){Nagar}, {Wilson}, {Falcke}, {Veilleux}, \&
  {Maiolino}}]{Nagar2003}
{Nagar}, N.~M., {Wilson}, A.~S., {Falcke}, H., {Veilleux}, S., \& {Maiolino},
  R. 2003, \aap, 409, 115

\bibitem[{{Nardini} {et~al.}(2008){Nardini}, {Risaliti}, {Salvati}, {Sani},
  {Imanishi}, {Marconi}, \& {Maiolino}}]{Nardini2008}
{Nardini}, E., {Risaliti}, G., {Salvati}, M., {et~al.} 2008, \mnras, 385, L130

\bibitem[{{Nardini} {et~al.}(2009){Nardini}, {Risaliti}, {Salvati}, {Sani},
  {Watabe}, {Marconi}, \& {Maiolino}}]{Nardini2009}
{Nardini}, E., {Risaliti}, G., {Salvati}, M., {et~al.} 2009, \mnras, 399, 1373

\bibitem[{{Nardini} {et~al.}(2010){Nardini}, {Risaliti}, {Watabe}, {Salvati},
  \& {Sani}}]{Nardini2010}
{Nardini}, E., {Risaliti}, G., {Watabe}, Y., {Salvati}, M., \& {Sani}, E. 2010,
  \mnras, 405, 2505

\bibitem[{{Oh} {et~al.}(2018){Oh}, {Koss}, {Markwardt}, {Schawinski},
  {Baumgartner}, {Barthelmy}, {Cenko}, {Gehrels}, {Mushotzky}, {Petulante},
  {Ricci}, {Lien}, \& {Trakhtenbrot}}]{Oh2018}
{Oh}, K., {Koss}, M., {Markwardt}, C.~B., {et~al.} 2018, \apjs, 235, 4

\bibitem[{{Oteo} {et~al.}(2017){Oteo}, {Zwaan}, {Ivison}, {Smail}, \&
  {Biggs}}]{Oteo2017}
{Oteo}, I., {Zwaan}, M.~A., {Ivison}, R.~J., {Smail}, I., \& {Biggs}, A.~D.
  2017, \apj, 837, 182

\bibitem[{{Padoan} {et~al.}(2012){Padoan}, {Haugb{\o}lle}, \&
  {Nordlund}}]{Padoan2012}
{Padoan}, P., {Haugb{\o}lle}, T., \& {Nordlund}, {\r{A}}. 2012, \apjl, 759, L27

\bibitem[{{Pearson} {et~al.}(2016){Pearson}, {Rigopoulou}, {Hurley}, {Farrah},
  {Afonso}, {Bernard-Salas}, {Borys}, {Clements}, {Cormier}, {Efstathiou},
  {Gonzalez-Alfonso}, {Lebouteiller}, \& {Spoon}}]{Pearson2016}
{Pearson}, C., {Rigopoulou}, D., {Hurley}, P., {et~al.} 2016, \apjs, 227, 9

\bibitem[{{Pereira-Santaella} {et~al.}(2018){Pereira-Santaella}, {Colina},
  {Garc{\'{\i}}a-Burillo}, {Combes}, {Emonts}, {Aalto}, {Alonso-Herrero},
  {Arribas}, {Henkel}, {Labiano}, {Muller}, {Piqueras L{\'o}pez}, {Rigopoulou},
  \& {van der Werf}}]{Pereira2018}
{Pereira-Santaella}, M., {Colina}, L., {Garc{\'{\i}}a-Burillo}, S., {et~al.}
  2018, \aap, 616, A171

\bibitem[{{Pereira-Santaella} {et~al.}(2016){Pereira-Santaella}, {Colina},
  {Garc{\'{\i}}a-Burillo}, {Planesas}, {Usero}, {Alonso-Herrero}, {Arribas},
  {Cazzoli}, {Emonts}, {Piqueras L{\'o}pez}, \&
  {Villar-Mart{\'{\i}}n}}]{Pereira2016}
{Pereira-Santaella}, M., {Colina}, L., {Garc{\'{\i}}a-Burillo}, S., {et~al.}
  2016, \aap, 587, A44

\bibitem[{{Pereira-Santaella} {et~al.}(2017){Pereira-Santaella},
  {Gonz{\'a}lez-Alfonso}, {Usero}, {Garc{\'{\i}}a-Burillo},
  {Mart{\'{\i}}n-Pintado}, {Colina}, {Alonso-Herrero}, {Arribas}, {Cazzoli},
  {Rico}, {Rigopoulou}, \& {Storchi Bergmann}}]{Pereira2017Water}
{Pereira-Santaella}, M., {Gonz{\'a}lez-Alfonso}, E., {Usero}, A., {et~al.}
  2017, \aap, 601, L3

\bibitem[{{Perna} {et~al.}(2020){Perna}, {Arribas}, {Catal{\'a}n-Torrecilla},
  {Colina}, {Bellocchi}, {Fluetsch}, {Maiolino}, {Cazzoli}, {Hern{\'a}n
  Caballero}, {Pereira Santaella}, {Piqueras L{\'o}pez}, \& {Rodr{\'\i}guez del
  Pino}}]{Perna2020}
{Perna}, M., {Arribas}, S., {Catal{\'a}n-Torrecilla}, C., {et~al.} 2020, \aap,
  643, A139

\bibitem[{{Perna} {et~al.}(2021){Perna}, {Arribas}, {Pereira Santaella},
  {Colina}, {Bellocchi}, {Catal{\'a}n-Torrecilla}, {Cazzoli}, {Crespo
  G{\'o}mez}, {Maiolino}, {Piqueras L{\'o}pez}, \& {Rodr{\'\i}guez del
  Pino}}]{Perna2021}
{Perna}, M., {Arribas}, S., {Pereira Santaella}, M., {et~al.} 2021, \aap, 646,
  A101

\bibitem[{{Petric} {et~al.}(2015){Petric}, {Ho}, {Flagey}, \&
  {Scoville}}]{Petric2015}
{Petric}, A.~O., {Ho}, L.~C., {Flagey}, N. J.~M., \& {Scoville}, N.~Z. 2015,
  \apjs, 219, 22

\bibitem[{{Planck Collaboration} {et~al.}(2011){Planck Collaboration},
  {Abergel}, {Ade}, {Aghanim}, {Arnaud}, {Ashdown}, {Aumont}, {Baccigalupi},
  {Balbi}, {Banday}, {Barreiro}, {Bartlett}, {Battaner}, {Benabed},
  {Beno{\^i}t}, {Bernard}, {Bersanelli}, {Bhatia}, {Bock}, {Bonaldi}, {Bond},
  {Borrill}, {Bouchet}, {Boulanger}, {Bucher}, {Burigana}, {Cabella},
  {Cardoso}, {Catalano}, {Cay{\'o}n}, {Challinor}, {Chamballu}, {Chiang},
  {Chiang}, {Christensen}, {Clements}, {Colombi}, {Couchot}, {Coulais},
  {Crill}, {Cuttaia}, {Danese}, {Davies}, {Davis}, {de Bernardis}, {de
  Gasperis}, {de Rosa}, {de Zotti}, {Delabrouille}, {Delouis}, {D{\'e}sert},
  {Dickinson}, {Dobashi}, {Donzelli}, {Dor{\'e}}, {D{\"o}rl}, {Douspis},
  {Dupac}, {Efstathiou}, {En{\ss}lin}, {Eriksen}, {Finelli}, {Forni},
  {Frailis}, {Franceschi}, {Galeotta}, {Ganga}, {Giard}, {Giardino},
  {Giraud-H{\'e}raud}, {Gonz{\'a}lez-Nuevo}, {G{\'o}rski}, {Gratton},
  {Gregorio}, {Gruppuso}, {Guillet}, {Hansen}, {Harrison},
  {Henrot-Versill{\'e}}, {Herranz}, {Hildebrandt}, {Hivon}, {Hobson}, {Holmes},
  {Hovest}, {Hoyland}, {Huffenberger}, {Jaffe}, {Jones}, {Jones}, {Juvela},
  {Keih{\"a}nen}, {Keskitalo}, {Kisner}, {Kneissl}, {Knox}, {Kurki-Suonio},
  {Lagache}, {Lamarre}, {Lasenby}, {Laureijs}, {Lawrence}, {Leach}, {Leonardi},
  {Leroy}, {Linden-V{\o}rnle}, {L{\'o}pez-Caniego}, {Lubin},
  {Mac{\'{\i}}as-P{\'e}rez}, {MacTavish}, {Maffei}, {Mandolesi}, {Mann},
  {Maris}, {Marshall}, {Martin}, {Mart{\'{\i}}nez-Gonz{\'a}lez}, {Masi},
  {Matarrese}, {Matthai}, {Mazzotta}, {McGehee}, {Meinhold}, {Melchiorri},
  {Mendes}, {Mennella}, {Mitra}, {Miville-Desch{\^e}nes}, {Moneti}, {Montier},
  {Morgante}, {Mortlock}, {Munshi}, {Murphy}, {Naselsky}, {Natoli},
  {Netterfield}, {N{\o}rgaard-Nielsen}, {Noviello}, {Novikov}, {Novikov},
  {Osborne}, {Pajot}, {Paladini}, {Pasian}, {Patanchon}, {Perdereau},
  {Perotto}, {Perrotta}, {Piacentini}, {Piat}, {Plaszczynski}, {Pointecouteau},
  {Polenta}, {Ponthieu}, {Poutanen}, {Pr{\'e}zeau}, {Prunet}, {Puget}, {Reach},
  {Rebolo}, {Reinecke}, {Renault}, {Ricciardi}, {Riller}, {Ristorcelli},
  {Rocha}, {Rosset}, {Rubi{\~n}o-Mart{\'{\i}}n}, {Rusholme}, {Sandri},
  {Santos}, {Savini}, {Scott}, {Seiffert}, {Shellard}, {Smoot}, {Starck},
  {Stivoli}, {Stolyarov}, {Sudiwala}, {Sygnet}, {Tauber}, {Terenzi},
  {Toffolatti}, {Tomasi}, {Torre}, {Tristram}, {Tuovinen}, {Umana},
  {Valenziano}, {Verstraete}, {Vielva}, {Villa}, {Vittorio}, {Wade}, {Wandelt},
  {Yvon}, {Zacchei}, \& {Zonca}}]{Planck2011}
{Planck Collaboration}, {Abergel}, A., {Ade}, P.~A.~R., {et~al.} 2011, \aap,
  536, A25

\bibitem[{{Ricci} {et~al.}(2017){Ricci}, {Bauer}, {Treister}, {Schawinski},
  {Privon}, {Blecha}, {Arevalo}, {Armus}, {Harrison}, {Ho}, {Iwasawa},
  {Sanders}, \& {Stern}}]{Ricci2017}
{Ricci}, C., {Bauer}, F.~E., {Treister}, E., {et~al.} 2017, \mnras, 468, 1273

\bibitem[{{Riechers} {et~al.}(2017){Riechers}, {Leung}, {Ivison},
  {P{\'e}rez-Fournon}, {Lewis}, {Marques-Chaves}, {Oteo}, {Clements}, {Cooray},
  {Greenslade}, {Mart{\'\i}nez-Navajas}, {Oliver}, {Rigopoulou}, {Scott}, \&
  {Weiss}}]{Riechers2017}
{Riechers}, D.~A., {Leung}, T.~K.~D., {Ivison}, R.~J., {et~al.} 2017, \apj,
  850, 1

\bibitem[{{Rieke} {et~al.}(2004){Rieke}, {Young}, {Engelbracht}, {Kelly},
  {Low}, {Haller}, {Beeman}, {Gordon}, {Stansberry}, {Misselt}, {Cadien},
  {Morrison}, {Rivlis}, {Latter}, {Noriega-Crespo}, {Padgett}, {Stapelfeldt},
  {Hines}, {Egami}, {Muzerolle}, {Alonso-Herrero}, {Blaylock}, {Dole}, {Hinz},
  {Le Floc'h}, {Papovich}, {P{\'e}rez-Gonz{\'a}lez}, {Smith}, {Su}, {Bennett},
  {Frayer}, {Henderson}, {Lu}, {Masci}, {Pesenson}, {Rebull}, {Rho}, {Keene},
  {Stolovy}, {Wachter}, {Wheaton}, {Werner}, \& {Richards}}]{Rieke2004MIPS}
{Rieke}, G.~H., {Young}, E.~T., {Engelbracht}, C.~W., {et~al.} 2004, \apjs,
  154, 25

\bibitem[{{Sakamoto} {et~al.}(2017){Sakamoto}, {Aalto}, {Barcos-Mu{\~n}oz},
  {Costagliola}, {Evans}, {Harada}, {Mart{\'\i}n}, {Wiedner}, \&
  {Wilner}}]{Sakamoto2017}
{Sakamoto}, K., {Aalto}, S., {Barcos-Mu{\~n}oz}, L., {et~al.} 2017, \apj, 849,
  14

\bibitem[{{Sakamoto} {et~al.}(2013){Sakamoto}, {Aalto}, {Costagliola},
  {Mart{\'\i}n}, {Ohyama}, {Wiedner}, \& {Wilner}}]{Sakamoto2013}
{Sakamoto}, K., {Aalto}, S., {Costagliola}, F., {et~al.} 2013, \apj, 764, 42

\bibitem[{{Sanders} {et~al.}(2003){Sanders}, {Mazzarella}, {Kim}, {Surace}, \&
  {Soifer}}]{SandersRBGS}
{Sanders}, D.~B., {Mazzarella}, J.~M., {Kim}, D.-C., {Surace}, J.~A., \&
  {Soifer}, B.~T. 2003, \aj, 126, 1607

\bibitem[{{Sanders} {et~al.}(1988){Sanders}, {Soifer}, {Elias}, {Madore},
  {Matthews}, {Neugebauer}, \& {Scoville}}]{Sanders1988}
{Sanders}, D.~B., {Soifer}, B.~T., {Elias}, J.~H., {et~al.} 1988, \apj, 325, 74

\bibitem[{{Scoville} {et~al.}(2017){Scoville}, {Murchikova}, {Walter},
  {Vlahakis}, {Koda}, {Vanden Bout}, {Barnes}, {Hernquist}, {Sheth}, {Yun},
  {Sanders}, {Armus}, {Cox}, {Thompson}, {Robertson}, {Zschaechner}, {Tacconi},
  {Torrey}, {Hayward}, {Genzel}, {Hopkins}, {van der Werf}, \&
  {Decarli}}]{Scoville2017}
{Scoville}, N., {Murchikova}, L., {Walter}, F., {et~al.} 2017, \apj, 836, 66

\bibitem[{{Sliwa} {et~al.}(2017){Sliwa}, {Wilson}, {Aalto}, \&
  {Privon}}]{Sliwa2017}
{Sliwa}, K., {Wilson}, C.~D., {Aalto}, S., \& {Privon}, G.~C. 2017, \apjl, 840,
  L11

\bibitem[{{Soifer} {et~al.}(2000){Soifer}, {Neugebauer}, {Matthews}, {Egami},
  {Becklin}, {Weinberger}, {Ressler}, {Werner}, {Evans}, {Scoville}, {Surace},
  \& {Condon}}]{Soifer2000}
{Soifer}, B.~T., {Neugebauer}, G., {Matthews}, K., {et~al.} 2000, \aj, 119, 509

\bibitem[{{Spoon} {et~al.}(2013){Spoon}, {Farrah}, {Lebouteiller},
  {Gonz{\'a}lez-Alfonso}, {Bernard-Salas}, {Urrutia}, {Rigopoulou},
  {Westmoquette}, {Smith}, {Afonso}, {Pearson}, {Cormier}, {Efstathiou},
  {Borys}, {Verma}, {Etxaluze}, \& {Clements}}]{Spoon2013}
{Spoon}, H.~W.~W., {Farrah}, D., {Lebouteiller}, V., {et~al.} 2013, \apj, 775,
  127

\bibitem[{{Spoon} {et~al.}(2007){Spoon}, {Marshall}, {Houck}, {Elitzur}, {Hao},
  {Armus}, {Brandl}, \& {Charmandaris}}]{Spoon07}
{Spoon}, H.~W.~W., {Marshall}, J.~A., {Houck}, J.~R., {et~al.} 2007, \apjl,
  654, L49

\bibitem[{{Springel} {et~al.}(2005){Springel}, {Di Matteo}, \&
  {Hernquist}}]{Springel2005}
{Springel}, V., {Di Matteo}, T., \& {Hernquist}, L. 2005, \mnras, 361, 776

\bibitem[{{Stanghellini} {et~al.}(1998){Stanghellini}, {O'Dea}, {Dallacasa},
  {Baum}, {Fanti}, \& {Fanti}}]{Stanghellini1998}
{Stanghellini}, C., {O'Dea}, C.~P., {Dallacasa}, D., {et~al.} 1998, \aaps, 131,
  303

\bibitem[{{Sturm} {et~al.}(2011){Sturm}, {Gonz{\'a}lez-Alfonso}, {Veilleux},
  {Fischer}, {Graci{\'a}-Carpio}, {Hailey-Dunsheath}, {Contursi}, {Poglitsch},
  {Sternberg}, {Davies}, {Genzel}, {Lutz}, {Tacconi}, {Verma}, {Maiolino}, \&
  {de Jong}}]{Sturm2011}
{Sturm}, E., {Gonz{\'a}lez-Alfonso}, E., {Veilleux}, S., {et~al.} 2011, \apjl,
  733, L16

\bibitem[{{Teng} {et~al.}(2015){Teng}, {Rigby}, {Stern}, {Ptak}, {Alexander},
  {Bauer}, {Boggs}, {Brandt}, {Christensen}, {Comastri}, {Craig}, {Farrah},
  {Gandhi}, {Hailey}, {Harrison}, {Hickox}, {Koss}, {Luo}, {Treister}, \&
  {Zhang}}]{Teng2015}
{Teng}, S.~H., {Rigby}, J.~R., {Stern}, D., {et~al.} 2015, \apj, 814, 56

\bibitem[{{Thompson} {et~al.}(2005){Thompson}, {Quataert}, \&
  {Murray}}]{Thompson2005}
{Thompson}, T.~A., {Quataert}, E., \& {Murray}, N. 2005, \apj, 630, 167

\bibitem[{{Ueda} {et~al.}(2014){Ueda}, {Iono}, {Yun}, {Crocker}, {Narayanan},
  {Komugi}, {Espada}, {Hatsukade}, {Kaneko}, {Matsuda}, {Tamura}, {Wilner},
  {Kawabe}, \& {Pan}}]{Ueda2014}
{Ueda}, J., {Iono}, D., {Yun}, M.~S., {et~al.} 2014, \apjs, 214, 1

\bibitem[{{Veilleux} {et~al.}(2009){Veilleux}, {Rupke}, {Kim}, {Genzel},
  {Sturm}, {Lutz}, {Contursi}, {Schweitzer}, {Tacconi}, {Netzer}, {Sternberg},
  {Mihos}, {Baker}, {Mazzarella}, {Lord}, {Sanders}, {Stockton}, {Joseph}, \&
  {Barnes}}]{Veilleux2009}
{Veilleux}, S., {Rupke}, D.~S.~N., {Kim}, D.-C., {et~al.} 2009, \apjs, 182, 628

\bibitem[{{Weigel} {et~al.}(2017){Weigel}, {Schawinski}, {Caplar}, {Carpineti},
  {Hart}, {Kaviraj}, {Keel}, {Kruk}, {Lintott}, {Nichol}, {Simmons}, \&
  {Smethurst}}]{Weigel2017}
{Weigel}, A.~K., {Schawinski}, K., {Caplar}, N., {et~al.} 2017, \apj, 845, 145

\bibitem[{{Weinberger} {et~al.}(2018){Weinberger}, {Springel}, {Pakmor},
  {Nelson}, {Genel}, {Pillepich}, {Vogelsberger}, {Marinacci}, {Naiman},
  {Torrey}, \& {Hernquist}}]{Weinberger2018}
{Weinberger}, R., {Springel}, V., {Pakmor}, R., {et~al.} 2018, \mnras, 479,
  4056

\bibitem[{{Wilson} {et~al.}(2014){Wilson}, {Rangwala}, {Glenn}, {Maloney},
  {Spinoglio}, \& {Pereira-Santaella}}]{Wilson2014}
{Wilson}, C.~D., {Rangwala}, N., {Glenn}, J., {et~al.} 2014, \apjl, 789, L36

\bibitem[{{Wright} {et~al.}(1994){Wright}, {Griffith}, {Burke}, \&
  {Ekers}}]{Wright1994}
{Wright}, A.~E., {Griffith}, M.~R., {Burke}, B.~F., \& {Ekers}, R.~D. 1994,
  \apjs, 91, 111

\bibitem[{{Xu} {et~al.}(2015){Xu}, {Cao}, {Lu}, {Gao}, {Diaz-Santos},
  {Herrero-Illana}, {Meijerink}, {Privon}, {Zhao}, {Evans}, {K{\"o}nig},
  {Mazzarella}, {Aalto}, {Appleton}, {Armus}, {Charmandaris}, {Chu}, {Haan},
  {Inami}, {Murphy}, {Sanders}, {Schulz}, \& {van der Werf}}]{Xu2015}
{Xu}, C.~K., {Cao}, C., {Lu}, N., {et~al.} 2015, \apj, 799, 11

\end{thebibliography}
\end{document}